\documentclass[fleqn,usenatbib]{mnras}
\usepackage{newtxtext,newtxmath}
\usepackage[T1]{fontenc}
\usepackage{ae,aecompl}
\usepackage{bigints}

\usepackage{graphicx}	
\usepackage{amsmath}	
\usepackage{bigints}
\usepackage{natbib}
\usepackage{caption}
\usepackage{ragged2e}
\usepackage{placeins}
\usepackage{bigints} 
\usepackage{physics}
\usepackage{cleveref}
\usepackage[export]{adjustbox}
\usepackage{longtable}
\usepackage{xtab}
\usepackage{bm}
\usepackage[mathscr]{eucal}
\usepackage{comment}
\usepackage{multirow}
\usepackage{tabularx}
\usepackage[table,xcdraw]{xcolor}
\usepackage{setspace}
\usepackage{longtable,rotating}
\usepackage{threeparttable}
\usepackage{booktabs}
\usepackage{tcolorbox}
\usepackage{mathtools,leftindex,tensor}
\usepackage[version=4]{mhchem}
\usepackage{makecell}
\usepackage{listings}
\usepackage{anyfontsize}
\DeclareRobustCommand{\ttfamily}{\fontencoding{T1}\fontfamily{lmtt}\selectfont}
\usepackage[normalem]{ulem}
\usepackage[frozencache=true,cachedir=minted-cache]{minted}
\tcbuselibrary{minted,breakable,xparse,skins}

\definecolor{bg}{gray}{0.95}
\DeclareTCBListing{mintedbox}{O{}m!O{}}{%
  breakable=true,
  listing engine=minted,
  listing only,
  minted language=#2,
  minted style=default,
  minted options={%
    linenos,
    gobble=0,
    breaklines=true,
    breakafter=,,
    fontsize=\small,
    numbersep=8pt,
    #1},
  boxsep=0pt,
  left skip=0pt,
  right skip=0pt,
  left=25pt,
  right=0pt,
  top=3pt,
  bottom=3pt,
  arc=5pt,
  leftrule=0pt,
  rightrule=0pt,
  bottomrule=2pt,
  toprule=2pt,
  colback=bg,
  colframe=orange!70,
  enhanced,
  overlay={%
    \begin{tcbclipinterior}
    \fill[orange!20!white] (frame.south west) rectangle ([xshift=20pt]frame.north west);
    \end{tcbclipinterior}},
  #3}
    
\usepackage{hyperref}
\hypersetup{
    colorlinks=true,
    linkcolor=blue,
    filecolor=magenta,      
    urlcolor=cyan,
    pdftitle={Beyond radial profiles: probability distributions to model the CGM},
    pdfpagemode=FullScreen,
    }

\usepackage{CJKutf8}

\usepackage{tikz} 
\usetikzlibrary{shapes.geometric,arrows.meta} 

\usepackage{scalerel,tikz}
\usetikzlibrary{svg.path}
\definecolor{orcidlogocol}{HTML}{A6CE39}
\tikzset{orcidlogo/.pic={
 \fill[orcidlogocol] svg{M256,128c0,70.7-57.3,128-128,128C57.3,256,0,198.7,0,128C0,57.3,57.3,0,128,0C198.7,0,256,57.3,256,128z};
 \fill[white] svg{M86.3,186.2H70.9V79.1h15.4v48.4V186.2z}
 svg{M108.9,79.1h41.6c39.6,0,57,28.3,57,53.6c0,27.5-21.5,53.6-56.8,53.6h-41.8V79.1z M124.3,172.4h24.5c34.9,0,42.9-26.5,42.9-39.7c0-21.5-13.7-39.7-43.7-39.7h-23.7V172.4z}
 svg{M88.7,56.8c0,5.5-4.5,10.1-10.1,10.1c-5.6,0-10.1-4.6-10.1-10.1c0-5.6,4.5-10.1,10.1-10.1C84.2,46.7,88.7,51.3,88.7,56.8z};
}}
\newcommand\orcidicon[1]{\href{https://orcid.org/#1}{\mbox{\scalerel*{
\begin{tikzpicture}[yscale=-1,transform shape]
\pic{orcidlogo};
\end{tikzpicture}
}{|}}}}

\defcitealias{Faerman2017ApJ}{FSM17}
\defcitealias{Faerman2019ApJ}{FSM20}

\newcommand{\ps}[1]{{ #1}}

\newcommand{\alankar}[1]{{ #1}}

\title[Log-normal distributions to model the CGM]{Beyond radial profiles: Using log-normal distributions to model the multiphase circumgalactic medium}

\author[Dutta et al.]{
\newauthor
Alankar Dutta$^{\orcidicon{0000-0002-9287-4033}\,1}$\thanks{E-mail: alankardutta@iisc.ac.in (AD)}, 
Mukesh Singh Bisht$^{\orcidicon{0000-0002-1497-4645}\,2}$\thanks{E-mail: msbisht@rrimail.rri.res.in (MB)}, 
Prateek Sharma$^{\orcidicon{0000-0003-2635-4643}\,1}$\thanks{E-mail: prateek@iisc.ac.in (PS)}, 
Ritali Ghosh$^{\orcidicon{0000-0001-8643-7104}\,1}$, 
\newauthor
Manami Roy$^{\orcidicon{0000-0001-9567-8807}\,2}$, 
and 
Biman B. Nath$^{\orcidicon{0000-0003-1922-9406}\,2}$ 
\\
$^{1}$Department of Physics, Indian Institute of Science, Bangalore, KA 560012, India \\
$^{2}$Raman Research Institute, Bangalore, KA 560080, India\\
}

\date{Accepted XXX. Received YYY; in original form ZZZ}

\pubyear{2022}

\begin{document}
\label{firstpage}
\pagerange{\pageref{firstpage}--\pageref{lastpage}}
\maketitle

\begin{abstract}
Recent observations and simulations reveal that the circumgalactic medium (CGM) surrounding galaxies is multiphase, with the gas temperatures spanning a wide range at most radii, $\sim 10^4\ {\rm K}$ to the virial temperature ($\sim 10^6$ K for Milky Way). Traditional CGM models using simple density profiles are inadequate \ps{at reproducing observations that indicate a broad temperature range.} 
Alternatively, a model based on probability distribution functions (PDFs) with parameters motivated by 
\ps{simulations} can better match multi-wavelength observations. In this work, we use log-normal distributions, commonly seen in the simulations of the multiphase interstellar and circumgalactic media, to 
\ps{model the multiphase CGM}. 
We generalize the isothermal background model by Faerman et al. 2017 to include more general CGM profiles.
We extend the existing probabilistic models from 1D-PDFs in temperature to 2D-PDFs in density-temperature phase space and constrain its parameters using a Milky Way-like {\tt Illustris TNG50-1} halo. We generate various synthetic observables such as column densities of different ions, UV/X-ray spectra, and dispersion and emission measures. X-ray and radio (Fast Radio Burst) observations \alankar{mainly} constrain the hot gas properties. However, interpreting cold/warm phase diagnostics is not straightforward since these phases are 
patchy, with inherent variability in intercepting these clouds along arbitrary lines of sight. We provide a tabulated comparison of model predictions with observations and plan to expand this into a comprehensive compilation of models and data. 
Our modeling provides a simple analytic framework that is useful for describing important aspects of the multiphase CGM.
\end{abstract}




\section{Introduction}\label{sec:introduction}

Several independent observational probes over the last decade have uncovered the circumgalactic medium (CGM hereafter), the diffuse atmospheres around galaxies like our Milky Way (for a review, see \citealt{Tumlinson2017review} and \citealt{2023ARA&A..61..131F}). Being diffuse, the CGM is hard to detect but is the major baryonic component of the galactic halos, making up to a few times more mass than the combined mass of the stars and the interstellar medium (ISM; 
\citealt{Werk2014, Das2020}). 

Traditionally, the CGM is modeled with a parametric density profile/distribution of the volume-filling hot phase (\citealt{Maller2004, Henley_2010, Sharma2012, Gupta2012, Miller2013, Mathews2017, Yao2017, 10.1093/mnras/stz1859, Yamasaki2020, Faerman2019ApJ}). This approach is incomplete because observations (e.g., \citealt{Werk2014, Tumlinson_2013}) show that most of the sightlines intercept not only the hot phase but also the ions tracing the cold/warm phase ($10^4 - 10^{5.5}$ K). Traditional models that only include the hot phase, therefore, fail to explain the ubiquity of the cooler phases in observations. Presently, there are only piecemeal physical models to account for the cold/warm phases. For instance, to explain the observed OVI (O$^{+5}$) column densities, \citealt{Faerman2017ApJ} propose an ad hoc introduction of a $10^{5.5}$ K phase. Likewise, \citealt{Faerman2023} presume a fixed volume fraction for the cold phase in ionization/thermal equilibrium. Since the cold phase has low thermal pressure, it is necessary to have a large non-thermal support to maintain the total pressure balance with the hot phase. However, the conclusions drawn from such models depend heavily on the model assumptions. Therefore, generating multi-wavelength observables from a diverse range of physical models is imperative.

To address the limitation of simple profiles to describe a multiphase CGM, we generalize the model of \citet{Faerman2017ApJ} based on a log-normal volume distribution across the complete range of CGM temperatures. This model is physically well-motivated, since log-normal distribution of densities and temperatures are routinely inferred from observations and simulations of different phases of the ISM (\citealt{Kortgen2017, Chen2018}) and \alankar{the} CGM (\citealt{Das2021, Vijayan2022, Mohapatra2022characterising}). \ps{Further, a log-normal temperature distribution captures the ``core" (up to a quadratic order Taylor series expansion in $\ln T$) of a generic peaked temperature distribution.} Because of the central limit theorem, log-normal distributions are understood as a natural outcome of generic multiplicative random walk processes. Additionally, log-normal distributions are a useful choice for modeling the multiphase CGM, as the weighted integrals of log-normal PDFs (probability distribution functions) 
involved in calculating different synthetic observables are analytically tractable. 

Since each observational probe gives a somewhat biased view of the CGM, there is a need for a library of CGM models that can be tested against \ps{diverse} multi-wavelength observations, ranging from radio (\citealt{Bhardwaj2021, Wu2022}), mm (\citealt{Singh2018, Amodeo2021}), UV (\citealt{2003ASSL..281..183W, Wakker_2009, Werk2014, 10.1093/mnras/sty529, Burchett_2019, 10.1093/mnras/staa1773, 2020ApJ...900....9L, Cooper2021, Haislmaier2021, Sameer2021}), optical (\citealt{2015MNRAS.446.2861K, 2015ApJ...804...79L, 2017ApJ...834..148N, 2019MNRAS.485.1961Z, Peroux19, 10.1093/mnras/staa3534}) to X-rays (\citealt{Gupta2012, Kaaret2020, Ponti2022}).
Observations (e.g., \citealt{Werk2014, Kacprzak_2015, Das2021, Bluem2022, 2024arXiv240208016Q}) and numerical simulations (e.g., \citealt{2015MNRAS.446..521S, Oppenheimer18, 10.1093/mnrasl/sly190, 10.1093/mnras/stz1859, Hummels2019, Nelson2020, 10.1093/mnras/staa358, 10.1093/mnras/stab2896, DeFelippis_2021, Augustin21, 10.1093/mnras/stad2637}) reveal that the CGM has a complex \alankar{kinematic and} multiphase structure, with multiple temperature components \alankar{with different line of sight velocities} present along most sightlines. This highlights the need for models that go beyond simple \ps{radial} density\alankar{/temperature profiles. In this work, w}e demonstrate that a probabilistic description utilizing distribution functions can better describe the multiphase CGM (the existing work in this direction is rather limited; e.g., \citealt{Faerman2017ApJ, Vijayan2022}).

We generalize the \citet{Faerman2017ApJ} (\citetalias{Faerman2017ApJ} hereafter) model of the CGM to allow for different phases in any generic thermodynamic state.
In the \citetalias{Faerman2017ApJ} model, the baseline isothermal log-normal PDF of the hot phase is modified to include an additional warm log-normal component at $10^{5.5}$ K. To demonstrate the flexibility of our generalization, we replace the isothermal baseline model of \citetalias{Faerman2017ApJ} with an isentropic profile (following \citealt{Faerman2019ApJ}) but retain the same prescription for the modified warm component. Our generalization can incorporate any arbitrary thermodynamic relation between phases. Specifically, we consider the hot and warm phases to be either isochoric or isobaric with respect to each other. We then compare the observables from all these models (OVI/VII/VIII/NV column densities, X-ray emission measure [EM], emission spectra, and dispersion measure [DM]) with the observed data. 

We next construct a one-zone, three-phase model comprising hot ($\sim 10^6$ K), warm ($\sim 10^5$ K), and cold ($\sim 10^4$ K) gas. Each phase is modeled as a two-dimensional log-normal PDF in the density-temperature ($\rho-T$) phase space. We adjust the volume fraction and the median density and temperature along with their corresponding spreads for each phase to match the $\rho-T$ distribution measured for a Milky Way-like halo from the {\tt Illustris TNG50-1} cosmological simulation (\citealt{Nelson2020}). From this fitted three-phase model, we generate column densities of OVI and MgII (the warm and cold gas tracers, respectively), 
estimate the EM, DM, and X-ray surface brightness, and compare with the observations. Additionally, we formulate a simple, \alankar{approximate} prescription to move beyond the one-zone approximation in our three-phase model of the CGM. 

Our paper is organized as follows. Section \ref{sec:faerman-generalize} explains the two-phase \citetalias{Faerman2017ApJ} model and introduces our generalized framework and the integrals needed to generate observables from the PDFs. Section \ref{sec:threePhase} introduces the three-phase model with 2D (in density-temperature space) log-normal distribution for each phase, calibrated with a simulated Milky Way-like halo from {\tt Illustris TNG50-1} cosmological simulation. Section \ref{sec:discussion} discusses the implications of our work for the multiphase CGM, in particular, the influence of warm/cold clouds with a small volume filling fraction. Section \ref{sec:summary} concludes with a summary of our work.

\section{Generalized Faerman models}
\label{sec:faerman-generalize}

\begin{figure}
\includegraphics[width=1.0\columnwidth]{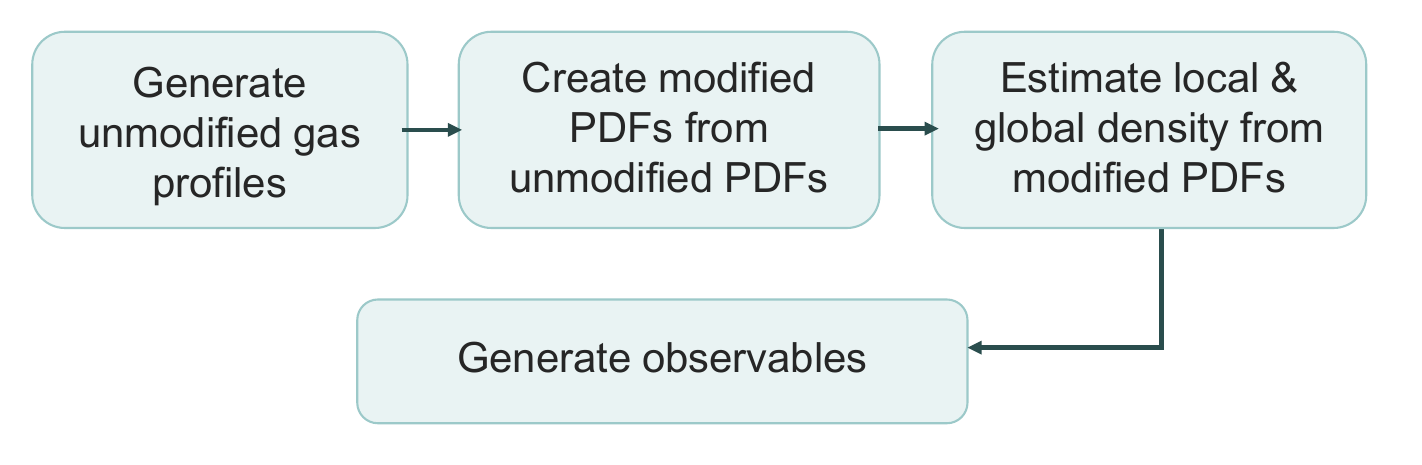}
\caption{Flowchart of the procedure proposed in \citetalias{Faerman2017ApJ} to produce a warm phase from an isothermal unmodified profile (see section \ref{subsec:FSM17_description}). We generalize the \citetalias{Faerman2017ApJ} procedure to model any generic thermodynamic profile of the 
CGM (see section \ref{subsec:isoThvsisoEnt}) with the flexibility to have an arbitrary pressure-density relation across the phases (see section \ref{subsec:isoBarvsisoChor}). 
Synthetic observables can be generated to match against observations after profiles are computed (see section \ref{subsec:gen_obs}).
}
\label{fig:algorithm}
\end{figure}

In this section, we elaborate on the model introduced by \citetalias{Faerman2017ApJ}. We list the symbols and notation used to reformulate the \citetalias{Faerman2017ApJ} model in Tab. \ref{tab:notation}, crucial for generalizing to a broad class of probabilistic CGM models. Fig. \ref{fig:algorithm} outlines the steps discussed in section \ref{subsec:FSM17_description} for generating these probabilistic models. As a specific example, in section \ref{subsec:isoThvsisoEnt} we replace the isothermal hot gas profile in \citetalias{Faerman2017ApJ} with an isentropic profile but retain the same prescription for the warm gas. In section \ref{subsec:isoBarvsisoChor}, we emphasize the implication of choosing isochoric or isobaric radiative cooling to generate the warm phase. The thermodynamic state of the hot and the warm gas used in these models can significantly alter synthetic observables, as discussed in section \ref{subsec:gen_obs}.

\subsection{FSM17 model for the CGM}
\label{subsec:FSM17_description}

Here we list the different steps involved in using the \citetalias{Faerman2017ApJ} model to study the CGM (see Fig. \ref{fig:algorithm}):
\begin{enumerate}
    \item {\em Create unmodified hydrostatic profiles} for the volume-filling gas in the CGM. We denote unmodified profiles with superscript $(u)$, such as $p^{(u)}$, $\rho^{(u)}$, $T^{(u)}$. Such unmodified profiles can be prescribed by any generic CGM model, like the precipitation model (\citealt{Sharma2012, Voit2019}) or the isentropic model (\citealt{Faerman2019ApJ}). \citetalias{Faerman2017ApJ} used isothermal gas in hydrostatic equilibrium as the unmodified profile. The total pressure $p_{\rm tot}$ considered by \citetalias{Faerman2017ApJ} has thermal, non-thermal (e.g., due to magnetic fields and cosmic rays), and turbulent components such that $p_{\rm tot} = p_{\rm th}+p_{\rm nth} + p_{\rm turb}$, where $p_{\rm nth} = (\alpha -1) p_{\rm th}$,  $p_{\rm turb}/p_{\rm th} = (\sigma_{\rm turb}/\sigma_{\rm th})^2$, and $p_{\rm th} = p_{\rm tot}/(\alpha + \sigma_{\rm turb}^2/\sigma_{\rm th}^2)$. 
    In this isothermal model, the total pressure is given by the simple hydrostatic solution,
    \begin{equation}
    \label{eq:1phaseisoT}
    p_{  \rm tot}^{(u)}(r) = p_{0,\rm tot}^{(u)} \exp( -\frac{[\Phi(r)-\Phi_0]}{c_t^2}),
    \end{equation}
    where $c_t$ is the effective isothermal sound speed, taking into account the contributions from the non-thermal and turbulent pressures, and the subscript $0$ refers to a reference radius with the total pressure $p_{0,\rm tot}^{(u)}$ and the potential $\Phi_0$. Therefore, the effective sound speed in Eq. (\ref{eq:1phaseisoT}) is $c_t^2 = \alpha \sigma _{\rm th}^2+\sigma _{\rm turb}^2 $ (here we use the same notation as  \citetalias{Faerman2017ApJ}). \ps{Tab. \ref{tab:fsm_params} lists our parameter values, which are taken from the fiducial models in \citetalias{Faerman2017ApJ} and \citetalias{Faerman2019ApJ}, chosen mainly to match the column density of OVI}. The non-thermal pressure support due to cosmic rays (e.g., \citealt{Salem2015, Ji2020MNRAS, Butsky2022}) or turbulence (easier to detect in clusters, e.g., \citealt{Li2020ApJ, Zuhone2018, Mohapatra2021MNRAS} 
    than in CGM\alankar{, e.g., }\citealt{Buie2020, Buie2020b, Chen2023}) \alankar{in the CGM can be significant.}
\begin{table}
\centering
\caption{{\fontsize{9pt}{9.2pt}\selectfont Symbols \& notation used for probabilistic models}} 
\label{tab:notation}
\setlength{\tabcolsep}{0.3em} 
{\renewcommand{\arraystretch}{1.1}
\begin{tabularx}{\linewidth}{cc}
\hline
{\fontsize{9pt}{9.2pt}\selectfont Symbol(s)} & {\fontsize{9pt}{9.2pt}\selectfont Meaning} \\ \hline
$r$ &  distance from the halo center \\
$\rho^{(u)},\ p^{(u)},\ T^{(u)}$ & unmodified profiles of hot CGM \\
\rule{0pt}{3ex}
\multirow{2}{*}{$\mathscr{P}_V^{(u)}(T),\ \mathscr{P}_M^{(u)}(T)$} &  \multirow{2}{*}{\begin{tabular}[c]{@{}c@{}} temperature-PDF (T-PDF) of unmodified gas \\ weighed by volume ($V$) or mass ($M$) \end{tabular}}\\
 &  \\
 \rule{0pt}{3ex}
\multirow{2}{*}{$\mathscr{P}_{V}^{(w)} (T),\ \mathscr{P}_{V}^{(h)}(T)$} &  \multirow{2}{*}{\begin{tabular}[c]{@{}c@{}} modified T-PDF weighed by $V$ of the warm \\ or hot phases\end{tabular}}\\
 &  \\
\rule{0pt}{3ex}
$\langle \rangle,\ \langle \rangle_M$ & average weighed by $V$ or $M$ \\ 
\rule{0pt}{3ex}
\multirow{2}{*}{$T_{{\rm  med},V}^{(u)},\ T_{{\rm  med},V}^{(w)}$} &  \multirow{2}{*}{\begin{tabular}[c]{@{}c@{}} median temperature $T$ of unmodified or warm gas \\ T-PDF weighed by $V$ \end{tabular}}\\
 &  \\
${\cal N}(x,\sigma)$ & Gaussian PDF, mean 0 and standard deviation $\sigma$ \\
${\cal N}(x,\mu,\sigma)$ & Gaussian PDF, mean $\mu$ and standard deviation $\sigma$ \\
\rule{0pt}{3ex}
\multirow{2}{*}{$\sigma _u$, $\sigma _w$} & \multirow{2}{*}{\begin{tabular}[c]{@{}c@{}} standard deviations of the unmodified or warm \\ gas T-PDFs  \end{tabular}}\\
 &  \\
\rule{0pt}{3ex}
\multirow{2}{*}{$T_c$} & \multirow{2}{*}{\begin{tabular}[c]{@{}c@{}}cutoff temperature below which unmodified PDF \\ is modified ($t_{\rm cool}(T_c)/t_{\rm ff} =$  cutoff )\end{tabular}} \\
 &  \\
 \rule{0pt}{3ex}
\multirow{2}{*}{$dM^{(i)},\ dV^{(i)}$} & \multirow{2}{*}{\begin{tabular}[c]{@{}c@{}} mass, volume for a phase in temperature \\ range $[T,T+dT]$  \end{tabular}} \\
 &  \\
\rule{0pt}{3ex}
$dM=\Sigma dM^{(i)}$ & total mass in range $[T,T+dT]$ \\
$dV=\Sigma dV^{(i)}$ & total volume in range $[T,T+dT]$ \\
$M^{(i)},\ V^{(i)}$ & total mass, volume for a phase across all $T$s \\
$M=\Sigma M^{(i)}$ & total mass including all phases across all $T$s \\
$V=\Sigma V^{(i)}$ & total volume including all phases across all $T$s \\[0.5ex]
$f_M^{(i)}=M^{(i)}/M$ & mass fraction in a phase \\[0.5ex]
$f_V^{(i)} = V^{(i)}/V$ & volume fraction in a phase \\[0.5ex]
$\langle \rho^{(i)} \rangle = M^{(i)}/V^{(i)}$ & local average density of a particular phase \\
$\langle \rho^{(i)} \rangle_g = M^{(i)}/V$ & global average density of a particular phase \tabularnewline[0.5ex]
\hline 
\end{tabularx}%
}
\end{table}
    
    The thermal pressure and density are related by the ideal gas equation of state $p_{\rm th}^{(u)} = \rho^{(u)} k_B T^{(u)} /\mu m_p $. The temperature $T^{(u)}$ here is the mass-weighted average temperature at each radius, which also determines the thermal broadening $\sigma _{\rm th}^2 = k_B T^{(u)}/\mu m_p$ of the unmodified gas. The temperature is assumed to be constant in the unmodified profile used in \citetalias{Faerman2017ApJ}. 
    We later also use an isentropic unmodified profile (see section \ref{subsec:isoThvsisoEnt}) from \citealt{Faerman2019ApJ} (\citetalias{Faerman2019ApJ} hereafter) to illustrate the procedure \ps{for a general unmodified profile}.

    \item {\em Unmodified log-normal temperature-PDFs}: After obtaining the unmodified profiles, \citetalias{Faerman2017ApJ} assumes a log-normal volume distribution of temperature in each radial shell of the CGM. The motivation behind this is that the non-thermal and turbulent processes in the CGM cause fluctuations around the unmodified gas properties \ps{and results in a locally peaked temperature/density distribution}. Therefore, we describe the gas temperature distribution in each shell 
    by a log-normal PDF and multiphase gas exists co-spatially at \ps{all radii}. A perfectly uniform multiphase shell is an approximation, and the cooler phases are expected to be spatially inhomogeneous (e.g., see Figure \ref{fig:cartoon}).
    \begin{figure}
	\centering
	\includegraphics[width=1.00\columnwidth]{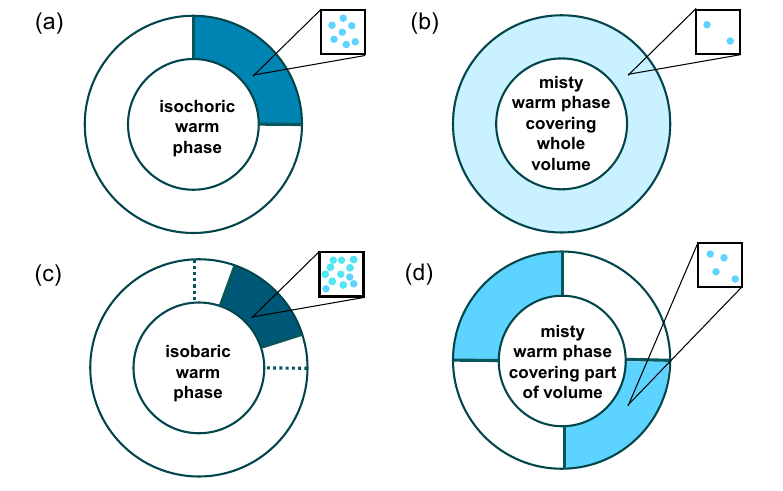}
	\caption{Cartoon illustrations of various kinds of modifications to introduce warm gas in any shell of our CGM model: the warm phase condenses out of the hot phase and is either (a) {\em isochoric} or (c) {\em isobaric} relative to the hot phase. Being isobaric with the hot phase, the warm phase has a higher density and occupies a smaller volume, as indicated by the darker shade of blue in (c). The warm gas may occupy only a part of the shell volume (as in (a), (c), and (d)) rather than being distributed uniformly (the {\em mist limit}; shown in (b)). The inset squares in the top right corner of each sub-figure indicate the densities of warm gas in different cases.
 }
	\label{fig:cartoon}
\end{figure}
    We quantify the amount of such co-spatial multiphase gas at any radius $r$ by the volume/mass fraction of the gas in the temperature range $[T, T+dT]$. The gas distribution is assumed to be log-normal (in volume and consequently \ps{also} in mass) at all radii from the center of the CGM and the volume \alankar{PDF} (see the light red line in Figure \ref{fig:isothermal_isochoric_mod}) can be expressed as 
    \begin{equation}
    \label{eq:V_u}
    \mathscr{P}_V^{(u)}(T) = \frac{1}{T} {\cal N}(x, \sigma _u), 
    \end{equation}
    where $x \equiv \ln (T/T_{{ \rm med}, V}^{(u)})$, $T_{{  \rm med}, V}^{(u)}$ is the median temperature in the volume PDF (since $\mathscr{P}_V^{(u)}(T) dT = {\cal N}(x, \sigma _u) dx$; see Tab.~\ref{tab:notation} for the notation), and
    \begin{equation}
    \label{eq:Normal}
        {\cal N}(x,\sigma _u) = \frac{1}{\sqrt{2\pi} \sigma _u} \exp(-\frac{x^2}{2 \sigma _u^2})
    \end{equation}
    is the normal distribution with zero mean and standard deviation $\sigma _u$. Note that the distribution in $\ln T$ space, $\mathscr{P}_V^{(u)}(\ln [T/T_{{\rm med},V}]) = T \mathscr{P}_V^{(u)}(T) = {\cal N}^{(u)}(x,\sigma_u)$, is Gaussian. In the \citetalias{Faerman2017ApJ} prescription, the unmodified gas at every radius is assumed to be isobaric so that the product of density and temperature within a phase is constant. This is an approximation of the thermodynamic state of the gas. In general, density and temperature are independent, and 2D PDFs (say in density and temperature) are necessary to describe the gas distribution (cf. section \ref{sec:threePhase}). 

\begin{table}
    \centering
    \caption{{\fontsize{9pt}{9.2pt}\selectfont Parameters of the two-phase CGM models}}
    \setlength{\tabcolsep}{0.5em} 
    {\renewcommand{\arraystretch}{0.9}
    \begin{tabular}{lcc}
    \hline
    \multicolumn{1}{c}{Input Parameters} & $\gamma = 1$ & $\gamma = 5/3$ \\ \hline
    \addlinespace[1ex]
    \ps{halo mass} 
    $M_{\rm 200}$ & $\rm 10^{12}\ M_{\odot}$ & $\rm 10^{12}\ M_{\odot}$ \\
    \ps{CGM radius} $r_{\rm CGM}$ & $ 1.1 \times {}^{\dag} r_{\rm 200}$ & $ 1.1 \times r_{\rm 200}$ \\
    $p_{\rm 0,tot}^{(u)}$ & $\rm 4580\ {\rm K\ cm^{-3}}$ & - \\
    $r_0$ (reference radius) & $\rm 8.5\ {\rm kpc}$ & $\rm 8.5\ {\rm kpc}$ \\
    $\alpha \equiv 1+p_{\rm nt} / p_{\rm th}$ & 1.9 & 2.1 \\
    $T(r_{\rm CGM})$ & - & $\rm 2.4 \times 10^5 \ {\rm K}$ \\
    $n_H(r_{\rm CGM})$ & - & $\rm 1.1 \times 10^{-5} \ {\rm cm^{-3}}$ \\
    $\sigma_{\rm turb}$ & $\rm 60 \ {\rm km\ s^{-1}}$ & $\rm 60 \ {\rm km\ s^{-1}}$ \\
    $^{\dag\dag}Z(r_{\rm CGM})$ & 0.3 & 0.3 \\
    $Z_0\equiv Z(r_0)$ & 1 & 1 \\
    redshift & 0.2 & 0.2 \\
    cutoff ($t_{\rm cool}/t_{\rm ff}$) & 4 & 4 \\
    $^{\S\S}\sigma_u,\ \sigma_h,\ \sigma_w$ & 0.3 & 0.3 \\
    $T_{\rm med,V}^{(u)}$ & $\rm 1.5 \times 10^6\ K$ & - \\
    $T_{\rm med,V}^{(h)}$ & $\rm 1.5 \times 10^6 \ K$ & - \\
    $T_{\rm med,V}^{(w)}$ & $\rm 3 \times 10^5 \ K$ & $\rm 3 \times 10^5 \ K$ \\ \addlinespace[1ex] \hline
    \multicolumn{1}{c}{Output Parameters} & $\gamma = 1$ & $\gamma = 5/3$ \\ \hline  \addlinespace[1ex]
    ${}^{\ddag}f_V^{(w)}$  [percent] & 8.5 (IC), 2.8 (IB) & 28.7 (IC), 38.6 (IB) \\ 
    $M_{\rm CGM}^{(h)}$, $M_{\rm CGM}^{(w)}$ {[$M_\odot$]} &  $7.9,1.4 \ \times10^{10}$ & $2.5,1.6 \ \times 10^{10}$   \\ \addlinespace[1ex]
    \bottomrule
    \end{tabular}%
    }
    \begin{tablenotes}
    \item ${}^{\dag}$ {\fontsize{7pt}{7.2pt}\selectfont 
    $r_{\rm 200}$ is the radius within which mean density of the 
    dark matter halo \\
    is 200 times the critical density of the present Universe.}
    \item ${}^{\dag\dag}$ {\fontsize{7pt}{7.2pt}\selectfont We use the metallicity profile shown in Fig. 3 of \citetalias{Faerman2019ApJ}.}
    \item \alankar{ ${}^{\S\S}$ {\fontsize{7pt}{7.2pt}\selectfont 
    Temperature spread denoted as $s$ in \citetalias{Faerman2017ApJ}.}}
    \item ${}^{\ddag}$ {\fontsize{7pt}{7.2pt}\selectfont Total volume fraction in the warm phase within the CGM.}
    \end{tablenotes}
    \label{tab:fsm_params}
    \end{table}

    \par For a shell of volume $V$ and mass $M$ at a radius $r$, the volume fraction of gas in the temperature range $[T, T+dT]$ is $\mathscr{P}_V^{(u)}(T)dT = dV/V$, and the mass fraction is $\mathscr{P}_M^{(u)}(T)dT = dM/M$. The average density for the shell is $\langle \rho^{(u)} \rangle=M/V$ and $dM/dV = \rho^{(u)}$ is the gas density at radius $r$ in the temperature range $[T, T+dT]$. Since $\rho^{(u)} = dM/dV = (M/V)  \mathscr{P}_M^{(u)}(T)dT/(\mathscr{P}_V^{(u)}(T) dT)$, the unmodified mass-PDF $\mathscr{P}_M^{(u)}(T)$ is related to the volume PDF as, 
    \begin{equation} 
    \label{eq:p_M}
        \mathscr{P}_M^{(u)} = \frac{\rho^{(u)}}{\langle \rho^{(u)} \rangle} \mathscr{P}_V^{(u)} = \frac{\langle T^{(u)} \rangle_M}{T} \mathscr{P}_V^{(u)},
    \end{equation}
    where $\langle \rangle$ denotes volume-weighted average and $\langle \rangle_M$ denotes mass-weighted average. The middle expression in Eq. \ref{eq:p_M} relating the mass and volume PDF \alankar{is generic for any thermodynamic process of the internal perturbations within a phase.} However, the rightmost expression assumes that the perturbations within the unmodified gas are isobaric. This isobaric assumption implies $\rho^{(u)} (r) T^{(u)} (r) = \langle \rho ^{(u)} (r) T ^{(u)} (r) \rangle = \langle \rho^{(u)} \rangle (r) \langle T^{(u)} \rangle_M(r)$ and hence gives the rightmost expression in Eq. \ref{eq:p_M}. 
    
    The leftmost and the rightmost expressions in Eq. \ref{eq:p_M} can be integrated over all temperatures to obtain \footnote{{\fontsize{7pt}{7.2pt}\selectfont We use the notation ${\cal N}(x,\mu,\sigma)$ for a normal distribution with mean $\mu$ and standard deviation $\sigma$. With only two arguments, ${\cal N}(x,\sigma)$ stands for a normal distribution with zero mean. Here and elsewhere, we complete the square in expressions to write the product of an exponential and a Gaussian as algebraically convenient shifted Gaussian in the following form,
    \begin{equation}
    \label{eq:comp_sqr_norm}
    e^{ax} {\cal{N}}(x,\mu,\sigma) = e^{a(\mu + a \sigma^2/2)} {\cal N} (x, \mu+a\sigma^2, \sigma).
    \end{equation}
    }}
    $$\langle T^{(u)} \rangle_M ^{-1} = \left \langle \frac{1}{T^{(u)}} \right \rangle = T^{(u)\ \ -1}_{{\rm med},V} \bigintsss dx e^{-x} {\cal N}(x,\sigma_u),$$
    where we use $T = T_{\rm med, V}e^x$.
    Further, on using the square completion given by Eq. \ref{eq:comp_sqr_norm} in the footnote, the  \ps{RHS of the preceding expression} can be simplified to 
    $$T^{(u)\ \ -1}_{{\rm med},V} e^{\sigma_u^2/2} \bigintsss dx {\cal N}(x,-\sigma_u^2,\sigma_u),$$
    \ps{relating} the volume PDF and the mass-weighted average temperature as
    \begin{equation}
        \label{eq:Trelation}
        \langle T^{(u)} \rangle_M = e^{-\sigma _u^2/2} T_{{  \rm med}, V}^{(u)}.
    \end{equation} 

    Consistent with the isobaric assumption introduced earlier, we obtain the mass PDF by combining Eqs. \ref{eq:p_M} and \ref{eq:Trelation} (and using Eq. \ref{eq:comp_sqr_norm}), 
    \begin{equation}
    \label{eq:P_M_u}
    \mathscr{P}_M^{(u)}(T) = \frac{1}{T} \exp\left[-\left(x + \frac{\sigma _u^2}{2}\right)\right] {\cal N}(x,\sigma_u) = \frac{1}{T}{\cal N}(y, \sigma_u),
    \end{equation}
    where $y = x + \sigma _u^2$, $y \equiv \ln (T/T_{{  \rm med},M}^{(u)})$, and $T_{{ \rm med},M}^{(u)} = e^{-\sigma _u^2} T_{{ \rm med}, V}^{(u)}$. This is similar to Eq. \ref{eq:V_u} and is applicable only under the isobaric assumption.
    
    \item {\em Modifying the unmodified PDFs}: \citetalias{Faerman2017ApJ} modifies the unmodified distribution of gas at every radius to incorporate the effects of physical processes such as radiative cooling. \citetalias{Faerman2017ApJ} model considers two temperature phases, namely hot and warm, around which the temperature distribution is log-normal. Further, the assumption is that the coolest/densest unmodified gas at any radius, with the ratio of the cooling time to the free-fall time $t_{\rm cool}/t_{\rm ff}$ smaller than a threshold value (chosen to be 4 \ps{here}), cools isochorically to the warm phase. The cooling time is $t_{\rm cool} = (\gamma -1)^{-1} p/n_H^2 \Lambda [T]$ and the free-fall time is $t_{\rm ff} = \sqrt{2r/g[r]}$, where $p$ is gas thermal pressure, $n_H$ is the hydrogen number density, $\Lambda[T]$ is the cooling function, $r$ is the shell radius and $g[r]$ is the gravitational acceleration.
    This model for the dropout of warm gas from the hot atmosphere is motivated by simulations of thermal instability in gravitationally stratified atmospheres (e.g., \citealt{Choudhury2019}). However, the warm gas is assumed not to cool further.
    The choice of the $t_{\rm cool}/t_{\rm ff}$ threshold determines the cut-off temperature $T_c$ below which the hot gas is assumed to be thermally unstable, and the cooling to the warm phase is assumed to be isochoric. Isochoric cooling preserves the area under the curve of the unmodified PDF undergoing modification (see Fig. \ref{fig:cartoon} for a cartoon and cyan dashed line in Fig. \ref{fig:isothermal_isochoric_mod}).

    The volume fraction of the warm phase at any distance $r$ from the halo center is
    \begin{equation}
        \label{eq:vfracW}
        f_V^{(w)} = \bigintsss_{-\infty }^{x_c} dx {\cal N}(x,\sigma _u) = \frac{1}{2} \left [ 1 + \erf\left (\frac{x_c}{\sqrt{2} \sigma _u} \right ) \right ],
    \end{equation}
    where $x_c=\ln (T_c/T_{{ \rm med}, V}^{(u)})$ corresponds to the temperature $T_c$ below which the unmodified gas cools to produce the warm phase, and we have introduced the integral of a normal distribution,
    $$
    \erf(x) = \frac{2}{\sqrt{\pi}}\bigintsss_0^x dt e^{-t^2}.
    $$
    Similarly, the mass fraction of the warm gas is
    \begin{equation}
        \label{eq:mfracW}
        f_M^{(w)} = \bigintsss_{-\infty}^{y_c} dy {\cal N}(y,\sigma _u) = \frac{1}{2} \left [ 1 + \erf\left (\frac{y_c}{\sqrt{2} \sigma _u} \right ) \right ],
    \end{equation}
    where $y_c= x_c + \sigma _u^2 = \ln (T_c/T_{{\rm  med}, M}^{(u)})$ corresponds to the cut-off temperature $T_c$. Note that $y_c > x_c$ implies $f_M^{(w)} > f_V^{(w)}$ (compare Eqs. \ref{eq:vfracW} \& \ref{eq:mfracW}), which is expected since the warm gas is denser.
    
    Now, we need to decide the thermodynamic condition of the re-distributed gas. The warm phase is assumed to attain a new log-normal distribution about a specified median temperature $T^{(w)}_{{ \rm med}, V}$ (see the vertical blue dotted line in Fig.  \ref{fig:isothermal_isochoric_mod}). Since the warm phase has a short cooling time, it is assumed to be maintained in a steady state, at $T^{(w)}_{{ \rm med}, V}$ by heating due to feedback and/or turbulent mixing. The warm gas volume distribution at any radius $r$ is given by
   \begin{equation}
       \label{eq:Pwarm}
    \mathscr{P} ^{(w)}_V(T) = \frac{f_V^{(w)}}{T} {\cal N}(x, \sigma _w),
   \end{equation}
   where $x \equiv \ln (T/T_{{\rm  med}, V}^{(w)})$, $T_{{  \rm med}, V}^{(w)}$ is the median temperature of the warm gas, and ${\cal N}(x,\sigma_w)$ is the Gaussian distribution with standard deviation $\sigma_w$ (see Eq. \ref{eq:Normal}). Similarly, the mass-PDF for the warm gas is given by
   $$
    \mathscr{P} ^{(w)}_M(T) = \frac{f_M^{(w)}}{T} {\cal N}(y, \sigma_w),
   $$
   where $y = x + \sigma_w^2$, $y \equiv \ln(T/T_{{\rm med},M}^{(w)})$, and $T_{{\rm med}, M}^{(w)} = e^{-\sigma_w^2}T_{{\rm med}, V}^{(w)}$. Note that these expressions are analogous to unmodified PDFs (Eqs. \ref{eq:V_u} to \ref{eq:P_M_u}) since the underlying PDF is the same, namely log-normal.

    The modified hot phase is assumed to maintain the unmodified PDF above the cutoff temperature and is given by 
       \begin{equation}
       \label{eq:Phot}
       \mathscr{P} ^{(h)}_{V}(T) = \frac{1}{T}{\cal N}(x, \sigma _u) \mathscr{H}(x-x_c),
   \end{equation}
   where $x_c$ corresponds to the (radius dependent) cut-off temperature $T_c$ and $\mathscr{H}(x)$ is the Heaviside function (unity for $x>0$ and zero for $x<0$). 

\item {\em Local and global gas density:} It is useful to distinguish between the local and global average mass densities. We define the local gas density for a phase as $\langle \rho^{(i)} \rangle \equiv M^{(i)}/V^{(i)}$ (without subscript $g$ in $\langle \rangle$; see Tab. \ref{tab:notation} for notation) as the gas mass in phase $i$ divided by the volume occupied by this phase. The global average gas density in phase $i$ is $\langle \rho^{(i)} \rangle_g \equiv M^{(i)}/V$, defined as the gas mass in $i$th phase divided by the total volume $V=\Sigma V^{(i)}$ ($\Sigma$ is sum over all phases). Thus, the local warm gas density is the {\em physical} density of the warm gas clumps. 
    \begin{figure}
	\centering
	\includegraphics[width=1.00\columnwidth]{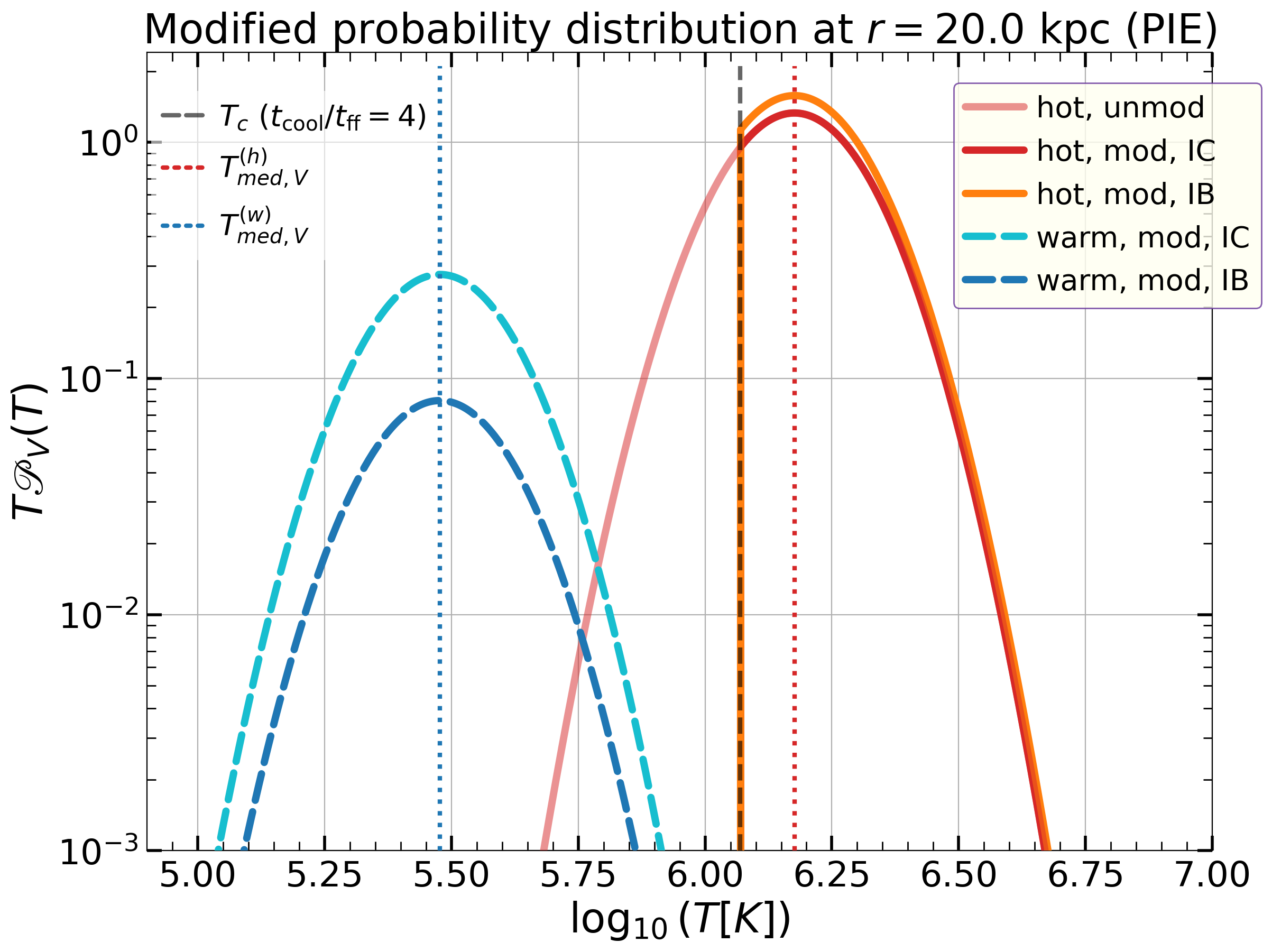}
	\caption{The volume probability distribution function (PDF) of temperature in a typical CGM shell at $r=20$ kpc for a $\gamma = 1$ polytrope (isothermal \citetalias{Faerman2017ApJ} model). The {\em light red} curve shows the unmodified log-normal PDF of the hot phase about a median temperature $T_{\rm med,V}^{(h)} = 1.5 \times 10^6 $ K (vertical {\em red} dotted line; see Eq. \ref{eq:Trelation}). The vertical {\em black} dashed line marks the temperature $T_c$ where the cooling time to the free-fall time ratio $t_{\rm cool}/t_{\rm ff}=4$ in this shell. Gas cooler than $T_c$ is assumed to be thermally unstable and populates a new (internally-) isobaric distribution to form a warm phase about a median temperature $T_{\rm med, V}^{(w)}=3 \times 10 ^ 5$ K ({vertical {\em blue} dotted line}). The {\em cyan} and {\em blue} dashed curves show the redistributed log-normal PDFs of the warm gas for isochoric and isobaric modification, respectively (see sections \ref{subsec:FSM17_description} \& \ref{subsec:isoBarvsisoChor}). Corresponding modified hot gas distributions are shown in {\em dark red} and {\em orange} curves. Being probability distributions, the sum of PDFs for warm and hot phases in both modifications ({\tt red+cyan}: {\it isochoric}; {\tt orange+blue}: {\it isobaric}) is normalized to unity.
 }
	\label{fig:isothermal_isochoric_mod}
\end{figure}

The global average density of a phase corresponds to the same mass being spread uniformly over the whole volume $V$. To calculate observables like column densities, we assume that the cooler phases are uniformly spread throughout the volume in the form of a mist (Fig. \ref{fig:cartoon}(b); the same figure also shows other possibilities where the warm clouds do not uniformly fill the whole spherical shell). \alankar{The reality is more complex than the mist limit which, by definition, gives an area-covering fraction of unity and is not consistent with observations.} The warm/cool gas is likely to \alankar{be} spread in the form of 
clouds that occupy a small volume and cover a projected area fraction $\lesssim 1$ (e.g., panels a, c, d in Fig. \ref{fig:cartoon}). \alankar{The CGM is expected to exhibit a patchy distribution of clouds with varying sizes and properties (also supported by cosmological simulations, e.g., \citealt{Nelson2020}). Consequently, different quasar sightlines probe a large variety of these cold and warm clouds, owing to the inherent stochasticity in their spatial distribution (see section \ref{sec:clouds} and also \citealt{2023arXiv231105691H} for a comprehensive discussion).}

For volume and mass fractions $f^{(i)}_V$ and, $f^{(i)}_M$ respectively, the expressions for the local and global average gas densities \alankar{in any phase $i$ (where $i= h,w$)} are given as
    \begin{equation}
        \label{eq:rhoL}
        \langle \rho^{(i)} \rangle \equiv \frac{M^{(i)}}{V^{(i)}} = \frac{f^{(i)}_M}{f^{(i)}_V} \langle \rho ^{(u)} \rangle
    \end{equation}
    and
     \begin{equation}
        \label{eq:rhoG}
        \langle \rho ^{(i)} \rangle_g \equiv \frac{M^{(i)}}{V} = f^{(i)}_M \langle \rho ^{(u)} \rangle,
    \end{equation}
    where $f^{(w)}_V+f^{(h)}_V= f^{(w)}_M+f^{(h)}_M=1$. Since each phase is assumed internally isobaric, we can define the density in the temperature range $[T, T+dT]$ as 
    $
    \rho^{(i)} = \langle \rho^{(i)} \rangle \langle T^{(i)} \rangle_M/T$.
    The ratio of Eqs. \ref{eq:rhoL} and \ref{eq:rhoG} gives,
    \begin{equation}
    \label{eq:rhoL_rhoG}
        \langle \rho ^{(i)} \rangle_g = f_V^{(i)} \langle \rho ^{(i)} \rangle,
    \end{equation}
    i.e., \alankar{the global average density of a phase equals the product of the volume fraction and the local average density of that phase.}

\begin{figure*}
\centering
\includegraphics[width=0.49\textwidth]{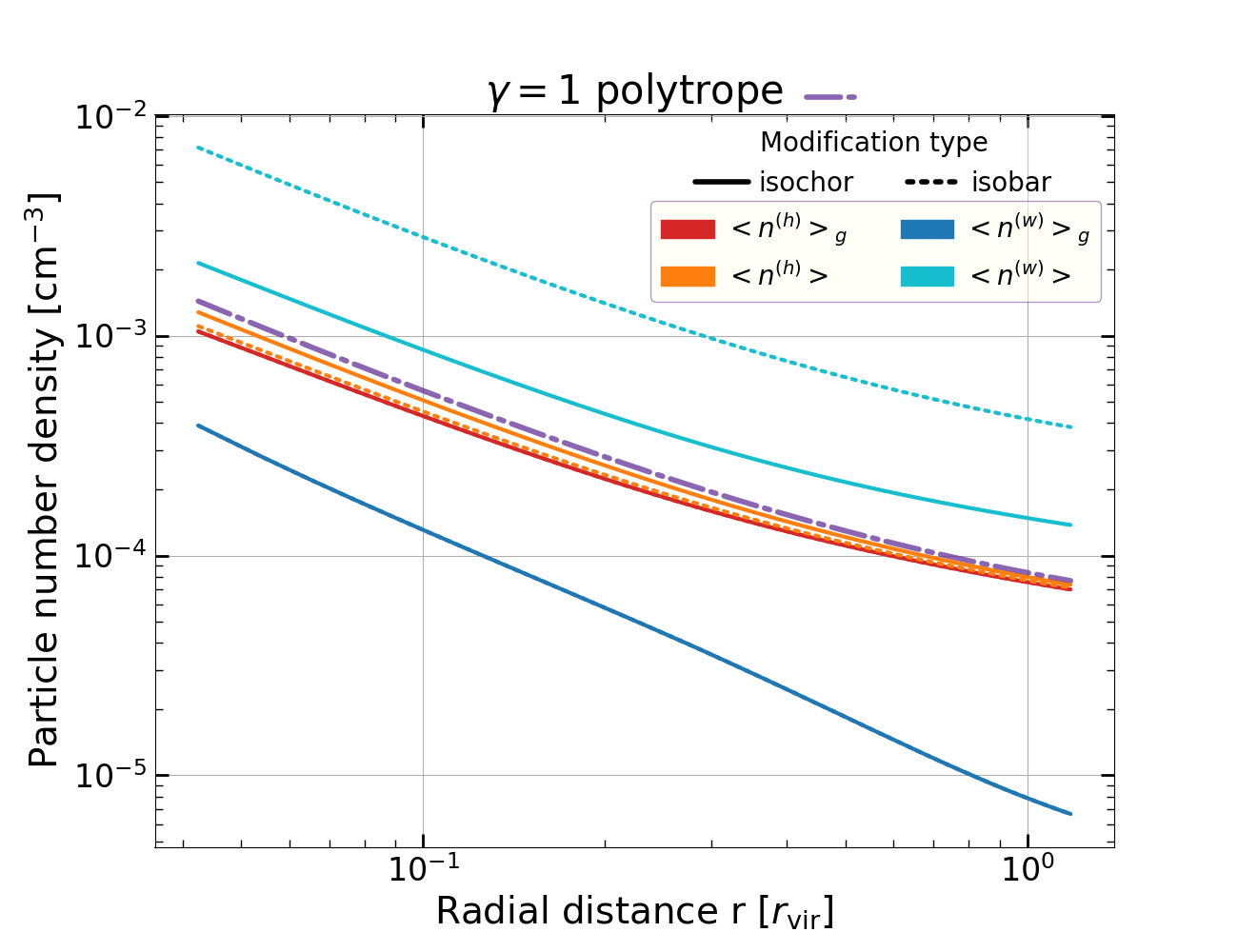}
\includegraphics[width=0.49\textwidth]{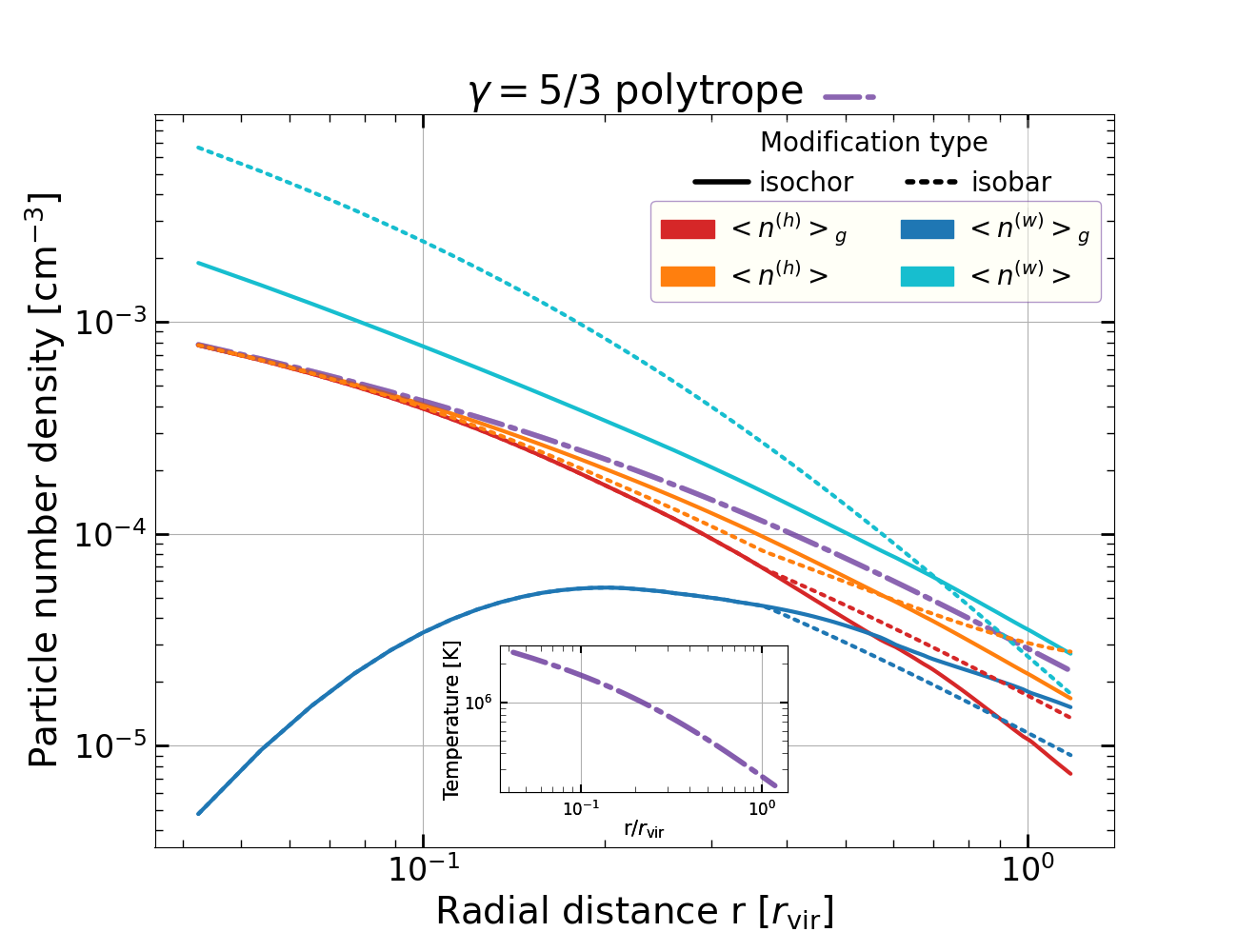}
\caption{The different number density profiles for our isothermal ($\gamma=1$ polytrope following \citetalias{Faerman2017ApJ}; {\em left panel}) and isentropic ($\gamma=5/3$ polytrope following \citetalias{Faerman2019ApJ}; {\em right panel}) CGM models (see section \ref{subsec:FSM17_description} \& \ref{subsec:isoThvsisoEnt}). 
 The {\em purple} dot-dashed lines in both panels show the unmodified 
 profiles. Additionally, the inset in the right panel shows the unmodified temperature profile; the median unmodified temperature for the isothermal model (left panel) is $1.5 \times 10^6$ K. The solid lines show density profiles with isochoric redistribution of the warm phase while isobaric redistribution is shown using dashed lines.
 The global average density profiles (indicated by $\langle \rangle_g$) are shown in {\em red} and {\em blue}, whereas the local average density profiles (indicated by $\langle \rangle$) are shown in {\em orange} and {\em cyan} (see point (iv) in section \ref{subsec:FSM17_description} for the definitions of local and global average densities). The average global density profile for the warm phase in the right panel shows a dip towards the center because $t_{\rm cool}/t_{\rm ff}$ is larger there and only a small amount of unmodified gas lies below our chosen $t_{\rm cool}/t_{\rm ff}=4$ threshold. In both the panels, the global density profiles for {\em isochoric} (solid {\tt red} + {\tt blue}) and {\em isobaric} (dashed {\tt red} + {\tt blue}) modifications coincide because the mass in each phase is the same in these cases. For $r/r_{\rm 200} \gtrsim 0.25$ the temperature of the unmodified isentropic profile (right panel) approaches the chosen warm phase temperature $T_{\rm med, V}^{(w)} = 10^{5.5}$ K (see inset in the right panel). 
 In this case, we assume that only a fraction $\le 0.4$ of the unmodified gas can condense into the warm phase. Such a high dropout fraction compared to the isothermal model explains the large variations in different densities at large radii in the isentropic model.
 }
 \label{fig:number_density}
\end{figure*}

    The solid lines in the left panel of Fig. \ref{fig:number_density} show the average global and local density profiles for the hot and warm phases for the isothermal CGM ($\gamma=1$ polytrope) from which warm gas condenses isochorically. As expected, the local density in all cases is larger than the global one. The global warm gas density at large radii is smaller because $t_{\rm cool}/t_{\rm ff}$ is larger and a smaller amount of gas drops out to the warm phase according to our prescription. The global and local densities of the hot phase are similar because most of the shell volume is occupied by the hot phase.
    The right panel is for an isentropic CGM ($\gamma = 5/3$ polytrope) discussed later in section \ref{subsec:isoThvsisoEnt}.

    \item {\em Calculating observables}: Having obtained the hot and warm gas temperature PDFs and their corresponding profiles, we can now calculate several observables such as OVI, OVII, OVIII, NV column densities, dispersion measure (DM), X-ray spectrum, and emission measure (EM). As calculating observables is independent of generating the model profiles, details on computing observables are postponed until section \ref{subsec:gen_obs}.
    
    As a demonstration of our \ps{general} description of the probabilistic CGM model, we now discuss specific modifications to the original \citetalias{Faerman2017ApJ} model.
    
\subsection{Isentropic unmodified profile}
\label{subsec:isoThvsisoEnt}
\citetalias{Faerman2017ApJ} used an isothermal ($\approx 1.5 \times 10^6$ K; comparable to the halo virial temperature) unmodified profile for the hot volume-filling CGM of Milky Way. Observations indicate the presence of OVI and NV ions in the Milky Way CGM (\citealt{Werk2013, Tumlinson2011ApJ}). Since these ions exist at a lower temperature ($\sim 10^{5.5}$ K), a CGM at the virial temperature is too hot to host sufficient OVI ions. To get around this, \citetalias{Faerman2017ApJ} proposed a thermal instability ansatz (discussed in section \ref{subsec:FSM17_description}) to introduce an additional warm phase at $10^{5.5}$ K. This results in a CGM  that can host OVI and NV columns consistent with the observations (see top panels of Fig. \ref{fig:column_densities}). On the other hand, \citetalias{Faerman2019ApJ} proposed an isentropic model without a spread in temperature at any radius. In contrast to the modified isothermal model, where every galactocentric distance hosts both \alankar{the} hot and warm phases, \citetalias{Faerman2019ApJ} has a \alankar{unique} temperature at every radius. Such an isentropic atmosphere in hydrostatic equilibrium (with reasonable boundary conditions) results in a transition from hot to warm temperatures at large radii approaching the virial radius (see inset in the right panel of Fig. \ref{fig:number_density}). This naturally produces a CGM that can host enough OVI and NV ions at large radii, \alankar{in compliance with observations, without the necessity of} introducing a separate warm phase (see Fig. 10 of \citetalias{Faerman2019ApJ}).

As pointed out earlier, any reasonable unmodified profile 
can be modified based on some physical prescription. The isothermal model in \citetalias{Faerman2017ApJ} (presented in section \ref{subsec:FSM17_description}) or the isentropic model in \citetalias{Faerman2019ApJ} are examples of such unmodified hydrostatic atmospheres. In this section, we modify the unmodified isentropic profile of \citetalias{Faerman2019ApJ} below the $t_{\rm cool}/t_{\rm ff}=4$ threshold and introduce a warm phase. This highlights the general applicability of the procedure described in section \ref{subsec:FSM17_description}. Another possibility (which we do not explore further) is to choose an unmodified hydrostatic profile at the precipitation threshold (say with $t_{\rm cool}/t_{\rm ff} = 20$ everywhere; 
\citealt{Sharma2012, Voit2019}) and modify gas below $t_{\rm cool}/t_{\rm ff} = 4$ to account for the condensation of the denser gas (as motivated by \citealt{Choudhury2019}).

The right panel of Fig. \ref{fig:number_density} shows the unmodified isentropic profile (purple dot-dashed line) and the global and local number density profiles for the modified hot and warm phases. The global warm gas density profile at the center shows a dip because only a small fraction of gas condenses, as $t_{\rm cool}/t_{\rm ff}$ at small radii is large. If we conserve the total shell volume and the mass of the warm gas condensing out, we expect the global densities for both isochoric and isobaric modification to be the same. However, note that the separation between the isochoric and isobaric modified density profiles for both the hot and warm phases at $r/r_{\rm 200} \gtrsim 0.3$ happens because we restrict the mass fraction of the warm gas to $f_M^{(w)} \le 0.4$. \alankar{The motivation behind this choice is that} hydrostatic equilibrium assumes that most mass is in the `hot' volume-filling phase.\footnote{{\fontsize{7pt}{7.2pt}\selectfont \ps{This choice, although somewhat arbitrary, should not affect the observables because of the comparable `warm' and `hot' phase temperatures at these radii.}}} 
Once the dropout gas mass fraction exceeds this, we limit the dropout mass to this value. 

\subsection{Isobaric instead of isochoric cooling to warm phase}
\label{subsec:isoBarvsisoChor}

\citetalias{Faerman2017ApJ} creates the warm phase by cooling the unmodified PDF below a cutoff temperature and assumes that the cooling happens isochorically to the warm phase (which is internally isobaric). While cooling may happen isochorically at intermediate temperatures with cooling times shorter than the sound-crossing time across cooling clouds (e.g., see Fig. 8 in \citealt{Mohapatra2022characterising}), the warm/cold clouds are expected to achieve pressure equilibrium after a few sound crossing times. In any case, it is a useful generalization to relax the isochoric assumption. The other extreme assumption is to assume that cooling from unmodified temperatures to the warm phase occurs isobarically. In this case, the warm gas will occupy a smaller volume and the hot phase pressure can drop because of adiabatic cooling (see Fig. \ref{fig:cartoon}(c) which shows the warm phase volume in panel (a) compressed to a smaller volume). This can cause compression and reduction in \alankar{the} volume of each shell in the absence of additional heating sources. Motivated by the importance of AGN/supernova feedback heating, the required additional heating of the hot phase is assumed to be present to preserve the shell volume. 

Let $V^{(u)}$ be the volume of a spherical shell in the unmodified profile that we split into smaller volumes of hot and warm phases $V^{(h)}$ and $V^{(w)}$ respectively. Mass conservation gives, $\langle \rho^{(h)} \rangle V^{(h)} + \langle \rho^{(w)} \rangle V^{(w)} = \langle \rho^{(u)} \rangle V^{(u)}$ and pressure balance between phases implies $\langle \rho^{(h)} \rangle \langle T^{(h)} \rangle_M = \langle \rho^{(w)} \rangle \langle T^{(w)} \rangle_M = \langle \rho^{(u)} \rangle \langle T^{(u)} \rangle_M$. Now these are three equations for four unknowns, namely, $\langle \rho^{(h)} \rangle$, $\langle \rho^{(w)} \rangle$, $V^{(h)}$, $V^{(w)}$. The median temperatures $\langle T^{(h)} \rangle_M$ and $\langle T^{(w)} \rangle_M$ are known since we specify them to construct the hot and warm PDFs. Since the total volume remains unchanged and the {\em total} $PdV$ work done for isobaric modification is zero, the total internal energy before and after modification remains the same; i.e., $\langle \rho^{(h)} \rangle \langle T^{(h)} \rangle_M V^{(h)} + \langle \rho^{(w)} \rangle \langle T^{(w)} \rangle_M V^{(w)} = \langle \rho^{(u)} \rangle \langle T^{(u)} \rangle_M V^{(u)}$. Thus, we can uniquely determine our modified hot and warm gas profiles. In this ansatz, energy loss due to the cooling of the hot phase to form the warm gas \alankar{needs to be} exactly compensated by external heating \alankar{sources} (which can come from feedback heating) to prevent compression of the hot phase. This is similar to the global thermal balance ansatz used in the modeling of cool core clusters (e.g., \citealt{sharma2012thermal}).

\begin{figure*}
\centering
\includegraphics[width=1.00\textwidth]{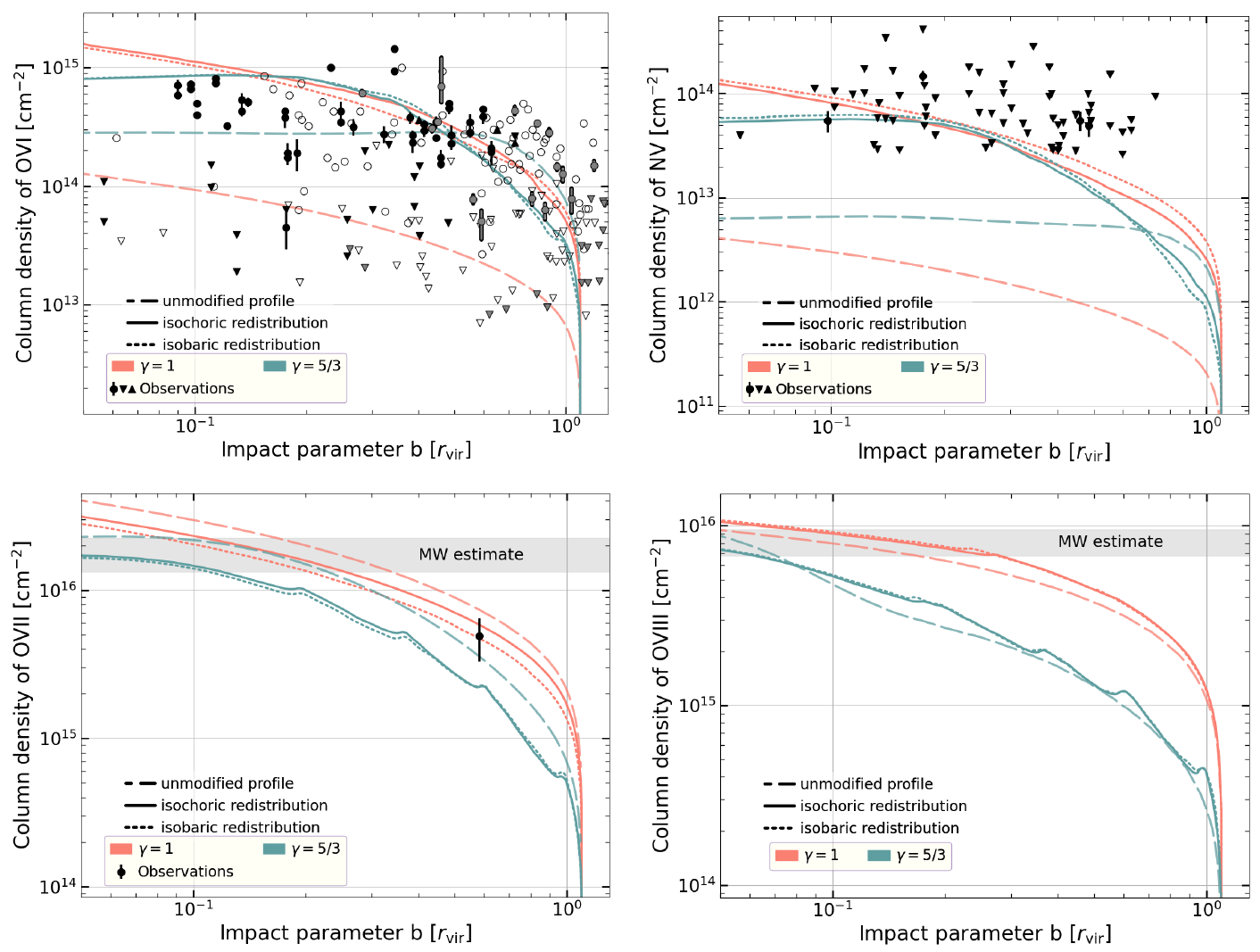}
 \caption{
 The 
 column density of different ions as a function of the impact parameter (normalized by the virial radius) from our isothermal ($\gamma=1$ polytrope) and isentropic ($\gamma=5/3$ polytrope) models (see Fig. \ref{fig:number_density} for density profiles). 
 In all the panels, the solid lines refer to isochoric modification, while the isobaric modification is shown using the \alankar{dotted lines. Using the dashed lines, we also show the column density estimates generated from the unmodified (isothermal and isentropic) profiles for reference.} {\em Orange} lines are for the isothermal 
 model and {\em cyan} for isentropic. 
 {\em Top panels}: The column densities of ions tracing the warm phase, OVI on the left and NV on the right. The data points are inferred from 
 absorption spectra of quasar sightlines through external galaxies (OVI: COS-Halos 
 [\citealt{Tumlinson2011Sci, Werk2016ApJ}] 
 and eCGM surveys [\citealt{Johnson2015}] in solid black markers, CGM${}^2$ survey [\citealt{Tchernyshyov_2022}] in open black markers, and CUBS VII survey [\citealt{2024arXiv240208016Q}] in solid gray markers; and NV: \citealt{Werk2013} in solid black markers). The (inverted)-triangles are (upper)-lower limits.
 {\em Bottom panels}: The column densities of ions tracing the hot phase, OVII on the left and OVIII on the right. The lone observation data point for OVII for an external galaxy is from \citealt{Mathur2023} \alankar{(cf. sections 2.1 \& 3.1 in their paper), which also produces OVIII column density of $7.8 \pm 2.6 \times 10^{15}\ \rm cm^{-2}$ (not marked; private communication, Sanskriti Das).} Due to limited observations for external galaxies, we only indicate the range estimated from the Milky Way CGM for the OVII and OVIII columns (in {\em gray} bands; from {\it Chandra} observations by \citealt{Gupta2012} and {\it XMM-Newton} observations by \citealt{Fang2015, Das2019ApJ}). \alankar{Since the sun is only $8 \ \rm kpc$ from the Galactic center, the column densities for the Milky Way ($\times 2$; shown in gray bands) provide an estimate on the upper limit of the ion columns in Milky Way-like external galaxies.}
 }
 \label{fig:column_densities}
\end{figure*}

The mass fraction in the warm phase $f^{(w)}_M$ is the same as for isochoric cooling to the warm phase, given by Eq. \ref{eq:mfracW} (since the same mass cools to the warm phase under both isochoric and isobaric assumptions). But this warm phase occupies a smaller volume under isobaric assumption \alankar{compared to isochoric}. The average warm gas density is, $\langle \rho^{(w)} \rangle = \langle \rho^{(u)} \rangle \langle T^{(u)}\rangle_M/\langle T^{(w)}\rangle_M$ since the warm phase and the unmodified gas have the same pressure. Thus, the warm phase volume fraction for this case is given by 
\begin{equation}
    \label{eq:fV_isobaric}
    f^{(w)}_V = \frac{V^{(w)}}{V} = \frac{M^{(w)}}{\langle \rho^{(w)} \rangle} \frac{\langle \rho^{(u)} \rangle}{M^{(u)}} = \frac{\langle T^{(w)} \rangle_M}{ \langle T^{(u)} \rangle_M} f^{(w)}_M.
\end{equation}
Recall that $\langle T^{(w)} \rangle_M/ \langle T^{(u)} \rangle_M = T_{{\rm med},V}^{(w)}e^{-(\sigma_w^2-\sigma_u^2)/2}/T_{{\rm med},V}^{(u)}$ (see Eq. \ref{eq:Trelation}). Fig. \ref{fig:number_density} shows that the local number density is higher for isobaric modification as compared to isochoric. However, since the mass of gas in all phases for both models is identical, the global densities are also, therefore, identical (shell volume is constant).
The volume (and also the area if not all sightlines are covered by clouds) filling factor for the isobaric warm phase is expected to be smaller than the isochoric case. While the column densities for isobaric and isochoric modifications are similar (see Fig. \ref{fig:column_densities}), the EM and luminosity are expected to be higher for isobaric modification because of a higher density in this case.

\subsection{Generating observables from probabilistic CGM models}
\label{subsec:gen_obs}
    Most of the observables are line-of-sight (LOS) integrals of various physical quantities. For example, column density is $\int n ds$ where $n$ is the local number density of a particular ion, and $ds$ is an infinitesimal path length along the LOS. Similarly, EM is proportional to $\int n^2 ds$. In our probabilistic model, every shell of the CGM is multiphase, and each phase contributes to these LOS integrals. We first discuss the column density estimate from our probabilistic model. Carrying this forward to other LOS integrals is straightforward.

\subsubsection{Column density in absorption}
\label{subsec:column_density}

    The contribution by phase $i$ in the temperature range $[T, T+dT]$ to the LOS integral is $\int n^{(i)} ds^{(i)}$, where $ds^{(i)}$ is the path length through phase $i$. Assuming clouds to be uniformly spread throughout the shell, i.e., the mist limit, implies $dV^{(i)} = dA_{\perp} ds^{(i)}$ and $ds^{(i)} = ds dV^{(i)}/V$ (see Tab. \ref{tab:notation} for notation), where $dA_{\perp}$ is an infinitesimal area perpendicular to the LOS and $dV^{(i)}$ is the differential volume of phase $i$. The column density is $\int  n^{(i)} ds^{(i)} = \int ds \int n^{(i)} dV^{(i)}/V = \int ds \int n^{(i)} {\cal P}_V^{(i)}(T) dT$, where the last expression follows from the definition of volume PDF. Note that $n^{(i)}$ is the local density of phase $i$ and, in general, $n^{(i)}$ depends on $(n, T)$, i.e., the thermodynamic state assumed within the phase. The expression for column density can be simplified further using the global average density of a particular phase $i$, i.e., $\langle n^{(i)} \rangle_g = \int n^{(i)} {\cal P}_V^{(i)}(T) dT$. Thus, the column density in the mist limit from all the phases can be written as $\Sigma \int \langle n^{(i)} \rangle_g ds $. The global number density is appropriate here because only a fraction of the available volume is occupied by the gas in a given phase. Similar expressions can be obtained for other LOS integrals. 
    
    Specifically, the expression for the column density of OVI at an impact parameter $b$ is
    \begin{equation}
    \label{eq:NOVIIcolden}
        N_{ \rm OVI}(b) = \bigintsss \langle n_{\rm OVI} \rangle _{g} (s) ds = 2 \bigintsss _b ^{r_{\rm CGM}}  \frac{\langle n_{\rm OVI} \rangle _{g}(r) r dr}{\sqrt{r^2-b^2}}, 
    \end{equation}
where $\langle n_{\rm OVI} \rangle_g$ is the global density of OVI ion in a shell at radius $r$, as motivated in the previous paragraphs. 
    \begin{figure*}
        \centering
        \includegraphics[width=1.02\textwidth]{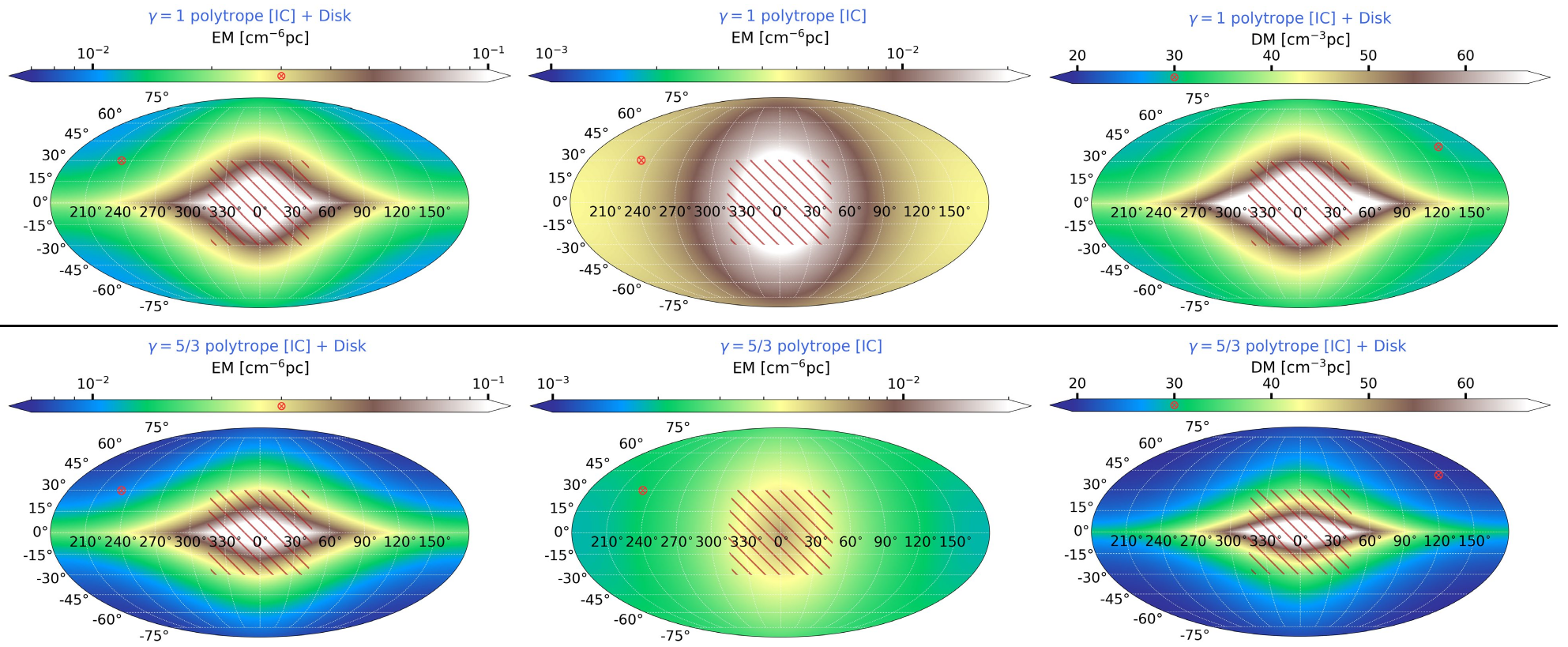}
     \caption{The Molleweide maps of observables in Galactic coordinates \alankar{$(l,b)$} from our models (namely, $\gamma=1$ [upper panels] and $\gamma=5/3$ [lower panels] polytropes with isochoric modification). The {\em left} and {\em middle} columns display our modeled Milky Way emission measure, with and without a coronal disk respectively, as observed from the position of the solar system. The {\em right} column shows the corresponding dispersion measure. These maps are generated from different CGM models discussed in section \ref{sec:faerman-generalize} (gas profiles shown in Fig. \ref{fig:number_density}). Eq. \ref{eq:exp_disk} models the coronal disk component. The {\it red crosshair} in the EM maps marks the eFEDS sightline ($l,b \approx 230^{\circ},30^{\circ}$) observed by \citealt{Ponti2022} and the estimated EM is $2.937 \times 10^{-2} \ {\rm pc \ cm^{-6}}$.
     The {\it red crosshair} in the DM maps in the {\em right} column marks the sightline ($l,b=142.19^{\circ}, 41.22^{\circ}$) of a nearby FRB, in the M81 galaxy (\citealt{Bhardwaj2021}). The DM estimated for the Milky Way halo along this sightline is $30 {\ \rm pc \ cm^{-3}}$. The region near the Galactic center is hatched in the maps to indicate that predictions would be unreliable there. This area is contaminated by eROSITA bubbles (\citealt{Predehl2020}) and other features, as well as the central cusp in the number density profile (Fig. \ref{fig:number_density}). 
     }
     \label{fig:DM_EM}
    \end{figure*}
The modified PDF and the global densities of the different gas phases can be used to estimate the global number density profiles of different ions (this assumes that all phases are uniformly mixed at each radius, also known as the mist approximation; see Fig. \ref{fig:cartoon}b).
The global average number density of OVI (to be plugged in Eq. \ref{eq:NOVIIcolden}) as a function of radius $r$ is given by (assuming photo+collisional ionization equilibrium; PIE) 
the following expression,
    \begin{equation}
    \label{eq:numdens_OVI}
         \langle n_{\rm OVI} \rangle _{g} (r) = a_{\rm  O} \left( \frac{Z(r)}{Z_\odot} \right) \left[\mathcal{I}^{(w)} + \mathcal{I}^{(h)} \right],\\
    \end{equation}
where \[ \mathcal{I}^{(i)} = \bigintsss dT
         \left(\frac{\mu^{(i)}}{\mu_H^{(i)}}\right)  n^{(i)} \mathscr{P}_{V}^{(i)}(T) f_{\rm OVI}(n^{(i)},T),\]
    for each component, where $\mu^{(i)}$  depends on $n^{(i)}$ and $T$, $a_{\rm O}$ is the number ratio of oxygen to hydrogen atoms in the sun (\citealt{Asplund2009}),  $Z(r)$ is the CGM metallicity (metal mass to total gas mass ratio) at that radius, and $f_{\rm OVI}$ is the OVI ion fraction. 
    We can use the equation of state $n^{(i)} T = \langle n^{(i)}\rangle \langle T^{(i)} \rangle_{M}$ to calculate the temperature integral above (since we assume the phases to be internally isobaric). We adopt the metallicity profile introduced by \citetalias{Faerman2019ApJ} \ps{(see their Eqs. 8 \& 9)}. 
    Our model parameters are listed in Tab. \ref{tab:fsm_params}. \alankar{We stick to \citetalias{Faerman2017ApJ} parameter values since we wish to test how models tuned for a particular observable fare against a broader range of multi-wavelength observables. 
    Further, the observables are estimated considering the mist limit, which gives a covering fraction of unity (see discussion at point (iv) in section \ref{subsec:FSM17_description} for details). It is anticipated that the measured column densities 
    will exceed our mist limit estimates because of the discrete nature of the clouds (discussed further in section \ref{sec:clouds}) and the ease of detecting higher columns.} 
    
    The top panels of Fig. \ref{fig:column_densities} show the OVI and NV column density profiles from our isothermal ($\gamma=1$) model as a function of the impact parameter. These ions trace the warm $\sim 10^{5.5}$ K phase (e.g., see Fig. 6 in \citealt{Tumlinson2017review}). The observational data are taken from \citealt{Werk2013, Werk2016ApJ, Tchernyshyov_2022} \& \citealt{2024arXiv240208016Q}. 
    The virial radii of the galaxies from the COS-Halos survey used in the normalization of the impact parameter are taken from \citealt{Tumlinson_2013}. \alankar{All the observed column densities were calculated from corresponding equivalent widths of ionic transitions in the absorption spectra using either Apparent Optical Depth analysis (\citealt{1991ApJ...379..245S}) or Voigt profile fitting (e.g. \citealt{ascl:VPFIT2014}; also see references from the corresponding surveys).} 
    Throughout this work, for the ionization models, we use {\tt CLOUDY 2017} (\citealt{2017RMxAA..53..385F}) and consider the CGM to be in photo+collisional ionization equilibrium (PIE) in the presence of Haardt-Madau extragalactic UV radiation (\citealt{Haardt2012}) at a redshift of 0.2 (matching COS-Halos galaxies). For some ion levels, the differences between the collisional (CIE) and photo+collisional ionization equilibrium (PIE) can be significant (e.g., see Fig. 6 in \citetalias{Faerman2019ApJ} \alankar{\& discussion in Appendix \ref{app:ion_cooling}}).
    
    Similarly, the bottom panels show the OVII and OVIII column density profiles. These ions trace the hot ($\gtrsim 10^{5.5}$ K) gas, and presently virtually no constraints exist on their column densities 
    in external galaxies, except for the one recent observation of OVII column by \citealt{Mathur2023}. Moreover, absorption/emission properties of ions like OVII may be significantly altered by resonant scattering (\citealt{Nelson2023}), not taken into account in our modeling. Observations also indicate that 
    \alankar{the absorption profiles of these ions can be highly saturated}. For example, the ratio of equivalent widths for K$\rm \alpha$ to K$\rm \beta$ transition for 
    OVII along multiple sightlines probing the Milky Way CGM deviates from a constant value expected ($\sim 0.15$) in the optically thin regime (\citealt{Gupta2012}). \ps{The poor spectral resolution in X-rays does not allow precise Voigt profile fitting. For OVII, we use the column density range indicated in Tab. 2 of \citet{Gupta2012}. From the OVIII equivalent width (EW) given in the same table, we obtain $N_{\rm OVIII}$ using the linear relation between the equivalent width (EW) and column density in the optically thin regime.} 
    In the absence of adequate constraints from external galaxies for these high ionization states, 
    we show the scaled OVII and OVIII columns 
    of the Milky Way CGM (\citealt{Gupta2012, Fang2015, Miller2015, Miller2013}), which are indicated in gray bands. 

\alankar{
To compare with our probabilistic models, in Fig. \ref{fig:column_densities} we also show the column density profiles for the unmodified profiles using dashed lines (red: isothermal; blue: isentropic).
The unmodified isentropic profile in \citetalias{Faerman2019ApJ} is cooler in the outskirts and can, therefore, produce higher OVI column density compared to the unmodified isothermal profile in \citetalias{Faerman2017ApJ}.
We modify the isentropic profile following the thermal instability ansatz in \citetalias{Faerman2017ApJ} 
with the threshold $t_{\rm cool}/t_{\rm ff} = 4$, and the warm phase has a median temperature of $10^{5.5}$ K. As expected, this modification does not significantly alter the column density profiles of OVI in the isentropic model. 
Since the unmodified isentropic profiles are cooler at large radii, the OVII and OVIII column densities are smaller farther out than for the isothermal model 
(see bottom rows of Fig. \ref{fig:column_densities}). 
}
\subsubsection{Emission \& Dispersion measures}
\label{sec:EM_DM}
    Just like the individual ions, we can calculate the electron number density to determine the dispersion and emission measures produced by our CGM model. \alankar{For every sightline ($l,b$) in the Molleweide maps in Fig. \ref{fig:DM_EM}, we sample $1000$ points along the line of sight (uniformly spaced from the location of the sun till 
    $r_{\rm CGM}$). For each of these points, we calculate the $(r,\theta,\phi)$ coordinates from the Galactic center and calculate the desired observable quantity interpolated from our models. 
    These values are then numerically integrated to obtain the observables at each $(l,b)$. 
    These observables can alternatively be (numerically) integrated directly in terms of the Galactocentric distance $r$, employing a change of variables as discussed in Appendix \ref{app:powerlaw} (Eq. \ref{eq:integral_MW3phase}). 
    }
    
    In the mist approximation, the contribution by a given phase to the emission measure integral $\int (n^{(i)})^2 ds^{(i)} = \int \langle (n^{(i)})^2 \rangle_g ds$, with
    $$
     \langle (n^{(i)})^2 \rangle_g =  \bigintsss \left( \frac{\rho^{(i)}}{{\mu_i m_p}} \right)^2 {\cal P}_V^{(i)}(T) dT,
    $$
    where $\rho^{(i)}$ can be related to $T$ by the assumed thermodynamic equation of state within the phase (isobaric in our case). Note that the ratio of emission measure and the square of column density contributed by a uniform volume is the clumping factor\footnote{{\fontsize{7pt}{7.2pt}\selectfont \ps{Clumping factor quantifies the spread in density, considered here in the misty limit. Note that multiphase CGM is both clumpy (showing spread in density) and patchy/cloudy (not filling the volume uniformly), and these two are distinct properties.}}}
    \begin{equation}
    \label{eq:clumping_factor}
        C = \frac{\Sigma \langle (n^{(i)})^2 \rangle_g}{[\Sigma \langle n^{(i)} \rangle_g]^2}.
    \end{equation}
    Since the ionization of hydrogen and helium is the dominant contributor of free electrons, the dispersion measure (${\rm DM}=\int n_e ds$) is mostly insensitive to the ionization state of the metals. This makes DM-based inferences robust and less sensitive to model parameters. The DM generated from our models can be compared with the DMs of the Fast Radio Bursts (FRBs) in nearby galaxies (\citealt{Bhardwaj2021, Ravi2023, Cook2023}). The emission measure ${\rm EM}=\int n_e n_H ds$) is another observable that we generate from our CGM models. The EM constraints are available from the observations of the continuum soft X-ray emission from the Milky Way CGM. Estimating EM requires the observed X-ray spectrum to be broken down into contributions from several components like the local hot bubble, cosmic X-ray background, solar wind charge exchange, MW halo, and MW Galactic disk.
   Many different surveys, till now, have estimated the EM from the Milky Way, e.g., {\it ROSAT} (\citealt{1992KlBer..35..803H}), {\it Suzaku} (\citealt{Gupta2014}), {\it XMM-Newton} (\citealt{Henley_2010, Henley_2013, Das_2019XMM_em, Bhattacharyya_2023}), {\it HaloSat} (\citealt{Kaaret_2019, Kaaret2020, Bluem2022}), and eROSITA eFEDS (\citealt{Ponti2023, Ponti2022}). We compare our models with these observations.
    \begin{figure}
	\centering
	\includegraphics[width=1.00\columnwidth]{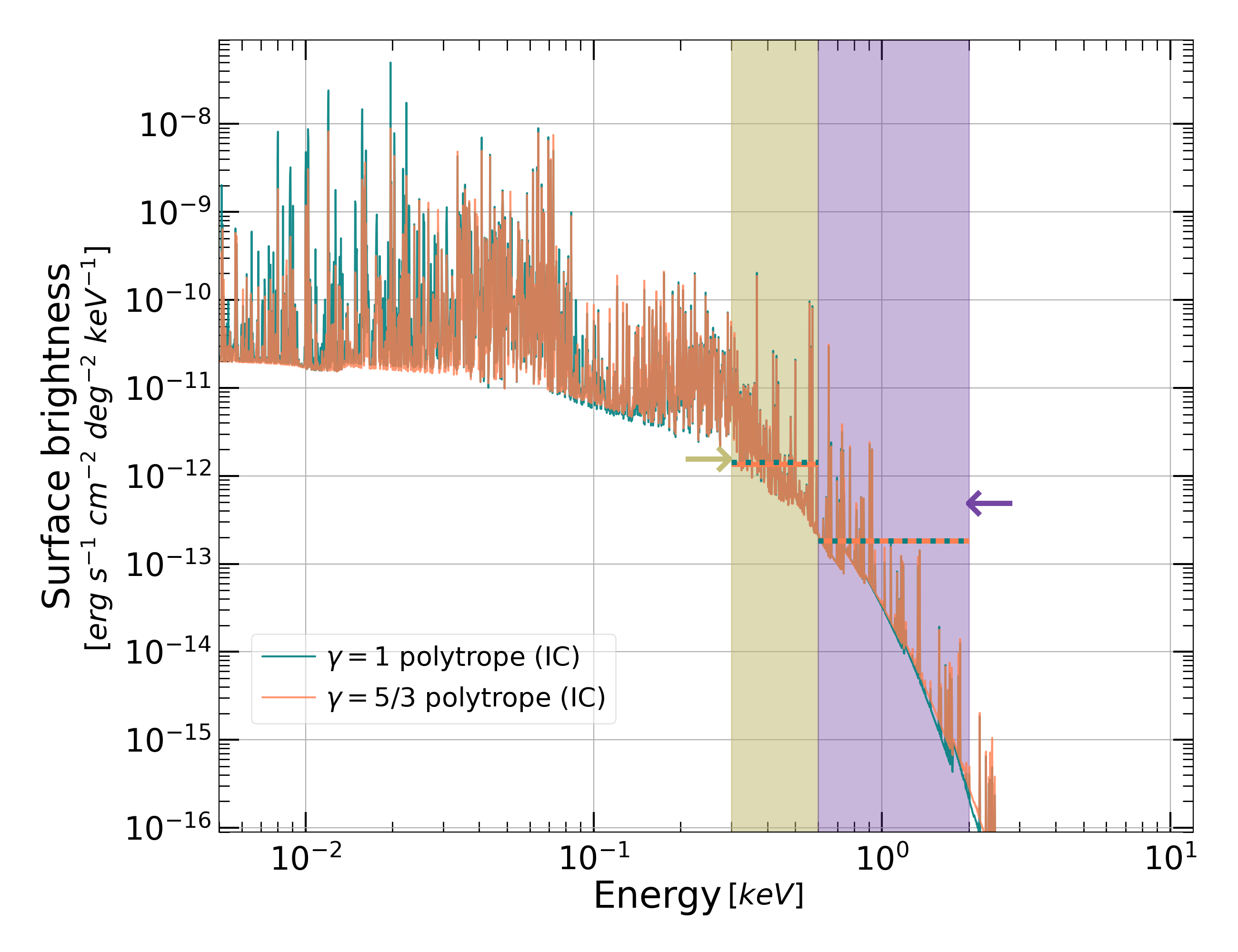}
	\caption{Synthetic surface brightness along $l,b=230^{\circ},30^{\circ}$ (eFEDS region) from our \ps{polytropic models of the CGM} (see Fig. \ref{fig:number_density} and section \ref{sec:faerman-generalize} for details). The {\em cyan} curve is for a $\gamma = 1$ polytrope while the {\em orange} curve is for $\gamma = 5/3$ polytrope, both of which consider isochoric modifications. The vertical {\em olive} ($0.3-0.6$ keV) and {\em purple} ($0.6-2.0$ keV) bands mark the energy bands used for the eFEDS survey of the Milky Way CGM in soft X-rays (\citealt{Ponti2022}). The observed surface brightness towards the eFEDS field of the CGM in $0.3-2$ keV is $2.05 \times 10^{-12}$ erg cm$^{-2}$ s$^{-1}$ deg$^{-2}$ (Tab. 4 in \citealt{Ponti2022}). The surface brightness in the same energy band ($0.3-2.0$ keV) obtained from $\gamma = 1$ and $\gamma = 5/3$ polytrope models with the addition of a coronal disk (see Eq. \ref{eq:exp_disk}) is $1.07$ $\times 10^{-12}$  and $9.4$ $\times 10^{-13}$ erg cm$^{-2}$ s$^{-1}$ deg$^{-2}$, respectively. The arrows indicate the surface brightness levels observed in each band, while the horizontal lines show the model prediction.
 }
    \label{fig:spectrum}
\end{figure}

\begin{figure}
	\centering
	\includegraphics[width=1.00\columnwidth]{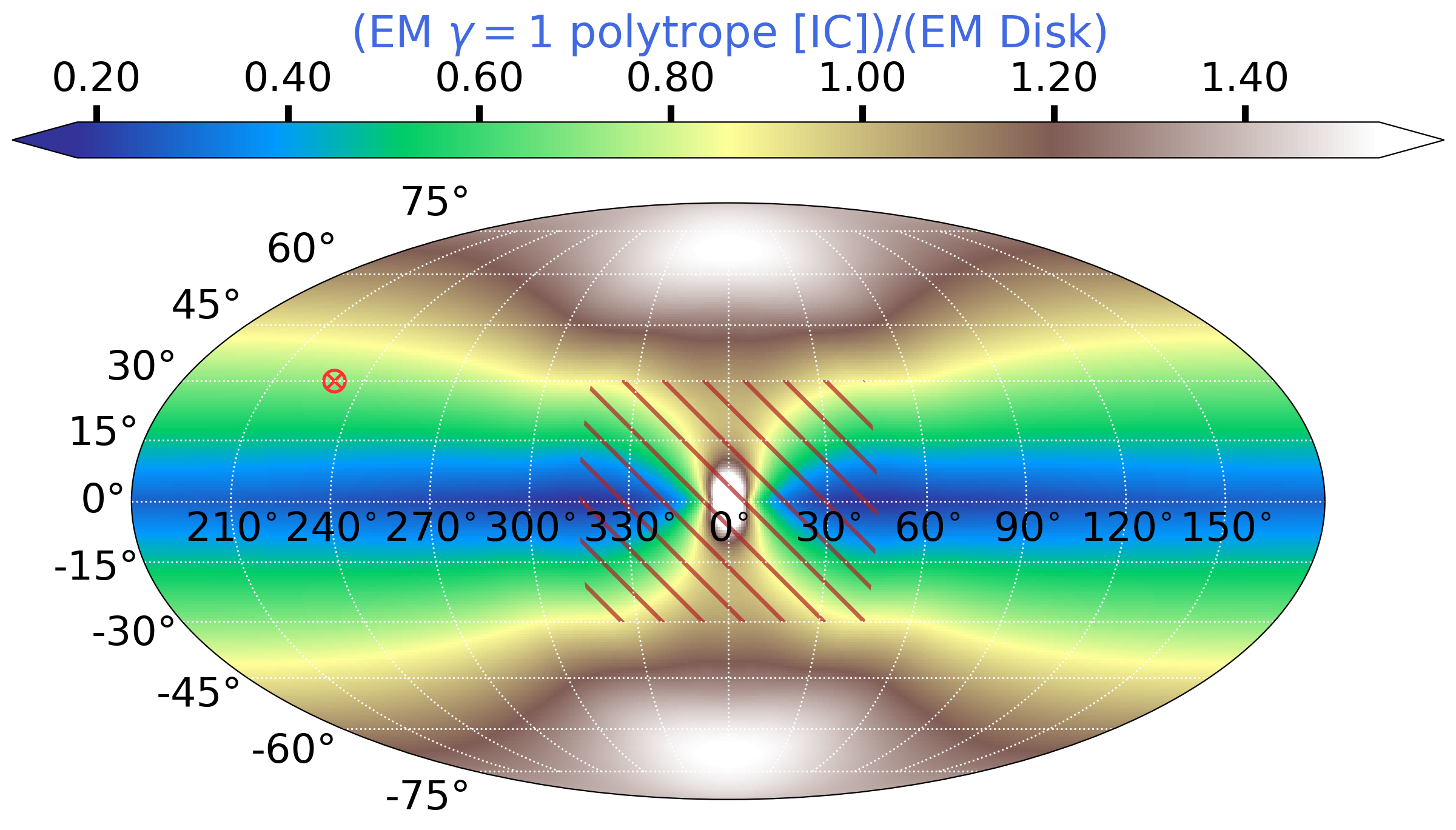}
    \includegraphics[width=1.00\columnwidth]{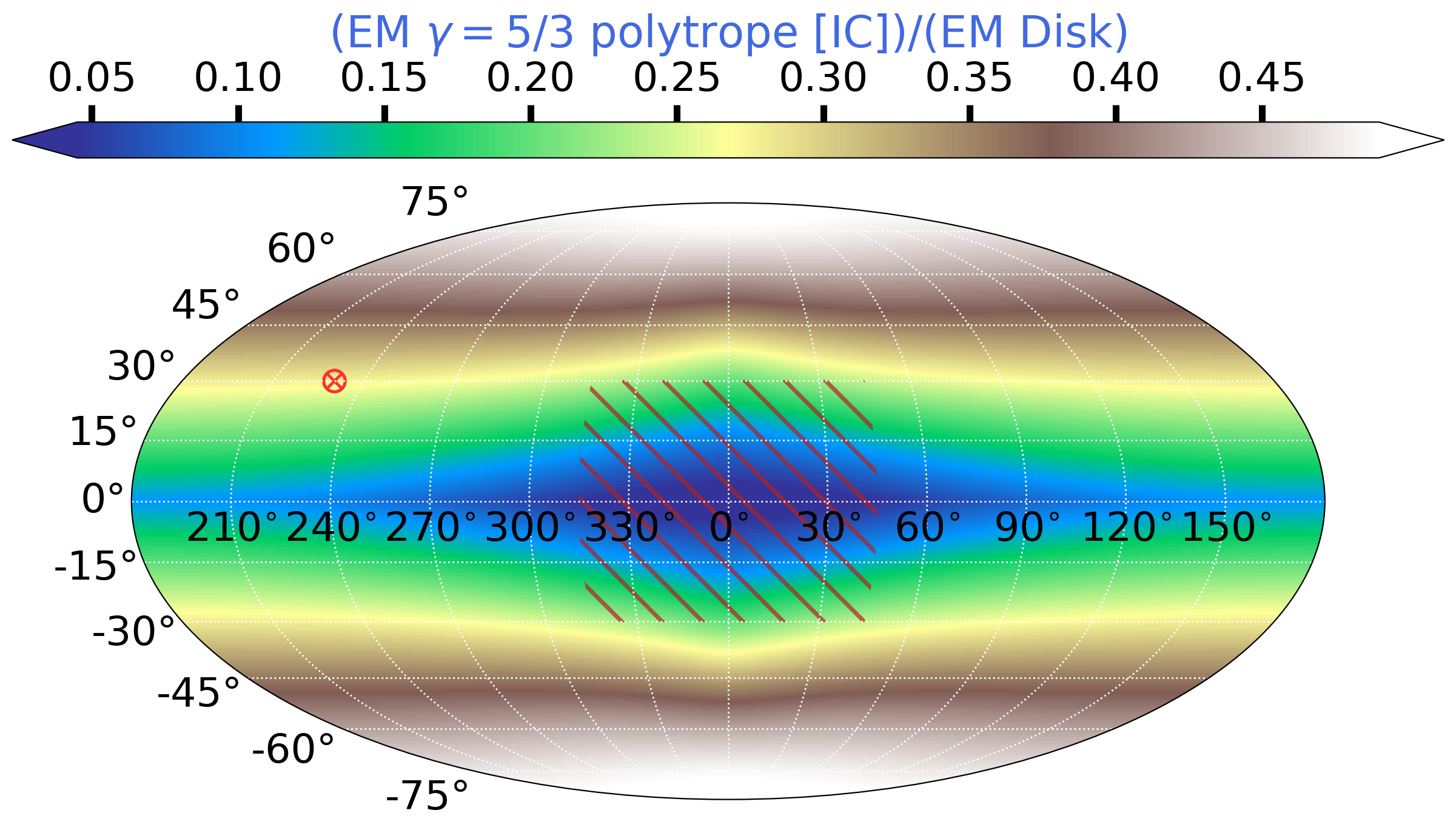}
	\caption{Molleweide projection \alankar{in Galactic coordinates $(l,b)$} of the ratio of CGM halo to disk emission measure (EM) for $\gamma =1$ (top panel) and $\gamma = 5/3$ (bottom panel) polytropes,  both with isochoric modification. The {\it red crosshair} marks the eFEDS sightline $l,b=230^{\circ},30^{\circ}$. Along this sightline, the halo to disk EM is $\approx$ 0.7 and 0.25 for the $\gamma=1$ and the $\gamma=5/3$ polytropes,  respectively. The neighborhood of the Galactic center is again hatched (like \ref{fig:DM_EM}) due to possible contamination by the eROSITA bubble. In the $\gamma = 1$ map, the CGM dominates towards the Galactic center because of the cusp-like density profile in this case (see left panel of Fig. \ref{fig:number_density}).}
    \label{fig:ratio_em}
\end{figure}

    The EM value with just the spherical CGM (middle panels of Fig. \ref{fig:DM_EM}) is of a few factors smaller than the value observed in the eFEDS field. Thus, we include an X-ray emitting disk adapted from (\citealt{Yamasaki2020}) having a density
    \begin{equation}
        \label{eq:exp_disk}
        n_H (R, z) = n_{H\rm ,0} \exp \left[-\left(\frac{R}{R_0} + \frac{|z|}{z_0}\right)\right],
    \end{equation}
    where $n_{H\rm ,0}$ is the hydrogen number density at the center of the coronal disk and $R_0 = 8.5 \ {\rm kpc}$ and $z_0 = 3.0 \ {\rm kpc}$ are the scale radius and height of the disk. 
    We set $n_{H,0} = 4.8 \times 10^{-3} {\rm \ cm^{-3}}$ and assume the disk to be isothermal at a temperature of $1.5 \times 10^6$ K. Our parameter values are different from \citealt{Yamasaki2020}, adjusted to match the recently observed X-ray surface brightness of the CGM in the eFEDS field (\citealt{Ponti2022}). 
    Simulations and theoretical considerations suggest that this coronal disk is expected and is maintained by heating from supernovae-driven outflows (\citealt{Weiner_2009, Rubin_2010}), which form the rising part of the Galactic fountain (\citealt{1980ApJ...236..577B, 10.1111/j.1365-2966.2010.16985.x, 2017ASSL..430..323F, Kim_2018, 10.1093/mnras/stz2928}). A coronal disk not only increases the X-ray surface brightness towards the eFEDS field but also increases its anisotropy towards and opposite to the Galactic center, which can be compared with observations (e.g., Fig. 6 in \citealt{Bluem2022}).\footnote{{\fontsize{7pt}{7.2pt}\selectfont We note that the CGM emission measure estimated in \citet{Ponti2022} is larger than in \citet{Bluem2022} mainly because the former estimates the CGM metallicity to be $\lesssim 0.1 Z_\odot$ from the data. In contrast, the latter assumes the standard value of $0.3 Z_\odot$.}} \alankar{We assume that the contribution from the disk (modeled by Eq. \ref{eq:exp_disk}) can simply be superimposed for all the observables.} 
    
    The top row of Fig. \ref{fig:DM_EM} shows the EM and DM maps \alankar{in Molleweide projection} for our isochorically modified isothermal ($\gamma=1$ polytrope) CGM model, while the bottom row shows the EM and DM maps for isochorically modified isentropic ($\gamma=5/3$ polytrope) CGM model (see section \ref{subsec:isoThvsisoEnt}). The middle column shows the EM maps without the disk to highlight the CGM contribution for the $\gamma=1$ and the $\gamma=5/3$ polytropes. Because of a higher density at larger radii in the isothermal model, the isothermal model has higher values of EM and DM (see Fig. \ref{fig:number_density}). 
     \alankar{The red crosshairs on these maps mark the observed sightlines. Along $l,b \approx 230^{\circ},30^{\circ}$, \citealt{Ponti2022} (the eFEDS survey) report the Milky Way EM to be $2.9-3.1 \times 10^{-2} \ {\rm pc \ cm^{-6}}$ (their Tab. 2). Along $l,b=142.19^{\circ}, 41.22^{\circ}$, \citealt{Bhardwaj2021} estimate the DM of the Milky Way (using a nearby FRB in the M81 galaxy) to be $30 {\ \rm pc \ cm^{-3}}$.} \ps{The isothermal models exhibit better agreement with the observed EM and DM values, as shown quantitatively in Tab. \ref{tab:comparison_summary}. The model EM values, dominated by the disk, are smaller by a factor of 2-3 than observations (perhaps due to additional ISM contribution along this sightline). The DM values, dominated by the spherical CGM, are in very good agreement. Because of a weaker sensitivity to the dense ISM, the DM is a more constraining probe of the CGM than EM.} 
    
    

    Recently, the {\it eROSITA} X-ray telescope has looked at an unobscured field to constrain the Milky Way CGM properties. 
    The two X-ray bands in the {\it eROSITA} eFEDS survey outlined by \citealt{Ponti2022} are $0.3-0.6$ keV and $0.6-2.0$ keV.  Fig. \ref{fig:spectrum} shows the emission spectrum from our isothermal ($\gamma=1$ polytrope) and isentropic ($\gamma=5/3$ polytrope; see section \ref{subsec:isoThvsisoEnt}) models, both with isochoric modification. The surface brightness is calculated towards the direction of the eFEDS field $(l,b)\equiv (230^{\circ}, 30^{\circ})$.
    The two models give very similar surface brightness spectra because they are dominated by the coronal disk component, which is the same in both cases. The observed X-ray surface brightness in 0.6-2 keV is \alankar{twice our model predictions} because 
    of ISM/CGM clouds along the sightline and an even hotter/super-virial coronal disk component, as suggested by recent observations (\citealt{Ponti2022, Bluem2022}). If needed, such a component can be added to our models, as described in section \ref{sec:threePhase}. 
    
    Fig. \ref{fig:ratio_em} maps out the ratio 
    of EM contributed by the spherical CGM component to the disk for \alankar{both} the isothermal and isentropic models. As expected, the spherical component dominates at high latitudes ($b \gtrsim 30^\circ$). All sky maps from eROSITA at high latitudes and away from the Galactic center can help us distinguish between different CGM+disk  models.

 \end{enumerate}

\section{More than two phases}
\label{sec:threePhase}
Observations and numerical simulations show that in addition to the volume-filling hot phase ($\sim 10^6$ K) and the intermediate warm phase ($\sim 10^5$ K), there is a cold ($\sim 10^4$ K) phase in the CGM. The hot and warm phases presumably cool to produce the cold gas. Cold gas from the dense ISM can also be introduced into the CGM by supernovae/AGN-driven winds. Ram pressure stripping of satellite galaxies can deposit the satellite's cold gas into the CGM of the host galaxy (\citealt{Eric2023}). Recent observations of many galaxies also seem to support this picture of three-phase gas distribution (\citealt{2024arXiv240305617S}).

Simulations show that the hot and cold phases have relatively narrow distributions in $\log T$ compared to a broader distribution of warm/intermediate phase (e.g., see the left panel of Fig. 6 in \citealt{Nelson2020}; middle panel of Fig. 5 in \citealt{Kanjilal2021MNRAS}; Fig. 4 in \citealt{Mohapatra2022characterising}).  
Similarly, there is a large spread in density for all phases. This motivates us to introduce 2D log-normal distributions (in $\rho$-$T$ space) with appropriate spreads in $\rho$ and $T$ for each phase. 

We introduce a uniform (one-zone) model for the CGM, where spatial variation is not considered for simplicity. Our formalism can be generalized to include variations with radius. We approximate the volume-PDF of the CGM to be the sum of a number of (three for specificity) 2D log-normal distributions centered at chosen median temperatures and densities. Namely (the summation indices are not explicitly mentioned later), 
\begin{eqnarray}
\label{eq:3phase_Vpdf}
\nonumber
\mathscr{P}^{2D}_V(\rho, T) d\rho dT &&= \sum_{i=h,w,c} \mathscr{P}_V^{2D(i)}(\rho, T) d\rho dT \\
\nonumber
&&= \sum_{i=h,w,c} \mathscr{P}_V^{2D(i)}(x, y) dx dy \\
&&= \sum_{i=h,w,c} f_V^{(i)} {\cal N}^{2D}({\bf x},{\bf x}_i, {\bf \Sigma}_i) dx dy,
\end{eqnarray}
where $x = \ln (\rho/\rho^{(u)}_{{\rm med}, V})$, $x_i = \ln (\rho^{(i)}_{{\rm med}, V}/\rho^{(u)}_{{\rm med}, V})$, $y = \ln (T/T^{(u)}_{{\rm med}, V})$, $y_i = \ln (T^{(i)}_{{\rm med}, V}/T^{(u)}_{{\rm med}, V})$, ${\bf x} = (x,y)$, ${\bf x}_i = (x_i,y_i)$, ${\bf \Sigma}_i$ is the $x-y$ covariance matrix of each phase, and $f_V^{(i)}$ is the volume fraction of each phase. Here $\rho_{\rm med,V}^{(i)}$, and $T_{\rm med,V}^{(i)}$ are the median density and temperatures of each phase and $x_i$,$y_i$ are the medians in $(x,y)$ space while $\rho_{\rm med,V}^{(u)}$, $T_{\rm med,V}^{(u)}$ are respectively the phase independent reference density, temperature for the volume-PDF. Note that the 2D volume-PDF $\mathscr{P}^{2D}_V$ is log-normal in ($\rho$,$T$) but a 2D Gaussian in the ($x$,$y$) space. \alankar{In the numerical implementation of our model, we used hydrogen number density $n_H$ as an independent variable instead of $\rho$.}
\begin{figure}
	\centering
	\includegraphics[width=1.00\columnwidth]{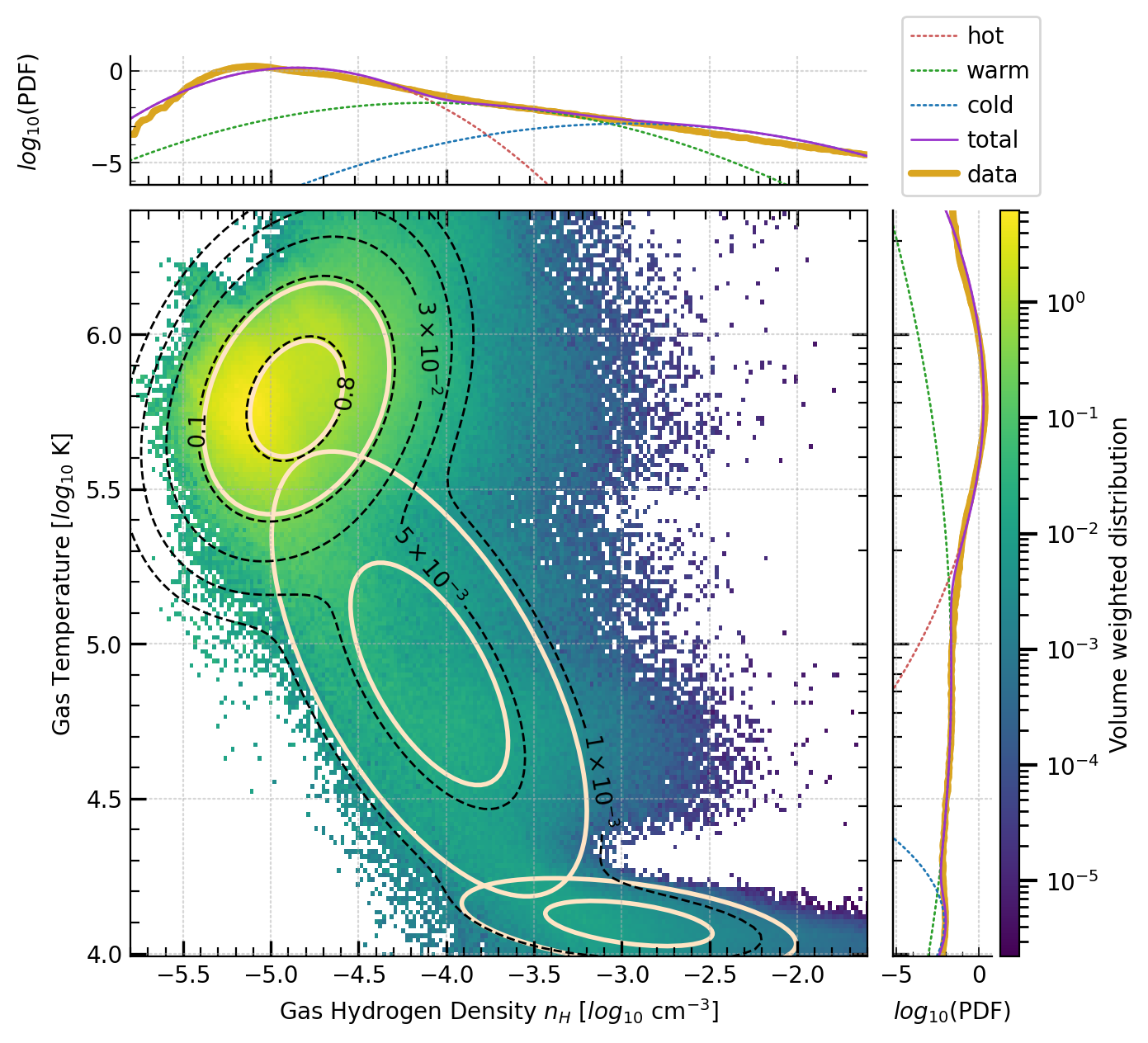}
	\caption{The central plot in this figure shows the volume-weighted 2D histogram of hydrogen number density and temperature of the CGM for our chosen halo from the {\tt Illustris TNG50-1} cosmological simulation (halo ID-110, snap-84; \alankar{only `non-star forming' gas within the virial radius is included}). The white rotated ellipses indicate $1\sigma$ and $2\sigma$ contours for each phase (cold, warm, and hot) obtained by fitting Eq. \ref{eq:3phase_Vpdf} \ps{and the black dashed contours correspond to the total PDF including all phases}. The top and right panels show the marginalized 1D PDFs in density and temperature, respectively. The thick solid line in {\em yellow} shows the marginalized PDFs from the simulation data. The thin solid lines in {\em purple} are from our best-fit model, with contributions from individual phases shown by dotted lines in colors listed in the legend. The best-fit parameters of our model are listed in Tab. \ref{tab:3p_params}.}
	\label{fig:2dPDFs}
\end{figure}
The 2D rotated-Gaussians ${\cal N}^{2D} ({\bf x},{\bf x}_i, {\bf \Sigma}_i)$ are  the 2D PDFs for each phase $i$ 
in the $x-y$ plane, and their explicit expression is 
\begin{eqnarray}
\label{eq:N2D}
\nonumber
    {\cal N}^{2D} ({\bf x},{\bf x}_i, {\bf \Sigma}_i) = \frac{1}{2\pi \sigma_{1,i} \sigma_{2,i}} \exp \left\{ - \left(A_i (x-x_i)^2 \right. \right. \\
    \left. \left. + B_i(x-x_i)(y-y_i) + C_i (y-y_i)^2\right) \right\},
\end{eqnarray}
where
$
A_i = ( {\cos^2 \alpha_i}/{\sigma_{1,i}^2} + {\sin^2 \alpha_i}/{\sigma_{2,i}^2} )/2,~
B_i = \sin 2 \alpha_i ( 1/\sigma_{1,i}^2 - 1/\sigma_{2,i}^2)/2,~
C_i =  ( {\sin^2 \alpha_i}/{\sigma_{1,i}^2} + {\cos^2 \alpha_i}/{\sigma_{2,i}^2} )/2
$ in terms of the standard deviations ($\sigma_{1,i}$, $\sigma_{2,i}$) along the principal axes of the individual Gaussians and the angle of rotation ($\alpha_i$) relative to the $x$ axis. Note that $4 A_i C_i - B_i^2 = 1/(\sigma_{1,i}^2 \sigma_{2,i}^2)$ is a useful simplification utilized later. Hence, the 2D volume-PDF can be fitted using the following parameters for each phase: median density $\rho_{\rm med,V}^{(i)}$, temperatures $T_{\rm med,V}^{(i)}$, angles of rotation $\alpha_i$, standard deviations along the two principal axes $\sigma_{1,i},\sigma_{2,i}$, and the volume fraction $f_V^{(i)}$. \alankar{Note that the combined volume fraction from all phases combined is unity by definition.}

We can obtain the marginalized PDFs by integrating the 2D PDFs along one of the axes. For example, 1D volume-PDF of temperature in log space is,
\begin{equation}
    {\cal P}_V(y) = \sum_i \frac{f_V^{(i)}}{2  \sigma_{1,i} \sigma_{2,i} \sqrt{\pi A_i}  } \exp \left\{ -  \frac{(y-y_i)^2 }{4 A_i \sigma_{1,i}^2 \sigma_{2,i}^2}  \right\},
    \label{eq:1Dvol}
\end{equation}
which is again a log-normal PDF \alankar{centered at $y_i$ with a standard deviation $\sqrt{2 A_i} \sigma_{1,i} \sigma_{2,i}$. For the other dimension, i.e., in density, similar log-normal PDF exists centered around $x_i$ with standard deviation $\sqrt{2 C_i} \sigma_{1,i} \sigma_{2,i}$.}

We can use our three-phase formalism to fit the simulation/observational data and constrain the free parameters for different phases. In the central plot of Fig. \ref{fig:2dPDFs} in color, we show the volume-weighted 2D histogram of hydrogen number density and temperature of the CGM gas (non-star forming) of one of the halos from the {\tt Illustris TNG50-1} cosmological simulation (halo ID-110, snap-84). \alankar{The selected halo has a virial mass of  $1.1 \times 10^{12}\ h^{-1} M_\odot$ (comparable to the Milky Way halo; \citealt{2006MNRAS.369.1688D}) and a star-formation rate of $3.63 \ M_\odot \rm yr^{-1}$ (similar to the Milky Way SFR of $2.0\pm 0.7\ M_\odot \rm yr^{-1}$; \citealt{2022ApJ...941..162E}). 
The median SFR of COS-Halos galaxies is $\approx 1.06\ M_\odot \rm yr^{-1}$; see Fig. 6 in \citealt{2012ApJS..198....3W}). The stellar mass of our chosen halo is $6.49 \times 10^{10}\ 
 h^{-1} M_\odot$ (slightly higher than the Milky Way stellar mass of $6.08 \pm 1.14 \times 10^{10} \ M_\odot$; \citealt{2015ApJ...806...96L})} We have \alankar{approximately} fitted the 2D histogram for the {\tt Illustris} halo with the 2D volume-PDFs of our three-phase model (Eq. \ref{eq:3phase_Vpdf}) \alankar{by eye}. 
The PDFs for all the expressions in this section use hydrogen number density $n_H$ as it is directly available from the simulation data and is independent of the ionization state. The above choice is convenient since plasma models, which are needed for producing observables, use $n_H$.

Tab. \ref{tab:3p_params} lists the best-fit parameters and the output parameters of our fitting. \alankar{We note that the parameters presented here are obtained manually by trial and error and are not quantitative statistical fits. 
We are working on an automated MCMC (Bayesian) fitting procedure to obtain a robust estimation of the model parameters and their corresponding uncertainties (which can then be propagated to the synthetic observables) by constraining our models with the simulation data.} 
The white rotated ellipses in the central plot are $1\sigma$ and $2\sigma$ contours of the best fit 2D volume PDFs $\mathscr{P}_V^{2D(i)}(\rho, T)$ for each phase (hot, warm, and cold) from our three-phase model.\footnote{{\fontsize{7pt}{7.2pt}\selectfont Unlike a 1-D Normal distribution for which 34\% and 95\% of random deviates lie within $1\sigma$ and $2\sigma$, respectively, the corresponding numbers in 2-D are 39\% and 86\%.}} \ps{The black dotted contours, indicating the 2-D PDF over all phases, show that the three log-normal PDFs capture the core of the three phases. Beyond the core, there is a deviation between the analytic model and the histograms from the simulation. In the future, one may explore going beyond log-normal distributions to include tails.} The hot phase is the volume-filling phase, whereas the warm and cold phases occupy a smaller volume. The temperature width of the warm phase is broader compared to the hot and cold phases. \ps{The cold phase is almost isothermal at $\sim 10^{4.1}$ K but with a large spread in density. Such broad spread for densities in the cold phase is also inferred from the line emission in the Slug nebula (\citealt{Cantalupo2019}) and is expected to be a robust feature of all multiphase CGMs.} The top and the right panels show the 1D volume PDFs marginalized over temperature and density, respectively. The thick yellow and thin purple solid lines show the simulation data and the corresponding best-fit 1D PDFs, respectively. The dotted colored lines show the contribution from individual phases (hot, warm, and cold) using the best-fit parameters. The marginalized PDFs match the simulation data well.

Similar to the 1D mass-PDF (Eq. \ref{eq:p_M}) discussed in previous sections, we can obtain the 2D mass-PDF in the $(x,y)$ space as
\begin{equation}
\label{eq:p_M_2D}
    {\cal P}_M^{2D}(x,y) = \frac{\rho_{{\rm med},V}^{(u)}}{\langle \rho \rangle} \sum_i e^x {\cal P}_V^{2D(i)} (x,y),
\end{equation}
which is again log-normal and $\langle \rho \rangle$ is the total average density that can be obtained by the normalization condition as (integrating above over $x$ and $y$),
$$
\frac{\langle \rho \rangle}{\rho_{{\rm med},V}^{(u)}} = \sum_i f_V^{(i)}  \exp\left\{ x_i + \sigma_{1,i}^2 \sigma_{2,i}^2 C_i  \right\}.
$$
The mass fraction $f_M^{(i)}$ of each phase is then given by
\begin{equation}
\label{eq:mf_2D}
    f_M^{(i)} = \frac{f_V^{(i)}  \exp\left\{ x_i + \sigma_{1,i}^2 \sigma_{2,i}^2 C_i  \right\}}{\sum_i f_V^{(i)}  \exp\left\{ x_i + \sigma_{1,i}^2 \sigma_{2,i}^2 C_i  \right\}}.
\end{equation}
Unlike volume fraction $f_V^{(i)}$ which is a model parameter, mass fraction $f_M^{(i)}$ is a derived quantity and depends on other free parameters. Marginalization of the mass PDF gives\footnote{{\fontsize{7pt}{7.2pt}\selectfont A useful identity is 
\begin{eqnarray}
\label{eq:identity_2D}
\nonumber
    & \int e^{p x} {\cal P}_V^{2D(i)}(x,y) dx = \frac{f_V^{(i)}}{2 \sigma_{1,i} \sigma_{2,i} \sqrt{\pi A_i}} \times \\ 
    & \exp \left \{  -\frac{\left[ y - y_i + p B_i \sigma_{1,i}^2 \sigma_{2,i}^2\right]^2}{4 A_i \sigma_{1,i}^2 \sigma_{2,i}^2}  + p^2 C_i \sigma_{1,i}^2 \sigma_{2,i}^2 + p x_i \right \}.
\end{eqnarray}
}}
\begin{equation}
{\cal P}_M(y) = \sum_i\frac{f_M^{(i)}}{2  \sigma_{1,i} \sigma_{2,i} \sqrt{\pi A_i}  }
    \exp \left\{ -\frac{\left( y - y_i + B_i\sigma_{1,i}^2\sigma_{2,i}^2\right)^2}{4A_i\sigma_{1,i}^2\sigma_{2,i}^2}\right\},
    \label{eq:1Dmass}
\end{equation}
which is again a log-normal PDF.

The 2D histogram in the central panel of Fig. \ref{fig:2dPDFs} shows that the hot phase is volume-filling and has a lower density in contrast to the cold phase, which is dense but occupies a minuscule fraction of the total volume. The luminosity of the gas at temperature $T$ is $L \propto n^2\Lambda[T] V$. Just like mass, it is important to know the luminosity contributed at different temperatures, especially when  CGM emission mapping is expected to be common in the near future (e.g., \citealt{Tuttle2019}). We obtain the 2-D luminosity PDF, 
\begin{equation}
    {\cal P}_L^{2D} (x, y) = \frac{(\rho_{{\rm med},V}^{(u)})^2}{\langle L \rangle} \sum_i e^{2 x} \Lambda (y) {\cal P}_V^{2D(i)} (x, y),
\end{equation}
where $\langle L \rangle$ is the volume-averaged luminosity that can be obtained from the normalization condition $\int {\cal P}_L^{2D} (x, y) dx dy = 1$. 

For the results in this section, we \alankar{assume the CGM metallicity to be temperature (phase) independent and fixed to a constant value of $0.3\ Z_{\odot}$. 
Further, we assume} that the cooling function only depends on temperature 
(general case can be treated numerically; see Appendix \ref{app:ion_cooling} for details). 
\alankar{This assumption is strictly valid 
in collisional ionization equilibrium (CIE). However, 
we adopt a temperature-dependent cooling function for plasma in photo+collisional-ionization equilibrium (PIE) 
in presence of Haardt-Madau extragalactic UV radiation (\citealt{Haardt2012}) at a redshift of 0.2. Our cooling function 
was generated using {\tt CLOUDY} for a plasma having hydrogen number density fixed to $2.0\times 10^{-5} \ {\rm cm^{-3}}$ (the average hydrogen density obtained by numerically integrating 
over our analytic volume PDF; Eq. \ref{eq:3phase_Vpdf}). 
This simplifying assumption 
allows us to obtain an analytic form for} the 1-D luminosity PDF marginalized over density (using Eq. \ref{eq:identity_2D} with $p=2$ and following the same procedure as for the mass PDF). The marginalized luminosity PDF is
\begin{equation}
    {\cal P}_L (y) = \sum_i \frac{f_L^{(i)} \Lambda(y, Z/Z_{\odot})}{2 \sigma_{1,i}\sigma_{2,i}\sqrt{\pi A_i}}  \exp \left\{ -\frac{\left( y - y_i + 2B_i\sigma_{1,i}^2\sigma_{2,i}^2\right)^2}{4A_i\sigma_{1,i}^2\sigma_{2,i}^2}\right\},
    \label{eq3:1Dlum}
\end{equation}
where 
$$ 
f_L^{(i)} = \frac{(\rho_{{\rm med},V}^{(u)})^2}{\langle L \rangle} f_V^{(i)} \exp(2 x_i +4 \sigma_{1,i}^2 \sigma_{2,i}^2 C_i)
$$
is the luminosity fraction of phase $i$, \alankar{and $Z/Z_{\odot}$ is the gas metallicity with respect to solar}. Note that the luminosity PDF \alankar{departs from} 
log-normal because of the cooling function.

\begin{table}
\centering
\caption{{\fontsize{9pt}{9.2pt}\selectfont Parameters of the three-phase CGM model${}^{*}$}}
\label{tab:3p_params}
\resizebox{\columnwidth}{!}{%
\setlength{\tabcolsep}{0.25em} 
{\renewcommand{\arraystretch}{1.4}
\begin{tabular}{@{}ll@{}}
\toprule
\multicolumn{1}{c}{\fontsize{9pt}{8.2pt}\selectfont Input Parameters}                              & \multicolumn{1}{l}{\fontsize{9pt}{8.2pt}\selectfont Value} \\ \midrule
$f_V^{(h)}$, $f_V^{(w)}$, $f_V^{(c)}$ {[}percent{]}         & \alankar{97.8, 2.04, 0.16}            \\
$T_{\rm med,V}^{(u)}$, $T_{\rm med,V}^{(h)}$, $T_{\rm med,V}^{(w)}$, $T_{\rm med,V}^{(c)}$ {[}$\log_{10} \ {\rm K}${]}                      & \alankar{5.0, 5.79, 4.90, 4.10}                   \\
$n_{H, \rm med,V}^{(u)}$, $n_{H, \rm med,V}^{(h)}$, $n_{H, \rm med,V}^{(w)}$, $n_{H, \rm med,V}^{(c)}$ {[}$\log_{10} \ {\rm cm^{-3}}${]}            & \alankar{-5.00, -4.85, -4.10, -2.96}               \\
($\sigma _1^{(h)}, \sigma_2^{(h)}$), ($\sigma _1^{(w)}, \sigma_2^{(w)}$), ($\sigma _1^{(c)}, \sigma_2^{(c)}$) & \alankar{(0.63,0.40), (0.56,1.20), (1.10,0.15)} \\
$\alpha ^{(h)}$, $\alpha ^{(w)}$, $\alpha ^{(c)}$ {[}deg{]} & \alankar{19, 55, 176}  \\
$Z/Z_{\odot}$ & 0.3         \\ \bottomrule
\multicolumn{1}{c}{\fontsize{9pt}{8.2pt}\selectfont Output Parameters}                              & \multicolumn{1}{l}{\fontsize{9pt}{8.pt}\selectfont Value} \\ \midrule
$f_L^{(h)}$, $f_L^{(w)}$, $f_L^{(c)}$ {[}percent{]}         &  \alankar{29.97, 25.25, 44.69} \\
$f_M^{(h)}$, $f_M^{(w)}$, $f_M^{(c)}$ {[}percent{]}         &  \alankar{73.32, 12.03, 11.90} \\
$M_{\rm CGM}$ \alankar{[$h^{-1} M_\odot$]} (gas mass) & \alankar{$9.92 \times 10^{10}$} \\
\alankar{$\langle n_H \rangle \ {\rm[ cm^{-3}]}$} & \alankar{$2.2 \times 10^{-5}$} \\
\bottomrule
\end{tabular}%
}}
\begin{tablenotes}
\item [] \alankar{{\fontsize{7pt}{7.2pt}\selectfont ${}^{*}$ This {\tt TNG50-1} halo at $z=0.2$ has a halo mass $M_{\rm 200} = 1.10 \times 10^{12}\ h^{-1} M_\odot$, stellar mass $M_* = 6.46 \times 10^{10}\ h^{-1} M_\odot$, star formation rate $\dot{M}_{\rm SFR} = 3.63\ M_\odot \rm yr^{-1}$ \& virial radius $r_{\rm 200} = 188.5\ h^{-1} \rm ckpc = 333.1\ \rm kpc$.}}
\end{tablenotes}
\end{table}

\begin{figure}
	\centering
	\includegraphics[width=1.0\columnwidth]{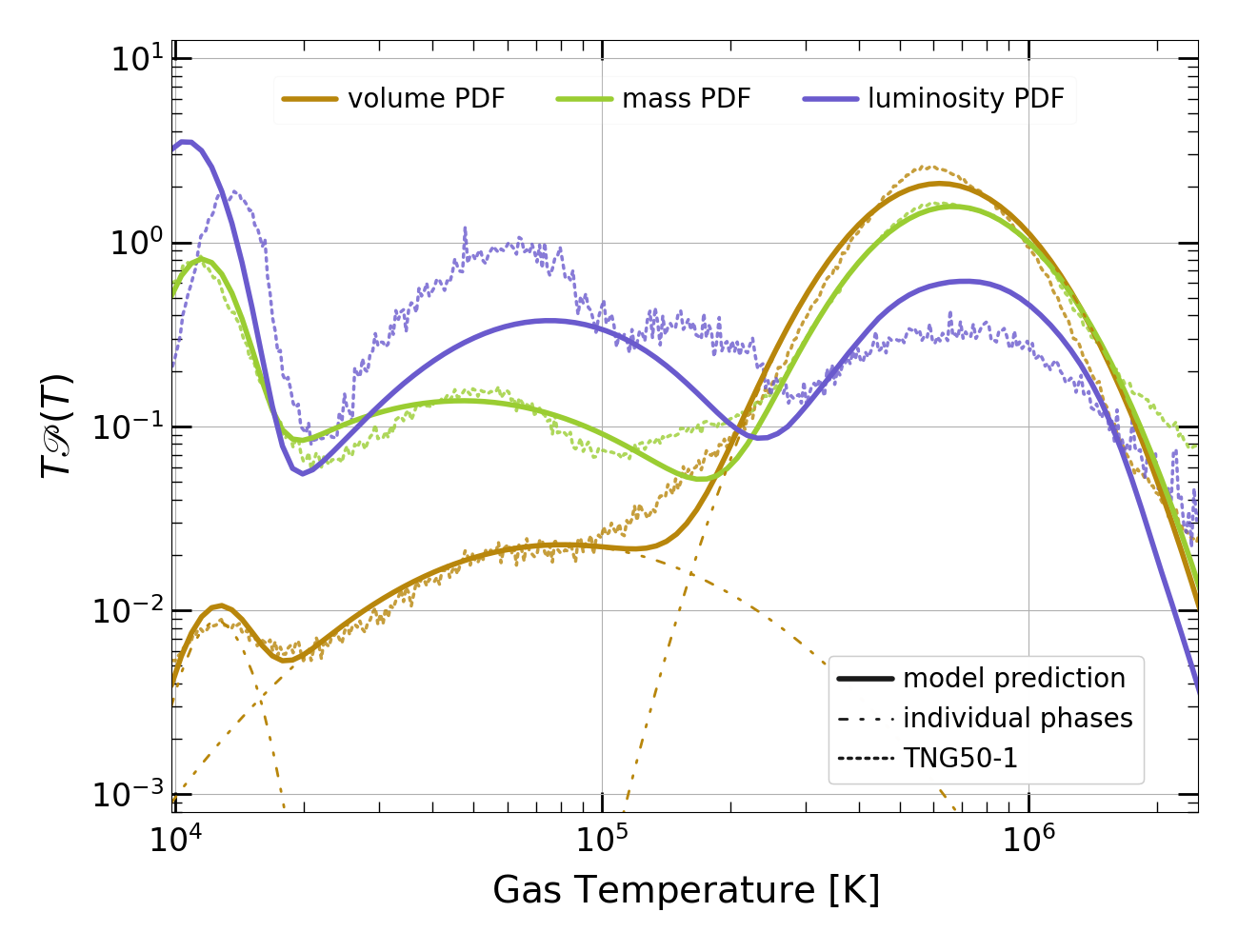}
	\caption{The 1D volume PDFs in temperature marginalized over density. The dashed lines are PDFs from the {\tt Illustris TNG50-1} halo (halo ID-110, snap-84). The solid lines show three-phase model PDFs (Eqs. \ref{eq:1Dvol}, \ref{eq:1Dmass} and \ref{eq3:1Dlum}) with best-fit parameters listed in Tab. \ref{tab:3p_params}. Note that the three-component Gaussian model captures the qualitative trends correctly. The hot, warm, and cold volume PDFs peak at $\sim (10^{5.8}, 10^{5.2},$ $10^{4.1})$ K respectively. 
 }
	\label{fig:PDFs}
\end{figure}

Fig. \ref{fig:PDFs} shows the 1D PDFs in temperature. The solid lines are from our best-fit model with the 2D PDFs in volume mass and luminosity marginalized in density (Eqs. \ref{eq:1Dvol}, \ref{eq:1Dmass} and \ref{eq3:1Dlum}). The dashed lines are 1D temperature PDFs from {\tt Illustris TNG50-1} data weighed by volume, mass, and luminosity of the simulation cells.\footnote{{\fontsize{7pt}{7.2pt}\selectfont The luminosity PDF of the warm 
phase is underestimated by our model, possibly due to the simple modeling of radiative cooling using a temperature-dependent cooling function with $Z=0.3Z_\odot$.}} Our three-phase model is able to capture the hot, warm, and cold phases for all three PDFs. The hot phase dominates the volume (and to a lesser extent mass), whereas luminosity has a dominant contribution from the cold and intermediate phases. 

\subsection{Synthetic observables from three phase model}
\label{subsec:syn_obs_3p}
We now move on to generate synthetic observables for our three-phase model. We extend the method to obtain observables with 1D PDFs in earlier sections to using 2D PDFs. For example, the expression for the column density of a particular ion (say OVI) remains unaltered from Eqs. \ref{eq:NOVIIcolden} and 
\ref{eq:numdens_OVI}, but ${\cal I}^{(i)}$ is now expressed as
\begin{equation}
    \mathcal{I}^{(i)} = \bigintsss d\rho dT 
         \left(\frac{\rho}{\mu_H^{(i)} m_p}\right) \mathscr{P}_{V}^{2D(i)}(\rho, T) f_{\rm OVI}(\rho,T).
\label{eq:Ndens_int_3p}
\end{equation}
Similarly, we can obtain the expressions for EM, DM, and other observables which now include 2D integrals. 

\begin{figure}
	\centering
	\includegraphics[width=1.0\columnwidth]{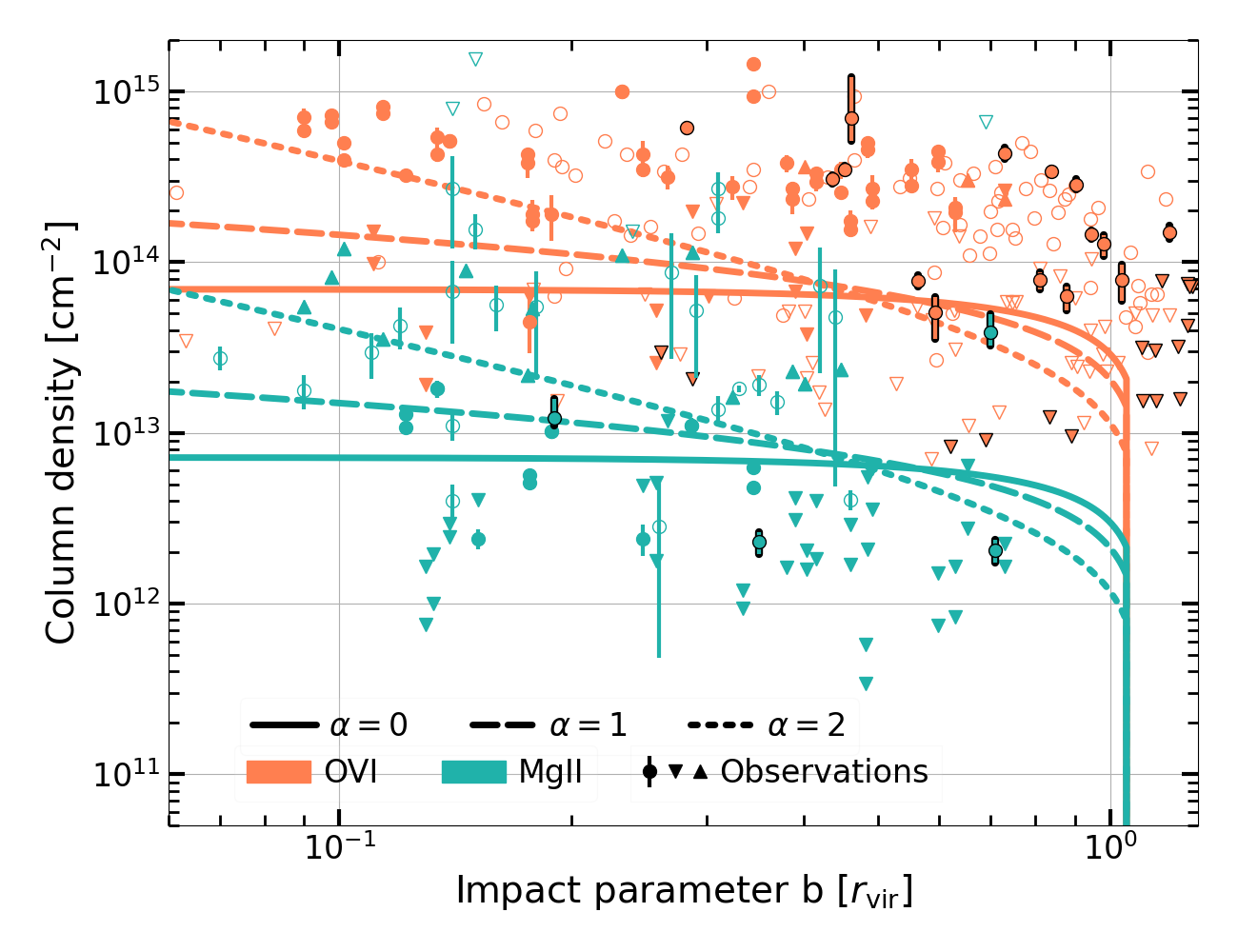}
	\caption{Column densities of OVI and MgII ions as a function of impact parameter predicted by our best-fit three-phase model. Observed columns from different surveys are shown with circles (detections), upper triangles (lower limits), and lower triangles (upper limits). Observations from different surveys that are shown here are from the COS-Halos survey \alankar{(Tab. 3 from \citealt{Werk2013})} in cyan/orange filled markers, CGM${}^2$ survey (Tab. 3 from \citealt{Tchernyshyov_2022}) in open orange markers, {\rm MAG}{\sc ii}CAT survey (Tab. 1 from \citealt{2016ApJ...818..171N}) in open cyan markers, and CUBS survey (Tab. A1 \& Tab. 1 from \citealt{2023arXiv230611274Q} for MgII and Tab. B1 from \citealt{2024arXiv240208016Q} for OVI) in filled cyan/orange markers with black borders. \alankar{The galaxies from the COS-Halos and the {\rm MAG}{\sc ii}CAT surveys have mass and size comparable to the Milky Way (c.f. Tab. 2 from \citealt{Tumlinson_2013} for COS-Halos \& Tab. 1 from \citealt{2016ApJ...818..171N} for {\rm MAG}{\sc ii}CAT).} 
    The one-zone assumption ({\em solid} lines) can be relaxed by assuming a power-law profile with index $\alpha$ (Eq. \ref{eq:n_pl_prof}; see Appendix \ref{app:powerlaw} for details), keeping the \alankar{total gas mass in the CGM} 
    constant (shown by the {\em dashed} [$\alpha=1$] and {\em dotted} [$\alpha=2$] lines). The column density of the OVI is better modeled with a steeper profile ($\alpha=2$). 
    \label{fig:1zone_CDens}
    }
\end{figure}

Fig. \ref{fig:1zone_CDens} shows the column density of MgII (cold $\sim 10^4$ K phase tracer) and OVI (warm phase $\sim 10^{5.5}$ K tracer) ions using solid, dotted and dashed lines for \alankar{different variants of} our three-phase model. 
\alankar{We 
use our fast plasma modeling code {\texttt AstroPlasma} 
to evaluate the ion fractions 
at each $\rho, T$ appearing in the integrals of the form presented in Eq. \ref{eq:Ndens_int_3p} 
and obtain the (volume-weighted) average ion density. The plasma is assumed to be in photo+collisional ionization equilibrium (PIE) 
at a redshift of 0.2 in the presence of Haardt-Madau extragalactic UV background (\citealt{Haardt2012}). We use the average ion density for this one-zone model to evaluate the column density profiles shown by the solid lines. 
Moving beyond the one-zone model, this average density 
is then used as the normalization factor $n_0$ appearing in Eq. \ref{eq:N_col_pl_prof} to obtain the column density 
for 
power-law profiles of number density in the CGM. The column densities of OVI and MgII 
for these power-law density profiles are shown using the dotted and dashed lines in Fig \ref{fig:1zone_CDens}.\footnote{\alankar{\fontsize{7pt}{7.2pt}\selectfont {We can relax the assumption of a constant density in the one-zone CGM model by assuming a power-law radial density profile with index $\alpha$ (Eq. \ref{eq:n_pl_prof}). The normalization is chosen to maintain the same CGM mass (see Appendix \ref{app:powerlaw} for details). Instead of self-consistently re-calculating the ionization in power-law profiles, which is computationally expensive, we simply rescaled the ion densities from the one-zone model.}}}} 
\alankar{{\it Circles (detections), upper triangles (lower limits), and lower triangles (upper limits)} indicate the observed column densities from different surveys. 
To mitigate bias, we plan to include observations from more surveys in the future, like the COS-Weak (\citealt{10.1093/mnras/sty529}) and the MAGG surveys (\citealt{10.1093/mnras/staa3147}), for which the virial radii of the foreground absorbing galaxies are not readily available and needs to independently determined from galaxy surveys, for example, by using halo abundance matching (\citealt{Churchill_2013}).   
} 

While our MgII column density profiles pass through the data points, the OVI columns are somewhat underestimated. This is also reflected in a lower X-ray surface brightness estimate (see Tab. \ref{tab:comparison_summary}). This is due to a lower hot/warm phase (median) density and temperature in our chosen {\tt TNG50-1} halo than the Milky Way estimates. This, however, is not too surprising since cosmological simulations show a large scatter in CGM properties at a given mass (\citealt{Ramesh2023}) \alankar{due to feedback, mergers, and different evolution histories of different halos. Detailed analysis of such variations and their implication on our models and synthetic observables is left for future}. Also note that the large variation in MgII columns as compared to OVI \alankar{in observed data} is a signature of the \ps{patchiness} of cold clouds, making them frequent only along a few sightlines as compared to 
a more area-filling warm phase (see discussion in section \ref{sec:clouds}). \alankar{These CGM models, therefore, provide a baseline estimate and need to be complemented with a model of discrete clouds to achieve better match with different multi-wavelength observations.} 

\section{Discussion}
\label{sec:discussion}
Two approaches to model the CGM are used commonly: (i) 1D profiles (hydrostatic or purely phenomenological) of the dominant hot phase and (ii) numerical simulations of the multiphase CGM that incorporate gas at all temperatures $\gtrsim 10^4$ K. While approach (i) may be satisfactory for the hot phase and can be confronted with X-ray observations, it does not incorporate warm and cooler phases routinely observed in quasar absorption line surveys of the CGM. Approach (ii), while simulating gas with a range of temperatures, also suffers from drawbacks such as \ps{computational expense}, insufficient resolution to resolve the structure of cooler phases, dependence on subgrid physics, and a limited statistical sample. 

In this context, we present a flexible analytic framework for modeling multiphase gas with log-normal distributions that are not only analytically tractable but also motivated by numerical simulations. Our approach can incorporate inputs constrained by numerical simulations and can quickly produce synthetic observables that can be tested against multi-wavelength observations. We can study not just trends with halo mass and environment, but this formalism also provides a baseline prediction for tracers of cold/warm gas, which can be made more realistic by incorporating clouds with \ps{non-trivial} area covering fraction discussed later in this section \ref{sec:clouds}.

\subsection{Interpreting X-ray observations}
\label{sec:X-ray_confusion}
Interpretation of X-ray emission \alankar{spectra} involves breaking up the observed spectrum into emission from different 
sources such as the CGM, the local hot bubble (\citealt{Liu_2017}), cosmic X-ray background (\citealt{2007A&A...463...79G}), and solar wind charge exchange (\citealt{Ponti2022}). 

In addition to these sources, many recent observations of X-ray emission from the Milky Way CGM (\citealt{Das_2019XMM_em, Bluem2022, Ponti2022}) consider two isothermal components (APEC models; \citealt{Smith2001}). The dominant contribution among these two components comes from the gas at the virial temperature of the Galaxy ($\sim 0.2 \ {\rm keV}$) while the sub-dominant (about an order of magnitude lower in emission) contribution is from an additional phase at a higher temperature $\sim 0.7 \ {\rm keV}$. In some of these works, the physical origin of this $\sim 0.7 \ {\rm keV}$ gas is 
the Galactic coronal disk maintained by supernovae-driven outflows (e.g., \citealt{1980ApJ...236..577B}).
However, the emission from the physical disk contributes not only to the $\sim 0.7 \ {\rm keV}$ APEC component in the X-ray emission but also to the $\sim 0.2 \ {\rm keV}$ component at virial temperature (\citealt{Kaaret_2019}). Some observations attribute the physical origin of the X-ray emission from the entire $\sim 0.2 \ {\rm keV}$ component ($\sim 0.2 \ {\rm keV}$ isotherm in the APEC model) to the CGM (\citealt{Ponti2022}) but it might have a non-negligible contribution from the disk.

Spectral fitting of APEC models along any single sightline cannot distinguish among physically distinct multiple components \alankar{with the same temperature contributing to} the total emission. Considering the EM from only the CGM of the Milky Way, 
towards and away from the Galactic center, one expects a variation of $\lesssim 2$ for high Galactic latitudes ($|b| \gtrsim 30^\circ$; see middle column of Fig. \ref{fig:DM_EM}).
However, the variation 
as observed by \citealt{Bluem2022} at the same latitudes is $\gtrsim 3$, \alankar{which favors models that include a coronal disk component} (see also \citealt{Ponti2022}).
The presence of such a disk has also been pointed out in previous works (\citealt{Yamasaki2020, Kaaret_2019}). Therefore, we \alankar{highlight the possibility} that the $\sim 0.2 \ {\rm keV}$ isothermal APEC model termed as the CGM in some observations might actually be {\it disk+CGM} and all the gas at the virial temperature cannot be assigned to just the spherical halo of the Milky Way. X-ray surveys such as the eROSITA all-sky survey (eRASS; \citealt{Predehl2021}) are expected to map out the diffuse X-ray emission at a range of latitudes and longitudes and help us break the degeneracy between various physical components that produce the X-ray spectra along different directions. \ps{Our models assume smooth disk and CGM, but the true gas distribution can have large variations (e.g., \citealt{Das2021_mnras}), which will affect our predictions quantitatively. Moreover, a careful statistical estimate of parameters of even our smooth model is left for future.}

\begin{figure}
    \centering
	\includegraphics[width=1\columnwidth]{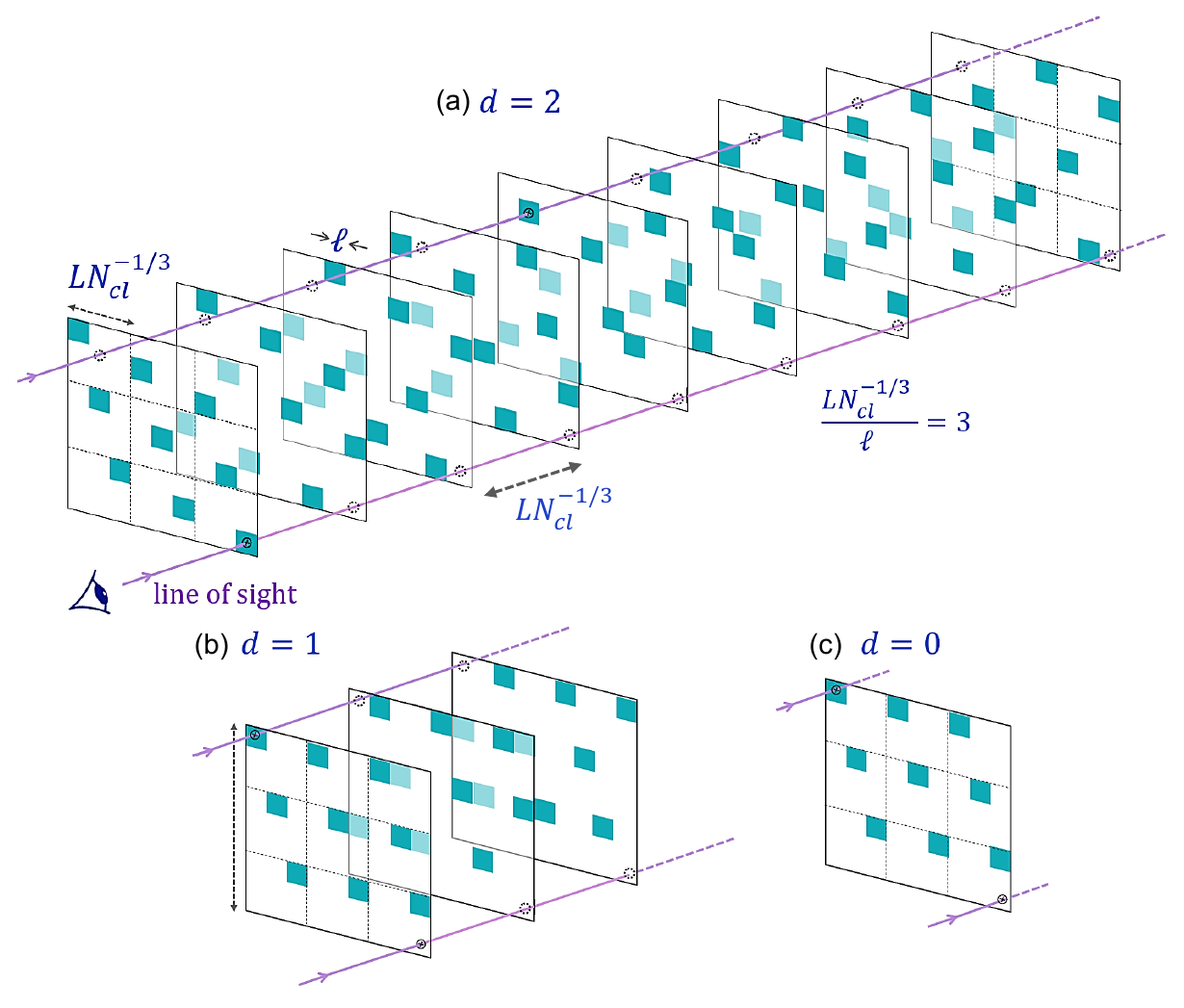}
 \caption{A portion of the CGM with clouds (a) spread out in two directions perpendicular to the LOS ($d=2$), (b) spread out in one direction perpendicular to LOS ($d=1$), and (c) no spread perpendicular to LOS ($d=0$). 
 Each blue square represents cubical clouds of size $l$. Each $3 \times 3$ layer shows the arrangement of cubes of side $L N_{\rm cl}^{-1/3}$ (mean distance between clouds for the maximally separated, most probable arrangement), each of which contains just one cloud. In all cases shown (a), (b), and (c), every sightline covered by a cloud will have overlapping clouds if the pattern is repeated along the LOS. 
 There are $f_V^{-d/3}$ number of clouds spread maximally in $d$ dimensions perpendicular to LOS (see section \ref{sec:clouds}). The number of clouds aligned along LOS with non-zero column is thus $N_{\rm cl}^{1/3}/f_V^{-d/3}$ and the column density is $N=nl \times N_{\rm cl}^{1/3}/f_V^{-d/3} = nL f_V^{(1+d)/3}$. From the $d=2$ (the mist limit) to $d=0$, the observed column would be enhanced by a factor of $f_V^{-2/3}$ ($\sim 100$ for our fiducial values!).}
 \label{fig:nonoverlapping_clouds}
\end{figure}
\subsection{Finite-size clouds \& observational implications}
\label{sec:clouds}
In this paper, all the LOS integrals were computed in the mist limit (Fig. \ref{fig:cartoon}b) \alankar{where a small volume fraction is filled by the cold/warm phases, comprising an infinite number of clouds each of which is infinitesimally small.} However, real CGM clouds have a finite size, and the observed area filling fraction is smaller than unity (and different for different \ps{ions and} column density thresholds of detection) for the tracers of warm/cold phase (\citealt{Augustin21}). 
Thus, clouds do not cover all the sightlines, and this natural variability needs to be considered when predicting the column densities of ions. A comprehensive analysis of cloud \alankar{morphology and their distribution} is beyond the scope of the present paper. \alankar{Nevertheless, in this subsection, we discuss the `mist' limit of cold/warm clouds in the CGM and demonstrate how such a configuration provides the most probable baseline configuration for the column density of ions. 
Additionally,} we discuss some qualitative effects of a more realistic spatial distribution of clouds (departing from the mist limit) on the observables, \alankar{which can potentially lower as well as increase the observed column densities of different ions, resulting in large scatter. Cloud distribution also affects the observed area covering fractions of the absorbing clouds.}

The column densities and covering fractions of the clouds (corresponding to the phases with small volume filling fractions) depend on their arrangement within the CGM volume. For simplicity, let us consider $N_{\rm cl}$ non-overlapping clouds, each assumed to be a cube of side $l$ with gas density $n$, arranged within a cube of side $L$ (representing the CGM) such that $l \ll L$. The volume fraction of clouds $f_V$ is given by $f_V = (l/L)^3 N_{\rm cl} $. We can express the cloud length $l$ in terms of $f_V$ and $N_{\rm cl}$ as $l = L f_V^{1/3} N_{\rm cl}^{-1/3}$. The number of MgII clouds identified in {\tt Illustris TNG50-1} halos is $N_{\rm cl} \gtrsim 10^4$ (\citealt{Nelson2020, Dutta2021}). We adopt a fiducial value of $N_{\rm cl} = 10^6$ to account for clouds smaller than the resolution limit of typical state-of-the-art cosmological simulations. The fiducial volume fraction of the cold phase is taken as $f_V=10^{-3}$, consistent with our {\tt Illustris TNG50-1} halo (see $f_V^{(c)}$ in Tab. \ref{tab:3p_params}); these give $l/L = 0.001$ or $l = 100$ pc for a CGM size of 100 kpc.

The maximum possible column density $N_{\rm max}$ is for the highly improbable arrangement of independent clouds {\em when all clouds lie along the line of sight} and the value, in this case, is $N_{\rm max} = n l N_{\rm cl} = n L f_V^{1/3} N_{\rm cl}^{2/3}$ ($= 1000 nL$ for fiducial parameters; $n$ is the local number density of the ion of interest within the cloud).\footnote{{\fontsize{7pt}{7.2pt}\selectfont Note that the combined length of the cloud, in this case, is $l N_{\rm cl} = L f_V^{1/3} N_{\rm cl}^{2/3} \gg L$. We consider this highly improbable case only for illustration of an extreme limit.}}
As the projected plane is covered by just one cloud, the area covering fraction for such a high column density is $f_A = l^2/L^2 = f_V^{2/3} N_{\rm cl}^{-2/3}$ (only $10^{-6}$ for fiducial parameters!). The number of clouds along this LOS is $N_{\rm cl, LOS} = N_{\rm cl}$ while any other parallel sightline encounters no cloud. 
The product of the column density and the area covering fraction is $N f_A = nL f_V$, which is fixed for all cloud arrangements because \alankar{$n f_V L^3  = N f_A L^2$} is just the total number of ions in the CGM, which is assumed to be a constant. The configuration where all clouds are lined up next to each other is however extremely unlikely. 

Fig. \ref{fig:nonoverlapping_clouds} shows specific examples of different arrangements of clouds. Panel (a) shows the {\em most probable separation of clouds} that corresponds to an equal separation between them; i.e., the mean distance between clouds is $(L^3/N_{\rm cl})^{1/3} = L N_{\rm cl}^{-1/3}$. This is because the volume of the CGM available for each non-overlapping cloud is $L^3/N_{\rm cl}$. In one of the most likely arrangements shown in (a) clouds are maximally spread out in all directions (along the LOS and also along both directions perpendicular to it). As an example, we choose the mean separation between the clouds to be three cloud sizes (i.e., $LN_{\rm cl}^{-1/3}/l=3$). 
Since there are $N_{\rm cl}^{1/3}$ clouds in a direction, and they are maximally separated in the perpendicular plane ($d=2$), the number of clouds overlapping along the LOS is $N_{\rm cl,LOS} = N_{\rm cl}^{1/3}/(LN_{\rm cl}^{-1/3}/l)^2 = f_V^{2/3} N_{\rm cl}^{1/3}$. The column density is, therefore, $N=  (n l) N_{\rm cl, LOS} =nLf_V$. Note that this is nothing but the integral of global number density along the LOS (see Eqs. \ref{eq:rhoL_rhoG} \& \ref{eq:NOVIIcolden}), \alankar{ which corresponds to the column density in the mist limit. {\it Thus, the integral of global number density along the LOS corresponds to the most probable arrangement of independent tiny clouds with minimal overlap 
along the way and can be taken as the baseline value for comparing with observations.}} Since every sightline (both along and perpendicular to the LOS) has equal column density in this case, \alankar{ the area covering fraction is 1} (assuming $N_{\rm cl} > f_V^{-2}$ so that all sightlines intercept at least one cloud on average). 
In this case, on moving the LOS in any direction by one cloud size, the encountered cloud configuration is statistically identical.

If clouds are spread out only in one direction in the plane perpendicular to the LOS ($d=1$; see Fig. \ref{fig:nonoverlapping_clouds}(b)), the column density is $N=nL f_V^{2/3}$ and the number of clouds along the LOS is $N_{\rm cl, LOS} =N_{\rm cl}^{1/3}/(LN_{\rm cl}^{-1/3}/l)= f_V^{1/3} N_{\rm cl}^{1/3}$, with area covering fraction $f_A=l N_{\rm cl}^{1/3}/L = f_V^{1/3}$. In all cases, the product of the area covering fraction and the column density is a constant equal to $nL f_V$, as explained in the previous paragraphs. 
The clouds overlap in all dimensions across the LOS in Fig. \ref{fig:nonoverlapping_clouds}(c) corresponding to $d=0$ and the column density is $nL f_V^{1/3}$,  such that every sightline with non-zero column has $N_{\rm cl}^{1/3}$ clouds.
Therefore, in general, for any configuration with spread in $d$ number of directions ($d=0/1/2$), the total number of clouds along the LOS is $N_{\rm cl, LOS} = N_{\rm cl}^{1/3}/(LN_{\rm cl}^{-1/3}/l)^d = N_{\rm cl}^{1/3} f_V^{d/3}$, which gives the column density $N = N_{\rm cl, LOS} \times n l = n L f_V^{(1+d)/3}$.

\alankar{{\it A smaller area covering fraction implies that the value of column density along \alankar{some} sightlines that encounter warm/cold clouds can be much larger than our mist estimate. 
\alankar{Similarly, some other sightlines would encounter a significantly lower column}. This \alankar{deviation from the mist estimate} is expected to be larger for phases with smaller volume fraction}} $f_V$, such as MgII tracing $10^4$ K clouds (see Fig. \ref{fig:PDFs}; Tab. \ref{tab:3p_params}). A large fluctuation in MgII columns (in comparison to OVI which traces a warmer phase) and a small covering fraction implied by data  in Fig. \ref{fig:1zone_CDens} can be explained by variable overlap of clouds along various sightlines. \alankar{The column density can, therefore,  be either enhanced or diminished by orders of magnitude compared to our mist estimate.} 
Observations provide a statistical sample of several LOS and contain upper limits \ps{that lie below the mist prediction} which can be interpreted with the baseline predictions of the mist limit; \ps{these sightlines simply do not encounter substantial cold clouds!}

The UV absorption spectra indicate the presence of several components along the LOS (e.g., \citealt{Stocke2013, Werk2014, Zahedy2019, Haislmaier2021}), indicating multiple clouds along a typical LOS. In the case of lensing of the background quasar, an estimate of the transverse extent of clouds (e.g., \citealt{Rudie2019, Augustin21}) can provide crucial information on cloud properties. Observational constraints such as these, combined with extensions of the toy cloud model presented here, can provide a wealth of information about the properties of the CGM clouds in different phases despite being unresolved.

\begin{table*}
\caption{{\fontsize{10pt}{10.2pt}\selectfont Comparison of models with observations ${}^{*}$}}
\label{tab:comparison_summary}
\begin{threeparttable}
\centering
\resizebox{\textwidth}{!}{%
\setlength{\tabcolsep}{0.5em} 
{\renewcommand{\arraystretch}{1.5}
\begin{tabular}{|r|llllll|}
\hline
\multicolumn{1}{|c|}{} &
  \multicolumn{1}{c|}{} &
  \multicolumn{5}{c|}{Models} \\ \cline{3-7} 
\multicolumn{1}{|c|}{} &
  \multicolumn{1}{c|}{} &
  \multicolumn{1}{c|}{} &
  \multicolumn{1}{c|}{} &
  \multicolumn{3}{c|}{three-phase CGM} \\ \cline{5-7} 
\multicolumn{1}{|c|}{\multirow{-3}{*}{Observables}} &
  \multicolumn{1}{c|}{\multirow{-3}{*}{Observed value}} &
  \multicolumn{1}{c|}{\multirow{-2}{*}{$\gamma = 1 ^{**}$}} &
  \multicolumn{1}{c|}{\multirow{-2}{*}{$\gamma = 5/3 ^{**}$}} &
  \multicolumn{1}{c|}{$\alpha = 0$} &
  \multicolumn{1}{c|}{$\alpha = 1^{***}$} &
  \multicolumn{1}{c|}{$\alpha = 2$} \\ \hline
\multicolumn{1}{|c|}{\begin{tabular}[c]{@{}c@{}} Dispersion measure (l,b) [$\rm pc \ cm^{-3}$] \\ 
CHIME {[}Bhardwaj +21${}^{\S}${]} \end{tabular}} &
  \multicolumn{1}{c|}{($142^{\circ}$,$41^{\circ}$)  [30]} &
  \multicolumn{1}{c|}{\cellcolor[HTML]{B7DAB7} 30.11 (7.04) } &
  \multicolumn{1}{c|}{\cellcolor[HTML]{B7DAB7} 20.55 (7.04) } &
  \multicolumn{1}{c|}{\cellcolor[HTML]{B7DAB7} 13.20 (7.04) } &
  \multicolumn{1}{c|}{\cellcolor[HTML]{B7DAB7} 21.81 (7.04) } &
  \multicolumn{1}{c|}{\cellcolor[HTML]{D3A7A7} 75.37 (7.04) } \\ \hline
\multicolumn{1}{|c|}{\begin{tabular}[c]{@{}c@{}}Emission measure (l,b) [$\rm 10^{-2}\ pc \ cm^{-6}$] \\ 
eFEDS {[}Ponti +23${}^{\dag}${]}\end{tabular}} &
  \multicolumn{1}{c|}{($230^{\circ},30^{\circ}$) [$2.9 - 3.1$]} &
  \multicolumn{1}{c|}{\cellcolor[HTML]{B7DAB7} 1.33 (0.77) } &
  \multicolumn{1}{c|}{\cellcolor[HTML]{B7DAB7} 0.99 (0.77) } &
  \multicolumn{1}{c|}{\cellcolor[HTML]{B7DAB7} 0.79 (0.77) } &
  \multicolumn{1}{c|}{\cellcolor[HTML]{B7DAB7} 0.98 (0.77) } &
  \multicolumn{1}{c|}{\cellcolor[HTML]{D3A7A7} 18.45 (0.77) } \\ \hline
\multicolumn{1}{|c|}{\begin{tabular}[c]{@{}c@{}}Surface brightness [$\rm 10^{-13} erg\ cm^{-2}\ s^{-1}\ deg^{-2}$] \\ 
eFEDS {[}Ponti +23${}^{\ddag}${]}\end{tabular}} &
  \multicolumn{1}{c|}{\begin{tabular}[c]{@{}c@{}}($230^{\circ},30^{\circ}$) \\
0.3-0.6 keV: [15.6] \\
0.6-2.0 keV: [4.9]\end{tabular}} &
  \multicolumn{1}{c|}{\cellcolor[HTML]{B7DAB7}
  \begin{tabular}[c]{@{}c@{}}
   14.22 (11.04)\\
   1.78 (1.42)\end{tabular}} &
  \multicolumn{1}{c|}{\cellcolor[HTML]{B7DAB7}
  \begin{tabular}[c]{@{}c@{}}
   13.66 (11.04)\\
   1.84 (1.42)\end{tabular}} &
  \multicolumn{1}{c|}{\cellcolor[HTML]{EBEB8A}
  \begin{tabular}[c]{@{}c@{}}
   11.15 (11.04) \\
   1.44 (1.42) \end{tabular}} &
  \multicolumn{1}{c|}{\cellcolor[HTML]{EBEB8A}
  \begin{tabular}[c]{@{}c@{}}
   11.17 (11.04) \\
   1.45 (1.42) \end{tabular}} &
  \multicolumn{1}{c|}{\cellcolor[HTML]{B7DAB7}
  \begin{tabular}[c]{@{}c@{}}
   14.72 (11.04) \\
   1.55 (1.42) \end{tabular}} \\ \hline
\multicolumn{1}{|c|}{Ion columns ($\rm cm^{-2}$)} &
  \multicolumn{6}{l|}{} \\ \hline
OVIII &
  \multicolumn{1}{l|}{$2.4\times10^{15}-4.2\times10^{15}$} &
  \multicolumn{1}{l|}{\cellcolor[HTML]{B7DAB7}$2.6\times10^{14}-1.1\times10^{16}$} &
  \multicolumn{1}{l|}{\cellcolor[HTML]{B7DAB7}$7.6\times10^{13}-7.7\times10^{15}$} &
  \multicolumn{1}{l|}{\cellcolor[HTML]{B7DAB7}$5.3\times10^{14}-1.8\times10^{15}$} &
  \multicolumn{1}{l|}{\cellcolor[HTML]{B7DAB7}$3.6\times10^{14}-4.7\times10^{15}$} &
  \cellcolor[HTML]{B7DAB7}$1.9\times10^{14}-2.4\times10^{16}$ \\ \hline
OVII &
  \multicolumn{1}{l|}{$3.6\times10^{15}-9.6\times10^{15}$} &
  \multicolumn{1}{l|}{\cellcolor[HTML]{B7DAB7}$3.5\times10^{14}-3.3\times10^{16}$} &
  \multicolumn{1}{l|}{\cellcolor[HTML]{B7DAB7}$8.9\times10^{13}-1.7\times10^{16}$} &
  \multicolumn{1}{l|}{\cellcolor[HTML]{B7DAB7}$9.6\times10^{14}-3.2\times10^{15}$} &
  \multicolumn{1}{l|}{\cellcolor[HTML]{B7DAB7}$6.6\times10^{14}-8.6\times10^{15}$} &
  \cellcolor[HTML]{B7DAB7}$3.6\times 10^{14}-4.4\times10^{16}$ \\ \hline
OVI &
  \multicolumn{1}{l|}{$4.5\times10^{13}-1.4\times10^{15}$} &
  \multicolumn{1}{l|}{\cellcolor[HTML]{B7DAB7}$1.1\times10^{13}-1.7\times10^{15}$} &
  \multicolumn{1}{l|}{\cellcolor[HTML]{B7DAB7}$4.6\times10^{12}-8.7\times10^{14}$} &
  \multicolumn{1}{l|}{\cellcolor[HTML]{B7DAB7}$4.2\times10^{13}-1.4\times10^{14}$} &
  \multicolumn{1}{l|}{\cellcolor[HTML]{B7DAB7}$2.8\times10^{13}-3.7\times10^{14}$} &
  \cellcolor[HTML]{B7DAB7}$1.5\times10^{13}-1.9\times10^{15}$ \\ \hline
NV &
  \multicolumn{1}{l|}{$4.9\times10^{13}-1.5\times10^{14}$} &
  \multicolumn{1}{l|}{\cellcolor[HTML]{B7DAB7}$5.4\times10^{11}-1.5\times10^{14}$} &
  \multicolumn{1}{l|}{\cellcolor[HTML]{B7DAB7}$1.2\times10^{11}-6.2\times10^{13}$} &
  \multicolumn{1}{l|}{\cellcolor[HTML]{B7DAB7}$4.1\times10^{12}-1.4\times10^{13}$} &
  \multicolumn{1}{l|}{\cellcolor[HTML]{B7DAB7}$2.8\times10^{12}-3.7\times10^{13}$} &
  \cellcolor[HTML]{B7DAB7}$1.5\times10^{12}-1.9\times10^{14}$ \\ \hline
MgII &
  \multicolumn{1}{l|}{$2.4\times10^{12}-1.8\times10^{13}$} &
  \multicolumn{1}{l|}{\cellcolor[HTML]{D3A7A7}$4.1\times10^5-1.8\times10^9$} &
  \multicolumn{1}{l|}{\cellcolor[HTML]{D3A7A7}$2.4\times10^3-4.3\times10^8$} &
  \multicolumn{1}{l|}{\cellcolor[HTML]{B7DAB7}$5.7\times10^{12}-1.9\times10^{13}$} &
  \multicolumn{1}{l|}{\cellcolor[HTML]{B7DAB7}$3.9\times10^{12}-5.1\times10^{13}$} &
  \cellcolor[HTML]{B7DAB7}$2.1\times 10^{12}-2.6\times10^{14}$ \\ \hline
\end{tabular}
}}
\begin{tablenotes}
\item [] {\fontsize{7pt}{7.2pt}\selectfont Note that for ion columns, 
only detections are listed in the second column "Observed value". In case of just detections, only upper or lower limits exist and no specific \\observed values are available ({\em triangles} in Fig. \ref{fig:column_densities}). For NV there mostly exists upper limits in detection. \alankar{\textit {Values in parentheses show the disk contribution, when considered.}}} 
\item [*] {\fontsize{7pt}{7.2pt}\selectfont Colors indicate agreement with corresponding observation [{\em Green}: Good; {\em Yellow}: Moderate; {\em Red}: Unacceptable].}
\item [**] {\fontsize{7pt}{7.2pt}\selectfont For these, we consider isochoric modification.}
\item [***] {\fontsize{7pt}{7.2pt}\selectfont $\alpha = 1$ is a singularity in Eq. \ref{eq:N_col_pl_prof}), so we used $\alpha = 1.01$.}
\item [${}^{\S}$] {\fontsize{7pt}{7.2pt}\selectfont Tab. 2 in \citealt{Bhardwaj2021}}
\item [${}^{\dag}$] {\fontsize{7pt}{7.2pt}\selectfont Tab. 2 in \citealt{Ponti2022}}
\item [${}^{\ddag}$] {\fontsize{7pt}{7.2pt}\selectfont Tab. 4 in \citealt{Ponti2022}}
\end{tablenotes}
\end{threeparttable}
\end{table*}

\subsection{Future directions}

In this paper, we have only focused on the emission and dispersion measures and column densities, but observations provide a wealth of other diagnostics including kinematic information. Our models can be extended to include velocity distributions, which may be drawn from PDFs with 1-point and 2-point statistics consistent with observations and galaxy formation simulations. In the following, we briefly explain the utility of a library of different CGM models that can be compared with independent observational constraints.

\subsubsection{Towards a library of models}

In this paper, we introduced probabilistic models of the CGM, including a new three-phase model. We used these models to predict various observables and compared them with observations. A combination of various observational constraints helps us \ps{quantitatively} assess various models (see Tab. \ref{tab:comparison_summary}).
This motivates an effort towards developing a library of models that can be continuously expanded to include the existing and new CGM models of varying complexity, from simple 1D profiles to 2D axisymmetric rotating models (e.g., \citealt{Sormani2018}), and to models like ours based on PDFs. In addition to CGM observables highlighted in this paper, we aim to add more observables such as the scattering measure (e.g., \citealt{Ocker2021}) which is sensitive to the size of CGM clouds, and thermal Sunyaev-Zeldovich $y-$ parameter (e.g., \citealt{Bregman2022}) which is an excellent tracer of the CGM mass. We can also include models of magnetic fields and turbulence in the CGM. 
\alankar{These models can constrain the turbulent and magnetic support in the
CGM when compared against the Faraday rotation measures observed from the CGM of external galaxies (\citealt{Hafen2024, 2023arXiv230811391B}).}

Such a library of models and observable predictions would enable one to quickly infer the physical properties and chemical composition of the CGM from observations, constrain model parameters, break degeneracy across models, and rank models according to the number of independent observations that they can match. We aim to expand our public code repository {\tt MultiphaseGalacticHaloModel} (link provided in the Data availability section \ref{sec:data_avail}) into a large library of CGM models and observables. 

\alankar{We discuss the CGM model proposed in \citealt{Yamasaki2020} as an example to demonstrate the insights that can be gained by comparing models to multi-wavelength observations.} In this work, the disk emission at the virial temperature of the spherical CGM is approximately an order of magnitude higher than the emission from the CGM gas along any line of sight (e.g., see their Fig. 1). However, the disk contribution in our models is either \ps{slightly} higher or comparable to the spherical CGM (see Fig. \ref{fig:ratio_em}).
Considering only Milky Way observations, it is not possible to break the degeneracy between our models and the \citealt{Yamasaki2020} model. Low disk contribution to EM in the \citealt{Yamasaki2020} model is a result of a low CGM central density of, $3.7 \times 10^{-4}\ {\rm cm^{-3}}$ in contrast to our models ($\gamma=1,~5/3$ unmodified profiles) with the CGM density at 10 kpc  $\sim 10^{-3}$ cm$^{-3}$ (see Fig. \ref{fig:number_density}).
In the observations of external CGMs, for most sightlines the contribution of the central disk is negligible, and the density of the CGM in the \citealt{Yamasaki2020} model is too low to produce a sufficiently large column of ions like OVI, NV. 
We can, therefore, justify the choice of parameters favoring a higher density CGM as used in our models by considering UV absorption studies of the CGM of Milky Way-like external galaxies. 
We also have to bear in mind that the CGM for the same halo mass can show a large scatter in physical properties (\citealt{Ramesh2023}).

\subsubsection{Fast plasma modeling repository: {\tt AstroPlasma}}
Generating observables like dispersion/emission measures or column densities of ions from fluid fields like density and temperature requires plasma modeling. One of the goals of our code base is to be computationally efficient for quick exploration of the parameter space of various CGM models and to compare with different observations. {\tt AstroPlasma} is a public code repository written as a standalone package that can provide functions that generate ionization properties (e.g., various ion fractions, mean molecular weight) for conditions in the CGM.

{\tt AstroPlasma} uses a large database of pre-run {\tt CLOUDY} (\citealt{2017RMxAA..53..385F}) models to interpolate the ionization properties for a range of CGM plasma conditions. This database can be expanded or updated as needed. We are able to achieve a low computational cost for evaluating observables by eliminating the need for on-the-fly calculation of expensive chemical networks. {\tt AstroPlasma} instead looks up its database of {\tt CLOUDY} models to interpolate the plasma properties across densities, temperatures, and metallicities. Currently, our database has plasma conditions for collisional ionization equilibrium (CIE)  and photo-ionization equilibrium (PIE) in the presence of Haardt-Madau UV background (\citealt{Haardt2012}) at different cosmological redshifts. 
We assume optically thin conditions and do not consider any radiative transfer effects. Appendix \ref{app:AstroPlasma} illustrates some 
common usage of {\tt AstroPlasma}.

\subsection{Comparison with related works}
Single phase models (e.g., \citealt{Maller2004, Sharma2012, Miller2013, Nakashima2018, Voit2019}) typically used to model the hot phase are insufficient even qualitatively to reproduce the properties of the observed multiphase CGM. \citetalias{Faerman2017ApJ} made an advance towards modeling the warm phase traced by OVI by adding a phase at $10^{5.5}$ K with a log-normal volume distribution over temperature on top of the isothermal hot phase. Later, \citet{Faerman2019ApJ} reproduced OVI column densities by assuming an isentropic, hydrostatic hot gas profile (with significant non-thermal pressure support) for which the outer temperatures are small enough to reproduce OVI column densities. Since this model cannot produce cold gas, \citet{Faerman2023} recently extended their isentropic model with non-thermal pressure support (due to turbulence, cosmic rays, and magnetic fields) to include a cold component at $\sim 10^4$ K in photo-ionization and thermal equilibrium. The cold phase with a small volume fraction ($f_V^{(c)} \sim 0.01$) and a non-thermal pressure fraction higher than the hot phase could reproduce the ballpark values of the column densities of the ions tracing the cold phase. Because of the lack of $\sim 10^5$ K gas in this model, it generally under-predicts the columns of intermediate ions such as CIV and SiIV. Based on PDFs motivated by a large spread in CGM temperatures at most radii in observations and simulations, our approach is fundamentally different and more akin to \citetalias{Faerman2017ApJ}. 

\citet{Ramesh2023} have systematically analyzed the CGM properties of 132 Milky Way-like halos in {\tt Illustris TNG50-1} cosmological galaxy formation simulations. They find a large spread in CGM properties of their sample (e.g., mass fraction in various CGM phases). The CGM properties depend on the specific star formation rate of the galaxy. Similarly, there is a systematic increase in the CGM X-ray luminosity with an increasing stellar mass of the central galaxy. The temperature-density PDF (the colored 2D histograms shown in Fig. \ref{fig:2dPDFs}) of their CGMs is affected by ongoing AGN feedback (see their Fig. 9). Inputs from their statistical study can be incorporated into our formalism, largely motivated by a large spread of densities and temperatures in CGM  simulations (see also \citealt{Esmerian2021, Fielding2020}, etc.), and variations in CGM properties over a large range of halo masses and redshifts can be quickly calculated. 

In between the simple models of hot CGM and complex cosmological simulations lie the high-resolution idealized simulations that focus on the physics of flows and thermodynamics around cold gas moving through the hot CGM (e.g., \citealt{Armillotta2016, Gronke2018, Kanjilal2021MNRAS, Mohapatra2022characterising, Mohapatra2023, Yang2023}). The boundary layers around such clouds are also multiphase with a characteristic volume PDF covering a broad range of temperatures (e.g., see Fig. 5 in \citealt{Kanjilal2021MNRAS}, Fig. 4 in \citealt{Mohapatra2023}) from $\sim 10^4$ K to $\gtrsim 10^6$ K. The relation between density/temperature fluctuations and turbulent Mach number in turbulent cooling layers is fundamentally different from isotropic homogeneous turbulence. For the latter, the rms fluctuations in $\ln \rho/\langle \rho \rangle$ scale as ${\cal M}^2$ (${\cal M}$ is turbulent Mach number) but for radiative turbulence density fluctuations are much higher due to radiative cooling (e.g., \citealt{Mohapatra2019}). The multiphase CGM can be thought of as a superposition of several such clouds, with their radiative boundary layers at intermediate temperatures and the confining pressure decreasing away from the center. Our formalism based on PDFs is also capable of capturing the temperature distribution of such boundary layers. Thus, a description based on PDFs, which is consistent with both the small-scale structure of the CGM clouds and galaxy formation simulations, seems appropriate for studying the CGM as a whole.
 
\section{Summary}
\label{sec:summary}
Here, we summarize the most important results and implications of our work.
\begin{enumerate}
\item We have highlighted the need to move beyond simple profiles to model the CGM. Simulations, both idealized and cosmological, indicate the cospatial presence of cold, warm, and hot gas.
We have shown that a probabilistic model of the multiphase CGM can reliably explain results from multi-wavelength observations (see Figs. \ref{fig:column_densities}, \ref{fig:DM_EM}, \ref{fig:spectrum}, \ref{fig:1zone_CDens}). These models, despite being more complex, still remain largely analytic. The standout improvement over the existing CGM models like those introduced in \citealt{Faerman2017ApJ, Faerman2019ApJ, Faerman2023} is that our probabilistic model can simultaneously explain the presence of most ions observed in UV absorption spectra of quasar sightlines passing through the CGM of intervening galaxies.
Additionally, our probabilistic models match the observations of the Milky Way CGM well. These include dispersion measure from a nearby FRB (\citealt{Bhardwaj2021}) and X-ray emission measure in the soft X-ray bands (\citealt{Kaaret_2019, Das2019ApJ, Bluem2022, Ponti2022}). 

\item We clarify an apparent confusion in the identification of physical sources of diffuse soft X-ray emission spectra \ps{from the Milky Way}. We highlight that 
the observed X-ray emission from the gas at the virial temperature of the CGM ($\sim 0.2 \ {\rm keV}$) 
\alankar{may be dominated by 
a dense disk rather than the spherically symmetric CGM (see Fig. \ref{fig:ratio_em}).}

The emission from this dense disk in addition to the CGM is degenerate in the isothermal APEC modeling of the X-ray spectrum along a single sightline. This degeneracy can, however, be lifted if the spatial variation of emission measure across multiple sightlines is considered. The disk can contribute to both $\sim 0.2 \ {\rm keV}$ phase and the newly introduced phase at $\sim 0.7 \ {\rm keV}$ (see \citealt{Das2019ApJ, Bluem2022, Ponti2022}). Since only the $\sim 0.7 \ {\rm keV}$ component is referred to as the Galactic coronal disk in \citealt{Ponti2022}, we highlight the \alankar{possibility that} $\sim 0.7 \ {\rm keV}$ gas in the disk is but a minor contributor to the X-ray emission, and a significantly larger emission comes from the $\sim 0.2 \ {\rm keV}$ phase of the disk. 
It is also physically plausible to have a dense disk \ps{with a broad spread of} temperatures between $0.2 \ {\rm keV}$ and $0.7 \ {\rm keV}$, but this gets decomposed into emission from two components due to the specifics of APEC modeling (e.g., \citealt{Vijayan2022}). Further investigation on this uncertainty is beyond the scope of this work.

\item We address the \ps{plausible} reason for large variations in the observed cold ion column densities across multiple quasar sightlines through different external galaxies (e.g., see Fig. \ref{fig:1zone_CDens}). Starting from the warm to the cold phase, the gas becomes progressively less volume filling. We introduce scaling relations in terms of the number of clouds ($N_{\rm cl}$) and their volume filling fraction ($f_V$) that can be \ps{qualitatively} \alankar{motivated from the distribution of cold/warm clouds within the CGM (see section \ref{sec:clouds} \& Fig. \ref{fig:nonoverlapping_clouds}).} We make baseline predictions of column densities of ions in the \alankar{mist} limit. 

\item Our work motivates a library of CGM models, and synthetic observables matched against observations (e.g., Tab. \ref{tab:comparison_summary}). Given recent advances across diverse CGM observations across multiple wavelengths, directly comparing a wide range of models to different observations is warranted. \alankar{When confronted with new observations, a comparative framework can reveal model strengths, weaknesses, and biases.} Ultimately, this exercise will help us accurately estimate important physical parameters, such as the fraction of baryons in various phases of the CGM.

A library of models and observables will enable benchmarking against observations and elucidate model limitations and assumptions. We, therefore, create a publicly available code repository called {\tt MultiphaseGalacticHaloModel} and a computationally inexpensive plasma modeling database called {\tt AstroPlasma} to continually encompass existing and future CGM models and compare them with the latest observations.
\end{enumerate}

\section{Acknowledgements}

PS acknowledges a Swarnajayanti Fellowship (DST/SJF/PSA-03/2016-17) and a National Supercomputing Mission (NSM) grant from the Department of Science and Technology, India. Research of AD is supported by the Prime Minister's Research Fellowship (PMRF)  from the Ministry of Education (MoE), Govt. of India. AD acknowledges Gurkirat Singh for his efforts in building and testing {\tt AstroPlasma}. AD acknowledges Dylan Nelson for his help in processing and analyzing {\tt Illustris TNG50} data dumps. AD acknowledges Sayak Dutta, Zhijie Qu \begin{CJK*}{UTF8}{bsmi}(屈稚杰)\end{CJK*}, and Sowgat Muzahid for their help in accessing and analyzing the observational data. AD also acknowledges Yakov Faerman, Gabriele Ponti, Sanskriti Das, Hsiao-Wen Chen \begin{CJK*}{UTF8}{bsmi}(陳曉雯)\end{CJK*}, Andrea Afruni, Priyanka Singh, Kartick Sarkar, Sukanya Mallik, and our anonymous referee for their useful comments and discussions. We thank Gary Ferland and collaborators for the {\tt CLOUDY} code. 
This research also benefited from discussions at \href{https://www.kitp.ucsb.edu/activities/halo21}{Fundamentals of gaseous halos program (Halo21)}, which was supported in part by grant NSF PHY-2309135 to the Kavli Institute for Theoretical Physics (KITP).


\section{Data Availability}
\label{sec:data_avail}
We have made all the codes and data used in this paper public. The {\tt CLOUDY}-like plasma modeling tool that we developed is hosted in the {\it GitHub} repo \href{https://github.com/dutta-alankar/AstroPlasma}{\tt AstroPlasma}\footnote{{\fontsize{7pt}{7.2pt}\selectfont \href{https://github.com/dutta-alankar/AstroPlasma}{https://github.com/dutta-alankar/AstroPlasma}; the code for the accompanying web-server that hosts the {\tt AstroPlasma} database is in a separate {\it GitHub} repo: \href{https://github.com/tbhaxor/CloudyInterpolator}{https://github.com/tbhaxor/CloudyInterpolator}.}} for general use. All the CGM models and the observables used in this work are hosted as a part of an expanding library of CGM models in the {\it GitHub} repo \href{https://github.com/dutta-alankar/MultiphaseGalacticHaloModel}{{\tt MultiphaseGalacticHaloModel}}.\footnote{{\fontsize{7pt}{7.2pt}\selectfont \href{https://github.com/dutta-alankar/MultiphaseGalacticHaloModel}{https://github.com/dutta-alankar/MultiphaseGalacticHaloModel}}}
Any other relevant data associated with this article are available to the corresponding author upon reasonable request.



\bibliographystyle{mnras}
\bibliography{refs.bib} 

\begin{thebibliography}{}
\makeatletter
\relax
\def\mn@urlcharsother{\let\do\@makeother \do\$\do\&\do\#\do\^\do\_\do\%\do\~}
\def\mn@doi{\begingroup\mn@urlcharsother \@ifnextchar [ {\mn@doi@}
  {\mn@doi@[]}}
\def\mn@doi@[#1]#2{\def\@tempa{#1}\ifx\@tempa\@empty \href
  {http://dx.doi.org/#2} {doi:#2}\else \href {http://dx.doi.org/#2} {#1}\fi
  \endgroup}
\def\mn@eprint#1#2{\mn@eprint@#1:#2::\@nil}
\def\mn@eprint@arXiv#1{\href {http://arxiv.org/abs/#1} {{\tt arXiv:#1}}}
\def\mn@eprint@dblp#1{\href {http://dblp.uni-trier.de/rec/bibtex/#1.xml}
  {dblp:#1}}
\def\mn@eprint@#1:#2:#3:#4\@nil{\def\@tempa {#1}\def\@tempb {#2}\def\@tempc
  {#3}\ifx \@tempc \@empty \let \@tempc \@tempb \let \@tempb \@tempa \fi \ifx
  \@tempb \@empty \def\@tempb {arXiv}\fi \@ifundefined
  {mn@eprint@\@tempb}{\@tempb:\@tempc}{\expandafter \expandafter \csname
  mn@eprint@\@tempb\endcsname \expandafter{\@tempc}}}

\bibitem[\protect\citeauthoryear{{Amodeo} et~al.,}{{Amodeo}
  et~al.}{2021}]{Amodeo2021}
{Amodeo} S.,  et~al., 2021, \mn@doi [\prd] {10.1103/PhysRevD.103.063514}, \href
  {https://ui.adsabs.harvard.edu/abs/2021PhRvD.103f3514A} {103, 063514}

\bibitem[\protect\citeauthoryear{{Armillotta}, {Fraternali}  \&
  {Marinacci}}{{Armillotta} et~al.}{2016}]{Armillotta2016}
{Armillotta} L.,  {Fraternali} F.,   {Marinacci} F.,  2016, \mn@doi [\mnras]
  {10.1093/mnras/stw1930}, \href
  {https://ui.adsabs.harvard.edu/\#abs/2016MNRAS.462.4157A} {462, 4157}

\bibitem[\protect\citeauthoryear{{Asplund}, {Grevesse}, {Sauval}  \&
  {Scott}}{{Asplund} et~al.}{2009}]{Asplund2009}
{Asplund} M.,  {Grevesse} N.,  {Sauval} A.~J.,   {Scott} P.,  2009, \mn@doi
  [\araa] {10.1146/annurev.astro.46.060407.145222}, \href
  {https://ui.adsabs.harvard.edu/abs/2009ARA&A..47..481A} {47, 481}

\bibitem[\protect\citeauthoryear{Augustin, Péroux, Hamanowicz, Kulkarni,
  Rahmani  \& Zanella}{Augustin et~al.}{2021}]{Augustin21}
Augustin R.,  Péroux C.,  Hamanowicz A.,  Kulkarni V.,  Rahmani H.,   Zanella
  A.,  2021, \mn@doi [Monthly Notices of the Royal Astronomical Society]
  {10.1093/mnras/stab1673}, 505, 6195

\bibitem[\protect\citeauthoryear{{Bhardwaj} et~al.,}{{Bhardwaj}
  et~al.}{2021}]{Bhardwaj2021}
{Bhardwaj} M.,  et~al., 2021, \mn@doi [\apjl] {10.3847/2041-8213/abeaa6}, \href
  {https://ui.adsabs.harvard.edu/abs/2021ApJ...910L..18B} {910, L18}

\bibitem[\protect\citeauthoryear{Bhattacharyya, Das, Gupta, Mathur  \&
  Krongold}{Bhattacharyya et~al.}{2023}]{Bhattacharyya_2023}
Bhattacharyya J.,  Das S.,  Gupta A.,  Mathur S.,   Krongold Y.,  2023, \mn@doi
  [The Astrophysical Journal] {10.3847/1538-4357/acd337}, 952, 41

\bibitem[\protect\citeauthoryear{{Bluem} et~al.,}{{Bluem}
  et~al.}{2022}]{Bluem2022}
{Bluem} J.,  et~al., 2022, \mn@doi [\apj] {10.3847/1538-4357/ac8662}, \href
  {https://ui.adsabs.harvard.edu/abs/2022ApJ...936...72B} {936, 72}

\bibitem[\protect\citeauthoryear{{B{\"o}ckmann} et~al.,}{{B{\"o}ckmann}
  et~al.}{2023}]{2023arXiv230811391B}
{B{\"o}ckmann} K.,  et~al., 2023, \mn@doi [arXiv e-prints]
  {10.48550/arXiv.2308.11391}, \href
  {https://ui.adsabs.harvard.edu/abs/2023arXiv230811391B} {p. arXiv:2308.11391}

\bibitem[\protect\citeauthoryear{{Bregman}}{{Bregman}}{1980}]{1980ApJ...236..577B}
{Bregman} J.~N.,  1980, \mn@doi [\apj] {10.1086/157776}, \href
  {https://ui.adsabs.harvard.edu/abs/1980ApJ...236..577B} {236, 577}

\bibitem[\protect\citeauthoryear{{Bregman}, {Hodges-Kluck}, {Qu}, {Pratt}, {Li}
   \& {Yun}}{{Bregman} et~al.}{2022}]{Bregman2022}
{Bregman} J.~N.,  {Hodges-Kluck} E.,  {Qu} Z.,  {Pratt} C.,  {Li} J.-T.,
  {Yun} Y.,  2022, \mn@doi [\apj] {10.3847/1538-4357/ac51de}, \href
  {https://ui.adsabs.harvard.edu/abs/2022ApJ...928...14B} {928, 14}

\bibitem[\protect\citeauthoryear{{Buie}, {Fumagalli}  \& {Scannapieco}}{{Buie}
  et~al.}{2020a}]{Buie2020}
{Buie} Edward I.,  {Fumagalli} M.,   {Scannapieco} E.,  2020a, \mn@doi [\apj]
  {10.3847/1538-4357/ab65bc}, \href
  {https://ui.adsabs.harvard.edu/abs/2020ApJ...890...33B} {890, 33}

\bibitem[\protect\citeauthoryear{{Buie}, {Gray}, {Scannapieco}  \&
  {Safarzadeh}}{{Buie} et~al.}{2020b}]{Buie2020b}
{Buie} Edward I.,  {Gray} W.~J.,  {Scannapieco} E.,   {Safarzadeh} M.,  2020b,
  \mn@doi [\apj] {10.3847/1538-4357/ab9535}, \href
  {https://ui.adsabs.harvard.edu/abs/2020ApJ...896..136B} {896, 136}

\bibitem[\protect\citeauthoryear{Burchett et~al.,}{Burchett
  et~al.}{2019}]{Burchett_2019}
Burchett J.~N.,  et~al., 2019, \mn@doi [The Astrophysical Journal Letters]
  {10.3847/2041-8213/ab1f7f}, 877, L20

\bibitem[\protect\citeauthoryear{{Butsky} et~al.,}{{Butsky}
  et~al.}{2022}]{Butsky2022}
{Butsky} I.~S.,  et~al., 2022, \mn@doi [\apj] {10.3847/1538-4357/ac7ebd}, \href
  {https://ui.adsabs.harvard.edu/abs/2022ApJ...935...69B} {935, 69}

\bibitem[\protect\citeauthoryear{{Cantalupo} et~al.,}{{Cantalupo}
  et~al.}{2019}]{Cantalupo2019}
{Cantalupo} S.,  et~al., 2019, \mn@doi [\mnras] {10.1093/mnras/sty3481}, \href
  {https://ui.adsabs.harvard.edu/abs/2019MNRAS.483.5188C} {483, 5188}

\bibitem[\protect\citeauthoryear{{Carswell} \& {Webb}}{{Carswell} \&
  {Webb}}{2014}]{ascl:VPFIT2014}
{Carswell} R.~F.,  {Webb} J.~K.,  2014, {VPFIT: Voigt profile fitting program},
  Astrophysics Source Code Library (\mn@eprint {ascl} {1408.015})

\bibitem[\protect\citeauthoryear{{Chen}, {Burkhart}, {Goodman}  \&
  {Collins}}{{Chen} et~al.}{2018}]{Chen2018}
{Chen} H. H.-H.,  {Burkhart} B.,  {Goodman} A.,   {Collins} D.~C.,  2018,
  \mn@doi [\apj] {10.3847/1538-4357/aabaf6}, \href
  {https://ui.adsabs.harvard.edu/abs/2018ApJ...859..162C} {859, 162}

\bibitem[\protect\citeauthoryear{Chen et~al.,}{Chen
  et~al.}{2020}]{10.1093/mnras/staa1773}
Chen H.-W.,  et~al., 2020, \mn@doi [Monthly Notices of the Royal Astronomical
  Society] {10.1093/mnras/staa1773}, 497, 498

\bibitem[\protect\citeauthoryear{{Chen} et~al.,}{{Chen}
  et~al.}{2023}]{Chen2023}
{Chen} H.-W.,  et~al., 2023, \mn@doi [arXiv e-prints]
  {10.48550/arXiv.2309.05699}, \href
  {https://ui.adsabs.harvard.edu/abs/2023arXiv230905699C} {p. arXiv:2309.05699}

\bibitem[\protect\citeauthoryear{{Choudhury}, {Sharma}  \&
  {Quataert}}{{Choudhury} et~al.}{2019}]{Choudhury2019}
{Choudhury} P.~P.,  {Sharma} P.,   {Quataert} E.,  2019, \mn@doi [\mnras]
  {10.1093/mnras/stz1857}, \href
  {https://ui.adsabs.harvard.edu/abs/2019MNRAS.488.3195C} {488, 3195}

\bibitem[\protect\citeauthoryear{Churchill, Trujillo-Gomez, Nielsen  \&
  Kacprzak}{Churchill et~al.}{2013}]{Churchill_2013}
Churchill C.~W.,  Trujillo-Gomez S.,  Nielsen N.~M.,   Kacprzak G.~G.,  2013,
  \mn@doi [The Astrophysical Journal] {10.1088/0004-637X/779/1/87}, 779, 87

\bibitem[\protect\citeauthoryear{{Cook} et~al.,}{{Cook}
  et~al.}{2023}]{Cook2023}
{Cook} A.~M.,  et~al., 2023, \mn@doi [arXiv e-prints]
  {10.48550/arXiv.2301.03502}, \href
  {https://ui.adsabs.harvard.edu/abs/2023arXiv230103502C} {p. arXiv:2301.03502}

\bibitem[\protect\citeauthoryear{{Cooper} et~al.,}{{Cooper}
  et~al.}{2021}]{Cooper2021}
{Cooper} T.~J.,  et~al., 2021, \mn@doi [\mnras] {10.1093/mnras/stab2869}, \href
  {https://ui.adsabs.harvard.edu/abs/2021MNRAS.508.4359C} {508, 4359}

\bibitem[\protect\citeauthoryear{Crain, McCarthy, Frenk, Theuns  \&
  Schaye}{Crain et~al.}{2010}]{10.1111/j.1365-2966.2010.16985.x}
Crain R.~A.,  McCarthy I.~G.,  Frenk C.~S.,  Theuns T.,   Schaye J.,  2010,
  \mn@doi [Monthly Notices of the Royal Astronomical Society]
  {10.1111/j.1365-2966.2010.16985.x}, 407, 1403

\bibitem[\protect\citeauthoryear{{Das}, {Mathur}, {Nicastro}  \&
  {Krongold}}{{Das} et~al.}{2019a}]{Das2019ApJ}
{Das} S.,  {Mathur} S.,  {Nicastro} F.,   {Krongold} Y.,  2019a, \mn@doi
  [\apjl] {10.3847/2041-8213/ab3b09}, \href
  {https://ui.adsabs.harvard.edu/abs/2019ApJ...882L..23D} {882, L23}

\bibitem[\protect\citeauthoryear{Das, Mathur, Gupta, Nicastro  \& Krongold}{Das
  et~al.}{2019b}]{Das_2019XMM_em}
Das S.,  Mathur S.,  Gupta A.,  Nicastro F.,   Krongold Y.,  2019b, \mn@doi
  [The Astrophysical Journal] {10.3847/1538-4357/ab5846}, 887, 257

\bibitem[\protect\citeauthoryear{{Das}, {Mathur}  \& {Gupta}}{{Das}
  et~al.}{2020}]{Das2020}
{Das} S.,  {Mathur} S.,   {Gupta} A.,  2020, \mn@doi [\apj]
  {10.3847/1538-4357/ab93d2}, \href
  {https://ui.adsabs.harvard.edu/abs/2020ApJ...897...63D} {897, 63}

\bibitem[\protect\citeauthoryear{{Das}, {Mathur}, {Gupta}, {Nicastro}  \&
  {Krongold}}{{Das} et~al.}{2021a}]{Das2021_mnras}
{Das} S.,  {Mathur} S.,  {Gupta} A.,  {Nicastro} F.,   {Krongold} Y.,  2021a,
  \mn@doi [\mnras] {10.1093/mnras/staa3299}, \href
  {https://ui.adsabs.harvard.edu/abs/2021MNRAS.500..655D} {500, 655}

\bibitem[\protect\citeauthoryear{{Das}, {Mathur}, {Gupta}  \& {Krongold}}{{Das}
  et~al.}{2021b}]{Das2021}
{Das} S.,  {Mathur} S.,  {Gupta} A.,   {Krongold} Y.,  2021b, \mn@doi [\apj]
  {10.3847/1538-4357/ac0e8e}, \href
  {https://ui.adsabs.harvard.edu/abs/2021ApJ...918...83D} {918, 83}

\bibitem[\protect\citeauthoryear{DeFelippis, Bouché, Genel, Bryan, Nelson,
  Marinacci  \& Hernquist}{DeFelippis et~al.}{2021}]{DeFelippis_2021}
DeFelippis D.,  Bouché N.~F.,  Genel S.,  Bryan G.~L.,  Nelson D.,  Marinacci
  F.,   Hernquist L.,  2021, \mn@doi [The Astrophysical Journal]
  {10.3847/1538-4357/ac2cbf}, 923, 56

\bibitem[\protect\citeauthoryear{{Dehnen}, {McLaughlin}  \&
  {Sachania}}{{Dehnen} et~al.}{2006}]{2006MNRAS.369.1688D}
{Dehnen} W.,  {McLaughlin} D.~E.,   {Sachania} J.,  2006, \mn@doi [\mnras]
  {10.1111/j.1365-2966.2006.10404.x}, \href
  {https://ui.adsabs.harvard.edu/abs/2006MNRAS.369.1688D} {369, 1688}

\bibitem[\protect\citeauthoryear{Dutta et~al.,}{Dutta
  et~al.}{2020}]{10.1093/mnras/staa3147}
Dutta R.,  et~al., 2020, \mn@doi [Monthly Notices of the Royal Astronomical
  Society] {10.1093/mnras/staa3147}, 499, 5022

\bibitem[\protect\citeauthoryear{{Dutta}, {Sharma}  \& {Nelson}}{{Dutta}
  et~al.}{2022}]{Dutta2021}
{Dutta} A.,  {Sharma} P.,   {Nelson} D.,  2022, \mn@doi [\mnras]
  {10.1093/mnras/stab3653}, \href
  {https://ui.adsabs.harvard.edu/abs/2022MNRAS.510.3561D} {510, 3561}

\bibitem[\protect\citeauthoryear{{Elia} et~al.,}{{Elia}
  et~al.}{2022}]{2022ApJ...941..162E}
{Elia} D.,  et~al., 2022, \mn@doi [\apj] {10.3847/1538-4357/aca27d}, \href
  {https://ui.adsabs.harvard.edu/abs/2022ApJ...941..162E} {941, 162}

\bibitem[\protect\citeauthoryear{{Esmerian}, {Kravtsov}, {Hafen},
  {Faucher-Gigu{\`e}re}, {Quataert}, {Stern}, {Kere{\v{s}}}  \&
  {Wetzel}}{{Esmerian} et~al.}{2021}]{Esmerian2021}
{Esmerian} C.~J.,  {Kravtsov} A.~V.,  {Hafen} Z.,  {Faucher-Gigu{\`e}re} C.-A.,
   {Quataert} E.,  {Stern} J.,  {Kere{\v{s}}} D.,   {Wetzel} A.,  2021, \mn@doi
  [\mnras] {10.1093/mnras/stab1281}, \href
  {https://ui.adsabs.harvard.edu/abs/2021MNRAS.505.1841E} {505, 1841}

\bibitem[\protect\citeauthoryear{Faerman \& Werk}{Faerman \&
  Werk}{2023}]{Faerman2023}
Faerman Y.,  Werk J.~K.,  2023, \mn@doi [The Astrophysical Journal]
  {10.3847/1538-4357/acf217}, 956, 92

\bibitem[\protect\citeauthoryear{{Faerman}, {Sternberg}  \& {McKee}}{{Faerman}
  et~al.}{2017}]{Faerman2017ApJ}
{Faerman} Y.,  {Sternberg} A.,   {McKee} C.~F.,  2017, \mn@doi [\apj]
  {10.3847/1538-4357/835/1/52}, \href
  {https://ui.adsabs.harvard.edu/abs/2017ApJ...835...52F} {835, 52}

\bibitem[\protect\citeauthoryear{{Faerman}, {Sternberg}  \& {McKee}}{{Faerman}
  et~al.}{2020}]{Faerman2019ApJ}
{Faerman} Y.,  {Sternberg} A.,   {McKee} C.~F.,  2020, \mn@doi [\apj]
  {10.3847/1538-4357/ab7ffc}, \href
  {https://ui.adsabs.harvard.edu/abs/2020ApJ...893...82F} {893, 82}

\bibitem[\protect\citeauthoryear{{Fang}, {Buote}, {Bullock}  \& {Ma}}{{Fang}
  et~al.}{2015}]{Fang2015}
{Fang} T.,  {Buote} D.,  {Bullock} J.,   {Ma} R.,  2015, \mn@doi [\apjs]
  {10.1088/0067-0049/217/2/21}, \href
  {https://ui.adsabs.harvard.edu/abs/2015ApJS..217...21F} {217, 21}

\bibitem[\protect\citeauthoryear{{Faucher-Gigu{\`e}re} \&
  {Oh}}{{Faucher-Gigu{\`e}re} \& {Oh}}{2023}]{2023ARA&A..61..131F}
{Faucher-Gigu{\`e}re} C.-A.,  {Oh} S.~P.,  2023, \mn@doi [\araa]
  {10.1146/annurev-astro-052920-125203}, \href
  {https://ui.adsabs.harvard.edu/abs/2023ARA&A..61..131F} {61, 131}

\bibitem[\protect\citeauthoryear{Faucher-Giguère}{Faucher-Giguère}{2020}]{10.1093/mnras/staa302}
Faucher-Giguère C.-A.,  2020, \mn@doi [Monthly Notices of the Royal
  Astronomical Society] {10.1093/mnras/staa302}, 493, 1614

\bibitem[\protect\citeauthoryear{{Ferland} et~al.,}{{Ferland}
  et~al.}{2017}]{2017RMxAA..53..385F}
{Ferland} G.~J.,  et~al., 2017, \rmxaa, \href
  {https://ui.adsabs.harvard.edu/abs/2017RMxAA..53..385F} {53, 385}

\bibitem[\protect\citeauthoryear{{Fielding} et~al.,}{{Fielding}
  et~al.}{2020}]{Fielding2020}
{Fielding} D.~B.,  et~al., 2020, \mn@doi [\apj] {10.3847/1538-4357/abbc6d},
  \href {https://ui.adsabs.harvard.edu/abs/2020ApJ...903...32F} {903, 32}

\bibitem[\protect\citeauthoryear{{Fraternali}}{{Fraternali}}{2017}]{2017ASSL..430..323F}
{Fraternali} F.,  2017, in {Fox} A.,  {Dav{\'e}} R.,  eds,  Astrophysics and
  Space Science Library Vol. 430, Gas Accretion onto Galaxies. p.~323
  (\mn@eprint {arXiv} {1612.00477}), \mn@doi{10.1007/978-3-319-52512-9_14}

\bibitem[\protect\citeauthoryear{{Gilli}, {Comastri}  \& {Hasinger}}{{Gilli}
  et~al.}{2007}]{2007A&A...463...79G}
{Gilli} R.,  {Comastri} A.,   {Hasinger} G.,  2007, \mn@doi [\aap]
  {10.1051/0004-6361:20066334}, \href
  {https://ui.adsabs.harvard.edu/abs/2007A&A...463...79G} {463, 79}

\bibitem[\protect\citeauthoryear{Grand et~al.,}{Grand
  et~al.}{2019}]{10.1093/mnras/stz2928}
Grand R. J.~J.,  et~al., 2019, \mn@doi [Monthly Notices of the Royal
  Astronomical Society] {10.1093/mnras/stz2928}, 490, 4786

\bibitem[\protect\citeauthoryear{{Gronke} \& {Oh}}{{Gronke} \&
  {Oh}}{2018}]{Gronke2018}
{Gronke} M.,  {Oh} S.~P.,  2018, \mn@doi [\mnras] {10.1093/mnrasl/sly131},
  \href {https://ui.adsabs.harvard.edu/\#abs/2018MNRAS.480L.111G} {480, L111}

\bibitem[\protect\citeauthoryear{{Gupta}, {Mathur}, {Krongold}, {Nicastro}  \&
  {Galeazzi}}{{Gupta} et~al.}{2012}]{Gupta2012}
{Gupta} A.,  {Mathur} S.,  {Krongold} Y.,  {Nicastro} F.,   {Galeazzi} M.,
  2012, \mn@doi [\apjl] {10.1088/2041-8205/756/1/L8}, \href
  {https://ui.adsabs.harvard.edu/abs/2012ApJ...756L...8G} {756, L8}

\bibitem[\protect\citeauthoryear{Gupta, Mathur, Galeazzi  \& Krongold}{Gupta
  et~al.}{2014}]{Gupta2014}
Gupta A.,  Mathur S.,  Galeazzi M.,   Krongold Y.,  2014, \mn@doi [Astrophysics
  and Space Science] {10.1007/s10509-014-1958-z}, 352, 775

\bibitem[\protect\citeauthoryear{{Haardt} \& {Madau}}{{Haardt} \&
  {Madau}}{2012}]{Haardt2012}
{Haardt} F.,  {Madau} P.,  2012, \mn@doi [\apj] {10.1088/0004-637X/746/2/125},
  \href {https://ui.adsabs.harvard.edu/abs/2012ApJ...746..125H} {746, 125}

\bibitem[\protect\citeauthoryear{Hafen et~al.,}{Hafen et~al.}{2024}]{Hafen2024}
Hafen Z.,  et~al., 2024, \mn@doi [Monthly Notices of the Royal Astronomical
  Society] {10.1093/mnras/stad3889}, 528, 39

\bibitem[\protect\citeauthoryear{{Haislmaier}, {Tripp}, {Katz}, {Prochaska},
  {Burchett}, {O'Meara}  \& {Werk}}{{Haislmaier} et~al.}{2021}]{Haislmaier2021}
{Haislmaier} K.~J.,  {Tripp} T.~M.,  {Katz} N.,  {Prochaska} J.~X.,  {Burchett}
  J.~N.,  {O'Meara} J.~M.,   {Werk} J.~K.,  2021, \mn@doi [\mnras]
  {10.1093/mnras/staa3544}, \href
  {https://ui.adsabs.harvard.edu/abs/2021MNRAS.502.4993H} {502, 4993}

\bibitem[\protect\citeauthoryear{Henley \& Shelton}{Henley \&
  Shelton}{2010}]{Henley_2010}
Henley D.~B.,  Shelton R.~L.,  2010, \mn@doi [The Astrophysical Journal
  Supplement Series] {10.1088/0067-0049/187/2/388}, 187, 388

\bibitem[\protect\citeauthoryear{Henley \& Shelton}{Henley \&
  Shelton}{2013}]{Henley_2013}
Henley D.~B.,  Shelton R.~L.,  2013, \mn@doi [The Astrophysical Journal]
  {10.1088/0004-637X/773/2/92}, 773, 92

\bibitem[\protect\citeauthoryear{{Hirth}, {Herbstmeier}  \& {Mebold}}{{Hirth}
  et~al.}{1992}]{1992KlBer..35..803H}
{Hirth} W.,  {Herbstmeier} U.,   {Mebold} U.,  1992, Kleinheubacher Berichte,
  \href {https://ui.adsabs.harvard.edu/abs/1992KlBer..35..803H} {35, 803}

\bibitem[\protect\citeauthoryear{{Hummels} et~al.,}{{Hummels}
  et~al.}{2019}]{Hummels2019}
{Hummels} C.~B.,  et~al., 2019, \mn@doi [\apj] {10.3847/1538-4357/ab378f},
  \href {https://ui.adsabs.harvard.edu/abs/2019ApJ...882..156H} {882, 156}

\bibitem[\protect\citeauthoryear{{Hummels}, {Rubin}, {Schneider}  \&
  {Fielding}}{{Hummels} et~al.}{2023}]{2023arXiv231105691H}
{Hummels} C.~B.,  {Rubin} K. H.~R.,  {Schneider} E.~E.,   {Fielding} D.~B.,
  2023, \mn@doi [arXiv e-prints] {10.48550/arXiv.2311.05691}, \href
  {https://ui.adsabs.harvard.edu/abs/2023arXiv231105691H} {p. arXiv:2311.05691}

\bibitem[\protect\citeauthoryear{{Ji} et~al.,}{{Ji} et~al.}{2020}]{Ji2020MNRAS}
{Ji} S.,  et~al., 2020, \mn@doi [\mnras] {10.1093/mnras/staa1849}, \href
  {https://ui.adsabs.harvard.edu/abs/2020MNRAS.496.4221J} {496, 4221}

\bibitem[\protect\citeauthoryear{{Johnson}, {Chen}  \& {Mulchaey}}{{Johnson}
  et~al.}{2015}]{Johnson2015}
{Johnson} S.~D.,  {Chen} H.-W.,   {Mulchaey} J.~S.,  2015, \mn@doi [\mnras]
  {10.1093/mnras/stv553}, \href
  {https://ui.adsabs.harvard.edu/abs/2015MNRAS.449.3263J} {449, 3263}

\bibitem[\protect\citeauthoryear{Kaaret et~al.,}{Kaaret
  et~al.}{2019}]{Kaaret_2019}
Kaaret P.,  et~al., 2019, \mn@doi [The Astrophysical Journal]
  {10.3847/1538-4357/ab4193}, 884, 162

\bibitem[\protect\citeauthoryear{Kaaret et~al.,}{Kaaret
  et~al.}{2020}]{Kaaret2020}
Kaaret P.,  et~al., 2020, \mn@doi [Nature Astronomy]
  {10.1038/s41550-020-01215-w}, 4, 1072

\bibitem[\protect\citeauthoryear{{Kacprzak}, {Churchill}, {Murphy}  \&
  {Cooke}}{{Kacprzak} et~al.}{2015a}]{2015MNRAS.446.2861K}
{Kacprzak} G.~G.,  {Churchill} C.~W.,  {Murphy} M.~T.,   {Cooke} J.,  2015a,
  \mn@doi [\mnras] {10.1093/mnras/stu2324}, \href
  {https://ui.adsabs.harvard.edu/abs/2015MNRAS.446.2861K} {446, 2861}

\bibitem[\protect\citeauthoryear{Kacprzak, Muzahid, Churchill, Nielsen  \&
  Charlton}{Kacprzak et~al.}{2015b}]{Kacprzak_2015}
Kacprzak G.~G.,  Muzahid S.,  Churchill C.~W.,  Nielsen N.~M.,   Charlton
  J.~C.,  2015b, \mn@doi [The Astrophysical Journal]
  {10.1088/0004-637X/815/1/22}, 815, 22

\bibitem[\protect\citeauthoryear{{Kanjilal}, {Dutta}  \& {Sharma}}{{Kanjilal}
  et~al.}{2021}]{Kanjilal2021MNRAS}
{Kanjilal} V.,  {Dutta} A.,   {Sharma} P.,  2021, \mn@doi [\mnras]
  {10.1093/mnras/staa3610}, \href
  {https://ui.adsabs.harvard.edu/abs/2021MNRAS.501.1143K} {501, 1143}

\bibitem[\protect\citeauthoryear{Khaire \& Srianand}{Khaire \&
  Srianand}{2019}]{10.1093/mnras/stz174}
Khaire V.,  Srianand R.,  2019, \mn@doi [Monthly Notices of the Royal
  Astronomical Society] {10.1093/mnras/stz174}, 484, 4174

\bibitem[\protect\citeauthoryear{Kim \& Ostriker}{Kim \&
  Ostriker}{2018}]{Kim_2018}
Kim C.-G.,  Ostriker E.~C.,  2018, \mn@doi [The Astrophysical Journal]
  {10.3847/1538-4357/aaa5ff}, 853, 173

\bibitem[\protect\citeauthoryear{{K{\"o}rtgen}, {Federrath}  \&
  {Banerjee}}{{K{\"o}rtgen} et~al.}{2017}]{Kortgen2017}
{K{\"o}rtgen} B.,  {Federrath} C.,   {Banerjee} R.,  2017, \mn@doi [\mnras]
  {10.1093/mnras/stx2208}, \href
  {https://ui.adsabs.harvard.edu/abs/2017MNRAS.472.2496K} {472, 2496}

\bibitem[\protect\citeauthoryear{{Lehner}, {Howk}  \& {Wakker}}{{Lehner}
  et~al.}{2015}]{2015ApJ...804...79L}
{Lehner} N.,  {Howk} J.~C.,   {Wakker} B.~P.,  2015, \mn@doi [\apj]
  {10.1088/0004-637X/804/2/79}, \href
  {https://ui.adsabs.harvard.edu/abs/2015ApJ...804...79L} {804, 79}

\bibitem[\protect\citeauthoryear{{Lehner} et~al.,}{{Lehner}
  et~al.}{2020}]{2020ApJ...900....9L}
{Lehner} N.,  et~al., 2020, \mn@doi [\apj] {10.3847/1538-4357/aba49c}, \href
  {https://ui.adsabs.harvard.edu/abs/2020ApJ...900....9L} {900, 9}

\bibitem[\protect\citeauthoryear{{Li} et~al.,}{{Li} et~al.}{2020}]{Li2020ApJ}
{Li} Y.,  et~al., 2020, \mn@doi [\apjl] {10.3847/2041-8213/ab65c7}, \href
  {https://ui.adsabs.harvard.edu/abs/2020ApJ...889L...1L} {889, L1}

\bibitem[\protect\citeauthoryear{{Licquia} \& {Newman}}{{Licquia} \&
  {Newman}}{2015}]{2015ApJ...806...96L}
{Licquia} T.~C.,  {Newman} J.~A.,  2015, \mn@doi [\apj]
  {10.1088/0004-637X/806/1/96}, \href
  {https://ui.adsabs.harvard.edu/abs/2015ApJ...806...96L} {806, 96}

\bibitem[\protect\citeauthoryear{Liu et~al.,}{Liu et~al.}{2016}]{Liu_2017}
Liu W.,  et~al., 2016, \mn@doi [The Astrophysical Journal]
  {10.3847/1538-4357/834/1/33}, 834, 33

\bibitem[\protect\citeauthoryear{Lochhaas, Bryan, Li, Li  \& Fielding}{Lochhaas
  et~al.}{2020}]{10.1093/mnras/staa358}
Lochhaas C.,  Bryan G.~L.,  Li Y.,  Li M.,   Fielding D.,  2020, \mn@doi
  [Monthly Notices of the Royal Astronomical Society] {10.1093/mnras/staa358},
  493, 1461

\bibitem[\protect\citeauthoryear{{Maller} \& {Bullock}}{{Maller} \&
  {Bullock}}{2004}]{Maller2004}
{Maller} A.~H.,  {Bullock} J.~S.,  2004, \mn@doi [\mnras]
  {10.1111/j.1365-2966.2004.08349.x}, \href
  {https://ui.adsabs.harvard.edu/abs/2004MNRAS.355..694M} {355, 694}

\bibitem[\protect\citeauthoryear{Marra et~al.,}{Marra
  et~al.}{2021}]{10.1093/mnras/stab2896}
Marra R.,  et~al., 2021, \mn@doi [Monthly Notices of the Royal Astronomical
  Society] {10.1093/mnras/stab2896}, 508, 4938

\bibitem[\protect\citeauthoryear{{Mathews} \& {Prochaska}}{{Mathews} \&
  {Prochaska}}{2017}]{Mathews2017}
{Mathews} W.~G.,  {Prochaska} J.~X.,  2017, \mn@doi [\apjl]
  {10.3847/2041-8213/aa8861}, \href
  {https://ui.adsabs.harvard.edu/abs/2017ApJ...846L..24M} {846, L24}

\bibitem[\protect\citeauthoryear{Mathur, Das, Gupta  \& Krongold}{Mathur
  et~al.}{2023}]{Mathur2023}
Mathur S.,  Das S.,  Gupta A.,   Krongold Y.,  2023, \mn@doi [Monthly Notices
  of the Royal Astronomical Society: Letters] {10.1093/mnrasl/slad085}, 525,
  L11

\bibitem[\protect\citeauthoryear{{Miller} \& {Bregman}}{{Miller} \&
  {Bregman}}{2013}]{Miller2013}
{Miller} M.~J.,  {Bregman} J.~N.,  2013, \mn@doi [\apj]
  {10.1088/0004-637X/770/2/118}, \href
  {https://ui.adsabs.harvard.edu/abs/2013ApJ...770..118M} {770, 118}

\bibitem[\protect\citeauthoryear{Miller \& Bregman}{Miller \&
  Bregman}{2015}]{Miller2015}
Miller M.~J.,  Bregman J.~N.,  2015, \mn@doi [The Astrophysical Journal]
  {10.1088/0004-637X/800/1/14}, 800, 14

\bibitem[\protect\citeauthoryear{{Mohapatra} \& {Sharma}}{{Mohapatra} \&
  {Sharma}}{2019}]{Mohapatra2019}
{Mohapatra} R.,  {Sharma} P.,  2019, \mn@doi [\mnras] {10.1093/mnras/stz328},
  \href {https://ui.adsabs.harvard.edu/\#abs/2019MNRAS.484.4881M} {484, 4881}

\bibitem[\protect\citeauthoryear{{Mohapatra}, {Federrath}  \&
  {Sharma}}{{Mohapatra} et~al.}{2021}]{Mohapatra2021MNRAS}
{Mohapatra} R.,  {Federrath} C.,   {Sharma} P.,  2021, \mn@doi [\mnras]
  {10.1093/mnras/staa3564}, \href
  {https://ui.adsabs.harvard.edu/abs/2021MNRAS.500.5072M} {500, 5072}

\bibitem[\protect\citeauthoryear{{Mohapatra}, {Jetti}, {Sharma}  \&
  {Federrath}}{{Mohapatra} et~al.}{2022}]{Mohapatra2022characterising}
{Mohapatra} R.,  {Jetti} M.,  {Sharma} P.,   {Federrath} C.,  2022, \mn@doi
  [\mnras] {10.1093/mnras/stab3603}, \href
  {https://ui.adsabs.harvard.edu/abs/2022MNRAS.510.3778M} {510, 3778}

\bibitem[\protect\citeauthoryear{{Mohapatra}, {Sharma}, {Federrath}  \&
  {Quataert}}{{Mohapatra} et~al.}{2023}]{Mohapatra2023}
{Mohapatra} R.,  {Sharma} P.,  {Federrath} C.,   {Quataert} E.,  2023, \mn@doi
  [arXiv e-prints] {10.48550/arXiv.2302.09380}, \href
  {https://ui.adsabs.harvard.edu/abs/2023arXiv230209380M} {p. arXiv:2302.09380}

\bibitem[\protect\citeauthoryear{Muzahid, Fonseca, Roberts, Rosenwasser,
  Richter, Narayanan, Churchill  \& Charlton}{Muzahid
  et~al.}{2018}]{10.1093/mnras/sty529}
Muzahid S.,  Fonseca G.,  Roberts A.,  Rosenwasser B.,  Richter P.,  Narayanan
  A.,  Churchill C.,   Charlton J.,  2018, \mn@doi [Monthly Notices of the
  Royal Astronomical Society] {10.1093/mnras/sty529}, 476, 4965

\bibitem[\protect\citeauthoryear{{Nakashima}, {Inoue}, {Yamasaki}, {Sofue},
  {Kataoka}  \& {Sakai}}{{Nakashima} et~al.}{2018}]{Nakashima2018}
{Nakashima} S.,  {Inoue} Y.,  {Yamasaki} N.,  {Sofue} Y.,  {Kataoka} J.,
  {Sakai} K.,  2018, \mn@doi [\apj] {10.3847/1538-4357/aacceb}, \href
  {https://ui.adsabs.harvard.edu/abs/2018ApJ...862...34N} {862, 34}

\bibitem[\protect\citeauthoryear{Nateghi, Kacprzak, Nielsen, Muzahid,
  Churchill, Pointon  \& Charlton}{Nateghi
  et~al.}{2020}]{10.1093/mnras/staa3534}
Nateghi H.,  Kacprzak G.~G.,  Nielsen N.~M.,  Muzahid S.,  Churchill C.~W.,
  Pointon S.~K.,   Charlton J.~C.,  2020, \mn@doi [Monthly Notices of the Royal
  Astronomical Society] {10.1093/mnras/staa3534}, 500, 3987

\bibitem[\protect\citeauthoryear{{Nelson} et~al.,}{{Nelson}
  et~al.}{2020}]{Nelson2020}
{Nelson} D.,  et~al., 2020, \mn@doi [\mnras] {10.1093/mnras/staa2419}, \href
  {https://ui.adsabs.harvard.edu/abs/2020MNRAS.498.2391N} {498, 2391}

\bibitem[\protect\citeauthoryear{Nelson et~al.,}{Nelson
  et~al.}{2023}]{Nelson2023}
Nelson D.,  et~al., 2023, \mn@doi [Monthly Notices of the Royal Astronomical
  Society] {10.1093/mnras/stad1195}, 522, 3665

\bibitem[\protect\citeauthoryear{{Nielsen}, {Churchill}, {Kacprzak}, {Murphy}
  \& {Evans}}{{Nielsen} et~al.}{2016}]{2016ApJ...818..171N}
{Nielsen} N.~M.,  {Churchill} C.~W.,  {Kacprzak} G.~G.,  {Murphy} M.~T.,
  {Evans} J.~L.,  2016, \mn@doi [\apj] {10.3847/0004-637X/818/2/171}, \href
  {https://ui.adsabs.harvard.edu/abs/2016ApJ...818..171N} {818, 171}

\bibitem[\protect\citeauthoryear{{Nielsen}, {Kacprzak}, {Muzahid}, {Churchill},
  {Murphy}  \& {Charlton}}{{Nielsen} et~al.}{2017}]{2017ApJ...834..148N}
{Nielsen} N.~M.,  {Kacprzak} G.~G.,  {Muzahid} S.,  {Churchill} C.~W.,
  {Murphy} M.~T.,   {Charlton} J.~C.,  2017, \mn@doi [\apj]
  {10.3847/1538-4357/834/2/148}, \href
  {https://ui.adsabs.harvard.edu/abs/2017ApJ...834..148N} {834, 148}

\bibitem[\protect\citeauthoryear{{Ocker}, {Cordes}  \& {Chatterjee}}{{Ocker}
  et~al.}{2021}]{Ocker2021}
{Ocker} S.~K.,  {Cordes} J.~M.,   {Chatterjee} S.,  2021, \mn@doi [\apj]
  {10.3847/1538-4357/abeb6e}, \href
  {https://ui.adsabs.harvard.edu/abs/2021ApJ...911..102O} {911, 102}

\bibitem[\protect\citeauthoryear{Oppenheimer, Schaye, Crain, Werk  \&
  Richings}{Oppenheimer et~al.}{2018}]{Oppenheimer18}
Oppenheimer B.~D.,  Schaye J.,  Crain R.~A.,  Werk J.~K.,   Richings A.~J.,
  2018, \mn@doi [Monthly Notices of the Royal Astronomical Society]
  {10.1093/mnras/sty2281}, 481, 835

\bibitem[\protect\citeauthoryear{{Ponti} et~al.,}{{Ponti}
  et~al.}{2023a}]{Ponti2023}
{Ponti} G.,  et~al., 2023a, \mn@doi [\aap] {10.1051/0004-6361/202244430}, \href
  {https://ui.adsabs.harvard.edu/abs/2023A&A...670A..99P} {670, A99}

\bibitem[\protect\citeauthoryear{{Ponti} et~al.,}{{Ponti}
  et~al.}{2023b}]{Ponti2022}
{Ponti} G.,  et~al., 2023b, \mn@doi [\aap] {10.1051/0004-6361/202243992}, \href
  {https://ui.adsabs.harvard.edu/abs/2023A&A...674A.195P} {674, A195}

\bibitem[\protect\citeauthoryear{{Predehl} et~al.,}{{Predehl}
  et~al.}{2020}]{Predehl2020}
{Predehl} P.,  et~al., 2020, \mn@doi [\nat] {10.1038/s41586-020-2979-0}, \href
  {https://ui.adsabs.harvard.edu/abs/2020Natur.588..227P} {588, 227}

\bibitem[\protect\citeauthoryear{{Predehl} et~al.,}{{Predehl}
  et~al.}{2021}]{Predehl2021}
{Predehl} P.,  et~al., 2021, \mn@doi [\aap] {10.1051/0004-6361/202039313},
  \href {https://ui.adsabs.harvard.edu/abs/2021A&A...647A...1P} {647, A1}

\bibitem[\protect\citeauthoryear{Péroux et~al.,}{Péroux
  et~al.}{2019}]{Peroux19}
Péroux C.,  et~al., 2019, \mn@doi [Monthly Notices of the Royal Astronomical
  Society] {10.1093/mnras/stz202}, 485, 1595

\bibitem[\protect\citeauthoryear{{Qu} et~al.,}{{Qu}
  et~al.}{2023}]{2023arXiv230611274Q}
{Qu} Z.,  et~al., 2023, \mn@doi [arXiv e-prints] {10.48550/arXiv.2306.11274},
  \href {https://ui.adsabs.harvard.edu/abs/2023arXiv230611274Q} {p.
  arXiv:2306.11274}

\bibitem[\protect\citeauthoryear{{Qu} et~al.,}{{Qu}
  et~al.}{2024}]{2024arXiv240208016Q}
{Qu} Z.,  et~al., 2024, \mn@doi [arXiv e-prints] {10.48550/arXiv.2402.08016},
  \href {https://ui.adsabs.harvard.edu/abs/2024arXiv240208016Q} {p.
  arXiv:2402.08016}

\bibitem[\protect\citeauthoryear{{Ramesh}, {Nelson}  \& {Pillepich}}{{Ramesh}
  et~al.}{2023}]{Ramesh2023}
{Ramesh} R.,  {Nelson} D.,   {Pillepich} A.,  2023, \mn@doi [\mnras]
  {10.1093/mnras/stac3524}, \href
  {https://ui.adsabs.harvard.edu/abs/2023MNRAS.518.5754R} {518, 5754}

\bibitem[\protect\citeauthoryear{{Ravi} et~al.,}{{Ravi}
  et~al.}{2023}]{Ravi2023}
{Ravi} V.,  et~al., 2023, \mn@doi [arXiv e-prints] {10.48550/arXiv.2301.01000},
  \href {https://ui.adsabs.harvard.edu/abs/2023arXiv230101000R} {p.
  arXiv:2301.01000}

\bibitem[\protect\citeauthoryear{Rohr, Pillepich, Nelson, Zinger, Joshi  \&
  Ayromlou}{Rohr et~al.}{2023}]{Eric2023}
Rohr E.,  Pillepich A.,  Nelson D.,  Zinger E.,  Joshi G.~D.,   Ayromlou M.,
  2023, \mn@doi [Monthly Notices of the Royal Astronomical Society]
  {10.1093/mnras/stad2101}, 524, 3502

\bibitem[\protect\citeauthoryear{Rubin, Weiner, Koo, Martin, Prochaska, Coil
  \& Newman}{Rubin et~al.}{2010}]{Rubin_2010}
Rubin K. H.~R.,  Weiner B.~J.,  Koo D.~C.,  Martin C.~L.,  Prochaska J.~X.,
  Coil A.~L.,   Newman J.~A.,  2010, \mn@doi [The Astrophysical Journal]
  {10.1088/0004-637X/719/2/1503}, 719, 1503

\bibitem[\protect\citeauthoryear{{Rudie}, {Steidel}, {Pettini}, {Trainor},
  {Strom}, {Hummels}, {Reddy}  \& {Shapley}}{{Rudie} et~al.}{2019}]{Rudie2019}
{Rudie} G.~C.,  {Steidel} C.~C.,  {Pettini} M.,  {Trainor} R.~F.,  {Strom}
  A.~L.,  {Hummels} C.~B.,  {Reddy} N.~A.,   {Shapley} A.~E.,  2019, \mn@doi
  [\apj] {10.3847/1538-4357/ab4255}, \href
  {https://ui.adsabs.harvard.edu/abs/2019ApJ...885...61R} {885, 61}

\bibitem[\protect\citeauthoryear{Saeedzadeh et~al.,}{Saeedzadeh
  et~al.}{2023}]{10.1093/mnras/stad2637}
Saeedzadeh V.,  et~al., 2023, \mn@doi [Monthly Notices of the Royal
  Astronomical Society] {10.1093/mnras/stad2637}, 525, 5677

\bibitem[\protect\citeauthoryear{Salem, Bryan  \& Corlies}{Salem
  et~al.}{2015}]{Salem2015}
Salem M.,  Bryan G.~L.,   Corlies L.,  2015, \mn@doi [Monthly Notices of the
  Royal Astronomical Society] {10.1093/mnras/stv2641}, 456, 582

\bibitem[\protect\citeauthoryear{{Sameer} et~al.,}{{Sameer}
  et~al.}{2021}]{Sameer2021}
{Sameer} et~al., 2021, \mn@doi [\mnras] {10.1093/mnras/staa3754}, \href
  {https://ui.adsabs.harvard.edu/abs/2021MNRAS.501.2112S} {501, 2112}

\bibitem[\protect\citeauthoryear{{Sameer} et~al.,}{{Sameer}
  et~al.}{2024}]{2024arXiv240305617S}
{Sameer} et~al., 2024, \mn@doi [arXiv e-prints] {10.48550/arXiv.2403.05617},
  \href {https://ui.adsabs.harvard.edu/abs/2024arXiv240305617S} {p.
  arXiv:2403.05617}

\bibitem[\protect\citeauthoryear{{Sarkar}, {Nath}  \& {Sharma}}{{Sarkar}
  et~al.}{2015}]{Sarkar2015}
{Sarkar} K.~C.,  {Nath} B.~B.,   {Sharma} P.,  2015, \mn@doi [\mnras]
  {10.1093/mnras/stv1806}, \href
  {https://ui.adsabs.harvard.edu/abs/2015MNRAS.453.3827S} {453, 3827}

\bibitem[\protect\citeauthoryear{{Savage} \& {Sembach}}{{Savage} \&
  {Sembach}}{1991}]{1991ApJ...379..245S}
{Savage} B.~D.,  {Sembach} K.~R.,  1991, \mn@doi [\apj] {10.1086/170498}, \href
  {https://ui.adsabs.harvard.edu/abs/1991ApJ...379..245S} {379, 245}

\bibitem[\protect\citeauthoryear{{Schaye} et~al.,}{{Schaye}
  et~al.}{2015}]{2015MNRAS.446..521S}
{Schaye} J.,  et~al., 2015, \mn@doi [\mnras] {10.1093/mnras/stu2058}, \href
  {https://ui.adsabs.harvard.edu/abs/2015MNRAS.446..521S} {446, 521}

\bibitem[\protect\citeauthoryear{{Sharma}, {McCourt}, {Quataert}  \&
  {Parrish}}{{Sharma} et~al.}{2012a}]{sharma2012thermal}
{Sharma} P.,  {McCourt} M.,  {Quataert} E.,   {Parrish} I.~J.,  2012a, \mn@doi
  [\mnras] {10.1111/j.1365-2966.2011.20246.x}, \href
  {https://ui.adsabs.harvard.edu/\#abs/2012MNRAS.420.3174S} {420, 3174}

\bibitem[\protect\citeauthoryear{{Sharma}, {McCourt}, {Parrish}  \&
  {Quataert}}{{Sharma} et~al.}{2012b}]{Sharma2012}
{Sharma} P.,  {McCourt} M.,  {Parrish} I.~J.,   {Quataert} E.,  2012b, \mn@doi
  [\mnras] {10.1111/j.1365-2966.2012.22050.x}, \href
  {https://ui.adsabs.harvard.edu/abs/2012MNRAS.427.1219S} {427, 1219}

\bibitem[\protect\citeauthoryear{{Singh}, {Majumdar}, {Nath}  \&
  {Silk}}{{Singh} et~al.}{2018}]{Singh2018}
{Singh} P.,  {Majumdar} S.,  {Nath} B.~B.,   {Silk} J.,  2018, \mn@doi [\mnras]
  {10.1093/mnras/sty1276}, \href
  {https://ui.adsabs.harvard.edu/abs/2018MNRAS.478.2909S} {478, 2909}

\bibitem[\protect\citeauthoryear{{Smith}, {Brickhouse}, {Liedahl}  \&
  {Raymond}}{{Smith} et~al.}{2001}]{Smith2001}
{Smith} R.~K.,  {Brickhouse} N.~S.,  {Liedahl} D.~A.,   {Raymond} J.~C.,  2001,
  \mn@doi [\apjl] {10.1086/322992}, \href
  {https://ui.adsabs.harvard.edu/abs/2001ApJ...556L..91S} {556, L91}

\bibitem[\protect\citeauthoryear{{Sormani}, {Sobacchi}, {Pezzulli}, {Binney}
  \& {Klessen}}{{Sormani} et~al.}{2018}]{Sormani2018}
{Sormani} M.~C.,  {Sobacchi} E.,  {Pezzulli} G.,  {Binney} J.,   {Klessen}
  R.~S.,  2018, \mn@doi [\mnras] {10.1093/mnras/sty2500}, \href
  {https://ui.adsabs.harvard.edu/abs/2018MNRAS.481.3370S} {481, 3370}

\bibitem[\protect\citeauthoryear{Stern, Fielding, Faucher-Giguère  \&
  Quataert}{Stern et~al.}{2019}]{10.1093/mnras/stz1859}
Stern J.,  Fielding D.,  Faucher-Giguère C.-A.,   Quataert E.,  2019, \mn@doi
  [Monthly Notices of the Royal Astronomical Society] {10.1093/mnras/stz1859},
  488, 2549

\bibitem[\protect\citeauthoryear{{Stocke}, {Keeney}, {Danforth}, {Shull},
  {Froning}, {Green}, {Penton}  \& {Savage}}{{Stocke}
  et~al.}{2013}]{Stocke2013}
{Stocke} J.~T.,  {Keeney} B.~A.,  {Danforth} C.~W.,  {Shull} J.~M.,  {Froning}
  C.~S.,  {Green} J.~C.,  {Penton} S.~V.,   {Savage} B.~D.,  2013, \mn@doi
  [\apj] {10.1088/0004-637X/763/2/148}, \href
  {https://ui.adsabs.harvard.edu/abs/2013ApJ...763..148S} {763, 148}

\bibitem[\protect\citeauthoryear{Strawn, Roca-Fàbrega  \& Primack}{Strawn
  et~al.}{2022}]{10.1093/mnras/stac3567}
Strawn C.,  Roca-Fàbrega S.,   Primack J.,  2022, \mn@doi [Monthly Notices of
  the Royal Astronomical Society] {10.1093/mnras/stac3567}, 519, 1

\bibitem[\protect\citeauthoryear{Tchernyshyov et~al.,}{Tchernyshyov
  et~al.}{2022}]{Tchernyshyov_2022}
Tchernyshyov K.,  et~al., 2022, \mn@doi [The Astrophysical Journal]
  {10.3847/1538-4357/ac450c}, 927, 147

\bibitem[\protect\citeauthoryear{{Tumlinson} et~al.,}{{Tumlinson}
  et~al.}{2011a}]{Tumlinson2011Sci}
{Tumlinson} J.,  et~al., 2011a, \mn@doi [Science] {10.1126/science.1209840},
  \href {https://ui.adsabs.harvard.edu/abs/2011Sci...334..948T} {334, 948}

\bibitem[\protect\citeauthoryear{{Tumlinson} et~al.,}{{Tumlinson}
  et~al.}{2011b}]{Tumlinson2011ApJ}
{Tumlinson} J.,  et~al., 2011b, \mn@doi [\apj] {10.1088/0004-637X/733/2/111},
  \href {https://ui.adsabs.harvard.edu/abs/2011ApJ...733..111T} {733, 111}

\bibitem[\protect\citeauthoryear{Tumlinson et~al.,}{Tumlinson
  et~al.}{2013}]{Tumlinson_2013}
Tumlinson J.,  et~al., 2013, \mn@doi [The Astrophysical Journal]
  {10.1088/0004-637X/777/1/59}, 777, 59

\bibitem[\protect\citeauthoryear{{Tumlinson}, {Peeples}  \& {Werk}}{{Tumlinson}
  et~al.}{2017}]{Tumlinson2017review}
{Tumlinson} J.,  {Peeples} M.~S.,   {Werk} J.~K.,  2017, \mn@doi [Annual Review
  of Astronomy and Astrophysics] {10.1146/annurev-astro-091916-055240}, \href
  {https://ui.adsabs.harvard.edu/\#abs/2017ARA&A..55..389T} {55, 389}

\bibitem[\protect\citeauthoryear{{Tuttle} et~al.,}{{Tuttle}
  et~al.}{2019}]{Tuttle2019}
{Tuttle} S.,  et~al., 2019, \mn@doi [Astro2020: Decadal Survey on Astronomy and
  Astrophysics] {10.3847/25c2cfeb.a00742b4}, \href
  {https://ui.adsabs.harvard.edu/abs/2019astro2020T.403T} {54, 041}

\bibitem[\protect\citeauthoryear{{Vijayan} \& {Li}}{{Vijayan} \&
  {Li}}{2022}]{Vijayan2022}
{Vijayan} A.,  {Li} M.,  2022, \mn@doi [\mnras] {10.1093/mnras/stab3413}, \href
  {https://ui.adsabs.harvard.edu/abs/2022MNRAS.510..568V} {510, 568}

\bibitem[\protect\citeauthoryear{{Voit}}{{Voit}}{2019}]{Voit2019}
{Voit} G.~M.,  2019, \mn@doi [\apj] {10.3847/1538-4357/ab2bfd}, \href
  {https://ui.adsabs.harvard.edu/abs/2019ApJ...880..139V} {880, 139}

\bibitem[\protect\citeauthoryear{Wakker \& Savage}{Wakker \&
  Savage}{2009}]{Wakker_2009}
Wakker B.~P.,  Savage B.~D.,  2009, \mn@doi [The Astrophysical Journal
  Supplement Series] {10.1088/0067-0049/182/1/378}, 182, 378

\bibitem[\protect\citeauthoryear{{Wakker}, {Savage}, {Richter}, {Meade}  \&
  {Sembach}}{{Wakker} et~al.}{2003}]{2003ASSL..281..183W}
{Wakker} B.~P.,  {Savage} B.~D.,  {Richter} P.,  {Meade} M.,   {Sembach} K.~R.,
   2003, in {Rosenberg} J.~L.,  {Putman} M.~E.,  eds,  Astrophysics and Space
  Science Library Vol. 281, The IGM/Galaxy Connection. The Distribution of
  Baryons at z=0. p.~183 (\mn@eprint {arXiv} {astro-ph/0208009}),
  \mn@doi{10.1007/978-94-010-0115-1_33}

\bibitem[\protect\citeauthoryear{Weiner et~al.,}{Weiner
  et~al.}{2009}]{Weiner_2009}
Weiner B.~J.,  et~al., 2009, \mn@doi [The Astrophysical Journal]
  {10.1088/0004-637X/692/1/187}, 692, 187

\bibitem[\protect\citeauthoryear{{Werk}, {Prochaska}, {Thom}, {Tumlinson},
  {Tripp}, {O'Meara}  \& {Meiring}}{{Werk} et~al.}{2012}]{2012ApJS..198....3W}
{Werk} J.~K.,  {Prochaska} J.~X.,  {Thom} C.,  {Tumlinson} J.,  {Tripp} T.~M.,
  {O'Meara} J.~M.,   {Meiring} J.~D.,  2012, \mn@doi [\apjs]
  {10.1088/0067-0049/198/1/3}, \href
  {https://ui.adsabs.harvard.edu/abs/2012ApJS..198....3W} {198, 3}

\bibitem[\protect\citeauthoryear{{Werk}, {Prochaska}, {Thom}, {Tumlinson},
  {Tripp}, {O'Meara}  \& {Peeples}}{{Werk} et~al.}{2013}]{Werk2013}
{Werk} J.~K.,  {Prochaska} J.~X.,  {Thom} C.,  {Tumlinson} J.,  {Tripp} T.~M.,
  {O'Meara} J.~M.,   {Peeples} M.~S.,  2013, \mn@doi [\apjs]
  {10.1088/0067-0049/204/2/17}, \href
  {https://ui.adsabs.harvard.edu/abs/2013ApJS..204...17W} {204, 17}

\bibitem[\protect\citeauthoryear{{Werk} et~al.,}{{Werk}
  et~al.}{2014}]{Werk2014}
{Werk} J.~K.,  et~al., 2014, \mn@doi [\apj] {10.1088/0004-637X/792/1/8}, \href
  {https://ui.adsabs.harvard.edu/abs/2014ApJ...792....8W} {792, 8}

\bibitem[\protect\citeauthoryear{{Werk} et~al.,}{{Werk}
  et~al.}{2016}]{Werk2016ApJ}
{Werk} J.~K.,  et~al., 2016, \mn@doi [\apj] {10.3847/1538-4357/833/1/54}, \href
  {https://ui.adsabs.harvard.edu/abs/2016ApJ...833...54W} {833, 54}

\bibitem[\protect\citeauthoryear{{Wu} \& {McQuinn}}{{Wu} \&
  {McQuinn}}{2022}]{Wu2022}
{Wu} X.,  {McQuinn} M.,  2022, arXiv e-prints, \href
  {https://ui.adsabs.harvard.edu/abs/2022arXiv220904455W} {p. arXiv:2209.04455}

\bibitem[\protect\citeauthoryear{{Yamasaki} \& {Totani}}{{Yamasaki} \&
  {Totani}}{2020}]{Yamasaki2020}
{Yamasaki} S.,  {Totani} T.,  2020, \mn@doi [\apj] {10.3847/1538-4357/ab58c4},
  \href {https://ui.adsabs.harvard.edu/abs/2020ApJ...888..105Y} {888, 105}

\bibitem[\protect\citeauthoryear{{Yang} \& {Ji}}{{Yang} \&
  {Ji}}{2023}]{Yang2023}
{Yang} Y.,  {Ji} S.,  2023, \mn@doi [\mnras] {10.1093/mnras/stad264}, \href
  {https://ui.adsabs.harvard.edu/abs/2023MNRAS.520.2148Y} {520, 2148}

\bibitem[\protect\citeauthoryear{{Yao}, {Manchester}  \& {Wang}}{{Yao}
  et~al.}{2017}]{Yao2017}
{Yao} J.~M.,  {Manchester} R.~N.,   {Wang} N.,  2017, \mn@doi [\apj]
  {10.3847/1538-4357/835/1/29}, \href
  {https://ui.adsabs.harvard.edu/abs/2017ApJ...835...29Y} {835, 29}

\bibitem[\protect\citeauthoryear{{Zabl} et~al.,}{{Zabl}
  et~al.}{2019}]{2019MNRAS.485.1961Z}
{Zabl} J.,  et~al., 2019, \mn@doi [\mnras] {10.1093/mnras/stz392}, \href
  {https://ui.adsabs.harvard.edu/abs/2019MNRAS.485.1961Z} {485, 1961}

\bibitem[\protect\citeauthoryear{{Zahedy}, {Chen}, {Johnson}, {Pierce},
  {Rauch}, {Huang}, {Weiner}  \& {Gauthier}}{{Zahedy}
  et~al.}{2019}]{Zahedy2019}
{Zahedy} F.~S.,  {Chen} H.-W.,  {Johnson} S.~D.,  {Pierce} R.~M.,  {Rauch} M.,
  {Huang} Y.-H.,  {Weiner} B.~J.,   {Gauthier} J.-R.,  2019, \mn@doi [\mnras]
  {10.1093/mnras/sty3482}, \href
  {https://ui.adsabs.harvard.edu/abs/2019MNRAS.484.2257Z} {484, 2257}

\bibitem[\protect\citeauthoryear{{ZuHone}, {Miller}, {Bulbul}  \&
  {Zhuravleva}}{{ZuHone} et~al.}{2018}]{Zuhone2018}
{ZuHone} J.~A.,  {Miller} E.~D.,  {Bulbul} E.,   {Zhuravleva} I.,  2018,
  \mn@doi [\apj] {10.3847/1538-4357/aaa4b3}, \href
  {https://ui.adsabs.harvard.edu/abs/2018ApJ...853..180Z} {853, 180}

\bibitem[\protect\citeauthoryear{van de Voort, Springel, Mandelker,
  van den Bosch  \& Pakmor}{van de Voort
  et~al.}{2018}]{10.1093/mnrasl/sly190}
van de Voort F.,  Springel V.,  Mandelker N.,  van den Bosch F.~C.,
  Pakmor R.,  2018, \mn@doi [Monthly Notices of the Royal Astronomical Society:
  Letters] {10.1093/mnrasl/sly190}, 482, L85

\makeatother
\end{thebibliography}



\appendix
\renewcommand\thefigure{\thesection A\arabic{figure}} 
\setcounter{figure}{0}   
\setcounter{table}{0} 

\section{{\tt AstroPlasma} code repository}
\label{app:AstroPlasma}
{\tt AstroPlasma} is a publicly hosted astrophysical plasma modeling code that aims to model the physical conditions and absorption and emission properties of the CGM plasma. {\tt AstroPlasma} uses a database of pre-computed {\tt CLOUDY} models (\citealt{2017RMxAA..53..385F}) to interpolate plasma properties in typical CGM conditions. This provides a fast, computationally inexpensive interface to perform on-the-fly calculations of plasma properties needed to generate observables from different CGM models/simulation data.
{\tt AstroPlasma} is written in a flexible and user-friendly manner in {\tt Python}, and its database and range of applicability can be updated and enhanced as per the need. Here, we illustrate some of the basic usage of {\tt AstroPlasma}. Interested readers can find more about it at the  \href{https://github.com/dutta-alankar/AstroPlasma}{\tt AstroPlasma} {\it GitHub} repository. We aim to maintain and expand {\tt AstroPlasma}, making it a useful tool for the community. We welcome contributions in the form of pull requests and issues.

\subsection{Ionization modeling}
{\tt AstroPlasma} can be used to calculate the ionization state of ions for given conditions. As an example, in the following Python code snippet, we calculate the ionization fraction of OVI.

\begin{mintedbox}{python}
# Import AstroPlasma Ionization module
from astro_plasma import Ionization

fIon = Ionization.interpolate_ion_frac

nH = 1.2e-04 # Hydrogen number density in cm^-3
temperature = 4.2e+05 # Temperature in kelvin
metallicity = 0.99 # Metallicity wrt solar
redshift = 0.001 # Cosmological redshift
mode = "CIE" # Ionization equilibrium: CIE/PIE

fOVI = fIon(nH = nH,
            temperature = temperature,
            metallicity = metallicity,
            redshift = redshift,
            element = "OVI",
            ) # This value is in log10
print(f"f_OVI = {pow(10, fOVI):.3e}")
\end{mintedbox}

It is also possible to directly calculate the number density of a specific ion (up to Zn with atomic number 30), all ions, electrons, or all neutral species using {\tt AstroPlasma}. Equilibrium ionization is also taken into account while calculating the mean mass of different species in the plasma. The following code snippet illustrates the usage of {\tt AstroPlasma} to calculate these quantities. None of these calculations involve any real-time plasma calculations but are interpolated from the {\tt CLOUDY} database.
\begin{mintedbox}{python}
from astro_plasma import Ionization

num_dens = Ionization.interpolate_num_dens

ne = num_dens(nH = nH,
              temperature = temperature,
              metallicity = metallicity,
              redshift = redshift,
              mode = mode,
              part_type = "electron",
              )
n = num_dens(nH = nH,
              temperature = temperature,
              metallicity = metallicity,
              redshift = redshift,
              mode = mode,
              part_type = "all",
              )
ni = num_dens(nH = nH,
              temperature = temperature,
              metallicity = metallicity,
              redshift = redshift,
              mode = mode,
              part_type = "ion",
              )
              
print(f"Free electron density = {ne:.3e} cm^-3")
print(f"Total particle density = {n:.3e} cm^-3")
print(f"Total ion density = {ni:.3e} cm^-3")

mean_mass = Ionization.interpolate_mu

mu = mean_mass(nH = nH,
               temperature = temperature,
               metallicity = metallicity,
               redshift = redshift,
               mode = mode,
               part_type = "all",
               )
print(f"Mean particle mass = {mu:.2f} mp")
\end{mintedbox}
\alankar{
Presently, {\tt AstroPlasma} only supports purely collisional and photo+collisional ionization equilibrium (CIE/PIE) conditions, 
using the HM12 photo-ionizing extragalactic background radiation 
from \citet{Haardt2012}. In the future, we plan to incorporate effects of different photo-ionizing background radiation (like \citealt{10.1093/mnras/stz174, 10.1093/mnras/staa302}), and also non-equilibrium ionization 
in {\tt AstroPlasma}.} 

\subsection{Spectral energy distribution (SED) generation}
{\tt AstroPlasma} can also calculate the spectral energy distribution (SED) of the emitted radiation from a one-zone plasma. In order to generate the SED of the whole CGM, we add up emissions from multiple such one-zones weighed by their volume. The plasma is assumed to be optically thin, and we do not account for radiative transfer. Since no real-time plasma calculation is involved in spectrum generation, using {\tt AstroPlasma} is relatively fast compared to on-the-fly calculation of chemical networks.
\begin{mintedbox}{python}
from astro_plasma import EmissionSpectrum

gen_spectrum = EmissionSpectrum.interpolate_spectrum

spectrum = gen_spectrum(nH = nH,
               temperature = temperature,
               metallicity = metallicity,
               redshift = redshift,
               mode = mode
               ) # Generate spectrum
               
import matplotlib.pyplot as plt

plt.loglog(spectrum[:,0], spectrum[:,1])
plt.xlabel(r"Energy (keV)")
plt.ylabel(r"Emissivity $4 \pi \nu j_{\nu}$ ($erg\ cm^{-3} s^{-1}$)")
plt.xlim(xmin = 1.0e-10, xmax=3.2)
plt.show()
\end{mintedbox}

\section{power-law density profiles and column densities}
\label{app:powerlaw}
The three-phase, one-zone model of the CGM introduced in section \ref{sec:threePhase} assumes a constant density for all the CGM gas. However, this assumption can be relaxed easily if we assume the plasma distribution as a function of radius has a gradual variation. Here, we outline the method to include a radially symmetric power-law density profile for the CGM gas. 

We keep the CGM mass fixed and spread it with a power-law density profile. This allows us to write the radial number density of the CGM gas $n(r)$ as 
\begin{equation}
    \label{eq:n_pl_prof}
    n(r) = n_0 \left(1-\frac{\alpha}{3}\right) \frac{1-\left(\frac{r_0}{r_{\rm CGM}}\right)^3}{1-\left(\frac{r_0}{r_{\rm CGM}}\right)^{3-\alpha}} \left(\frac{r}{r_{\rm CGM}}\right)^{-\alpha},
\end{equation}
where $n_0$ is the constant number density of the CGM gas in the one-zone model, $r_0$ and $r_{\rm CGM}$ are the inner and outer radii of the CGM respectively, and $\alpha$ is the power-law index, i.e., $n(r) \propto r^{-\alpha}$. \alankar{We stick to the \citetalias{Faerman2017ApJ} prescription for $r_{\rm CGM}$, i.e., $r_{\rm CGM} = 1.1\ r_{\rm 200}$.}

Assuming that the ion densities also follow power-law profiles (\alankar{possibly} different power-law indices for different phases), it is straightforward to calculate the column density $N(b)$ of any ion using the following integral at any given impact parameter $b$,
\begin{equation}
    \label{eq:N_col_pl_prof_int}
    N(b) = 2 \bigintsss _{b} ^{r_{\rm CGM}} dr \frac{r n(r)}{\sqrt{r^2-b^2}}.
\end{equation}
Substituting Eq. \ref{eq:n_pl_prof} into Eq.  \ref{eq:N_col_pl_prof_int}, we get the following expression containing transcendental functions,
\begin{eqnarray}
    \label{eq:N_col_pl_prof}
    \nonumber 
         N(b) = n_0 r_{\rm CGM} \left(1-\frac{\alpha}{3}\right) \frac{1-\left(\frac{r_0}{r_{\rm CGM}}\right)^3}{1-\left(\frac{r_0}{r_{\rm CGM}}\right)^{3-\alpha}} \Gamma \left(\frac{\alpha-1}{2}\right) \times \\
         \left [\frac{\sqrt{\pi}}{\left(\frac{b}{r_{\rm CGM}}\right)^{\alpha -1} \Gamma \left(\frac{\alpha}{2}\right)} - {}_{2}\Tilde{F}_{1} \left(\frac{1}{2}, \frac{\alpha-1}{2};\frac{\alpha + 1}{2};\left(\frac{b}{r_{\rm CGM}}\right)^2\right)\right ],
\end{eqnarray}
where $\Gamma$ denotes gamma function and ${}_{2}\Tilde{F}_{1}$ is the {\it Regularized} hypergeometric function.\footnote{\alankar{{\fontsize{7pt}{7.2pt}\selectfont 
For reference, see \href{https://docs.scipy.org/doc/scipy/reference/generated/scipy.special.hyp2f1.html}{{\texttt scipy} documentation} on Hypergeometric function and \href{https://mathworld.wolfram.com/RegularizedHypergeometricFunction.html}{Wolfram MathWorld} article on Regularized hypergeometric function.}}}

The density \alankar{profile} introduced in Eq. \ref{eq:n_pl_prof} can also be thought of as a model for the Milky Way CGM. \alankar{Just like Eq. \ref{eq:N_col_pl_prof_int}, which 
applies for an external galaxy, we make equivalent estimates for Milky Way observables from the solar system location. These} observables include the emission and dispersion measures 
\alankar{generated from our model, which} can then be directly compared with Milky Way observations along any 
line of sight.

\begin{figure}
\centering
\includegraphics[width=1.0\columnwidth]{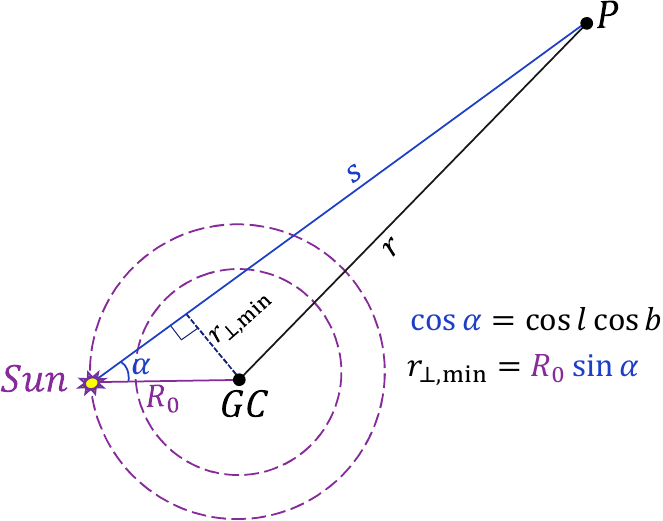}
\caption{\alankar{The relevant geometry to convert the LOS integral (along $s$, distance along LOS) to an integral involving $r$ (distance from Galactic center). 
For any Galactic coordinates ($l,b$), the angle $\alpha$ satisfies $\cos \alpha = \cos l \cos b$. 
}}
\label{fig:gal_coord}
\end{figure}

We denote $s$ as the distance between an observer at the position of the solar system and any point along a sightline in Galactic coordinates $(l, b)$. This same point has spherical coordinates $(r, \theta, \phi)$ with respect to the Galactic center. Converting between spherical and Galactic coordinates, $s\ \sin b = r \ \cos \theta $, and using the law of sines for a triangle 
(see Fig. \ref{fig:gal_coord} \& 
Fig. A1 in \citealt{Sarkar2015}), we can write
\begin{equation}
    \label{eq:triangles}
    \frac{R_0}{\sin (\phi - l)} = \frac{s\ \cos b}{\sin \phi } = \frac{r\ \sin \theta }{\sin l},
\end{equation}
where $R_0$ is the distance between the sun and the Galactic center ($\approx 8\ {\rm kpc}$). Eq. \ref{eq:triangles} combined with
\begin{equation}
    \label{eq:r_s_l_b}
    r^2 = s^2 + R_0^2 - 2 R_0 s \cos b \cos l,
\end{equation}
where $R_0$ is the distance between the sun and the Galactic center ($\approx 8\ {\rm kpc}$), 
can let us express $s$, $\theta$ and $\phi$ in terms of $l, b$, $r$ and $R_0$.

\alankar{Let the function} $f[n(r)]$ (\alankar{which implicitly depends on its 
distance from the Galactic center}) be any observable from our \ps{spherically symmetric} model integrated along a line of sight, e.g., $n_e(r) n_H(r)$ for emission measure or $n_e(r)$ for dispersion measure. Using Eq. \ref{eq:r_s_l_b}, we can write the line of sight integral as,\alankar{\footnote{\alankar{{\fontsize{7pt}{7.2pt}\selectfont We use Simpson's 
rule with $r$ uniformly sampled in log-space to numerically evaluate Eq. \ref{eq:integral_MW3phase}.}}}}
\begin{eqnarray}
 \label{eq:integral_MW3phase}
   && \mathcal{I}(l,b) = \bigintsss_{R_0}^{r_{\rm CGM}} dr \frac{rf[n(r)]}{\sqrt{r^2-r_{\rm \perp, min}^2}} \\
   \nonumber
  && + \begin{cases}
        2\bigintss_{r_{\rm \perp, min}}^{R_0}  dr \frac{r f[n(r)]}{\sqrt{r^2-r_{\rm \perp, min}^2}}; & l \in [0^{\circ}, 90^{\circ})\ \cup \ (270^{\circ}, 360^{\circ})\\
        0; &  l \in [90^{\circ}, 270^{\circ}]
    \end{cases}
\end{eqnarray}
where $r_{\rm \perp, min} = R_0 \sqrt{1-\cos ^2 l\ \cos ^2 b}$ is the minimum perpendicular distance of a sightline along $(l, b)$ from the Galactic center (see Fig. \ref{fig:gal_coord}). Note that the sightlines towards the Galactic center (first and fourth quadrants) have an additional contribution from smaller $r$. 
Eq. \ref{eq:integral_MW3phase} can be numerically integrated to estimate the emission and dispersion measures or surface brightness for different CGM profiles;\footnote{{\fontsize{7pt}{7.2pt}\selectfont \alankar{The terms (including the factor $2$ in the second term) in Eq. \ref{eq:integral_MW3phase} are only correct for a spherically symmetric profile. Otherwise, for any chosen $(l,b)$ sightline, the functional relations $\theta(r)$ and $\phi(r)$ using Eq. \ref{eq:triangles} need to be explicitly considered.}}} e.g., a power-law (with index $\alpha$) profile for our one-zone, three-phase model. A comparison of this model with observations is compiled in Tab. \ref{tab:comparison_summary}.

\section{Ionization \& Cooling function}
\label{app:ion_cooling}
\alankar{

\begin{figure}
    \includegraphics[width=1.12\columnwidth, center]{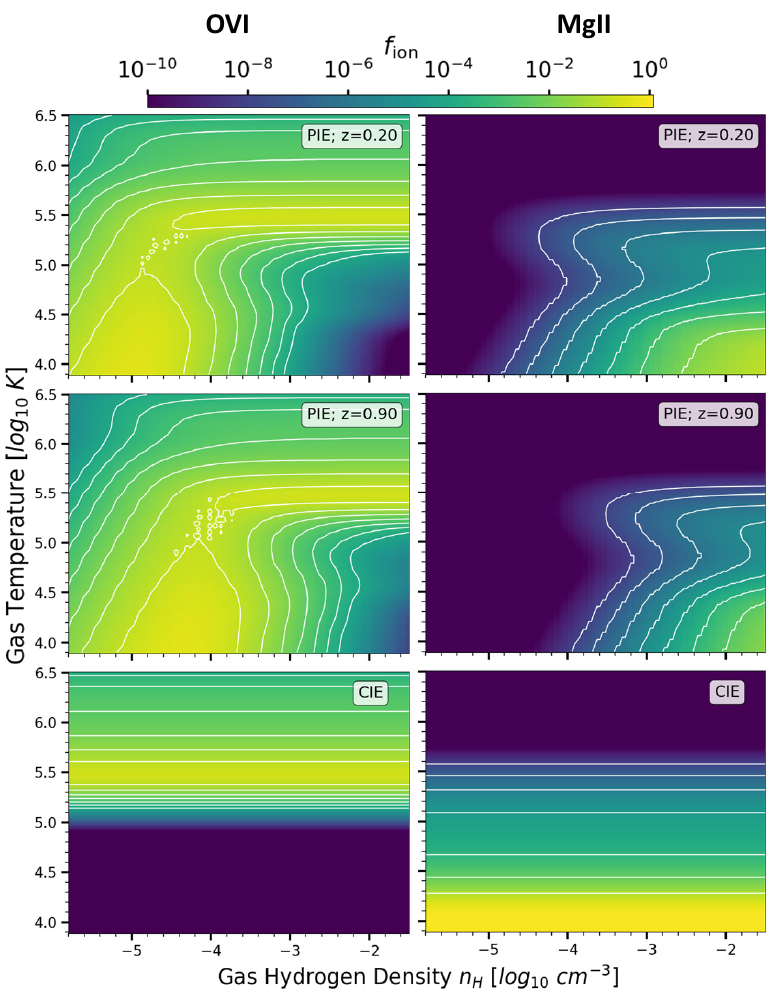}
	\caption{\alankar{The 
 ion fraction of $\rm OVI$ (left panels) and $\rm MgII$ (right panels) as a function of hydrogen number density $n_H$ and gas temperature generated using {\tt AstroPlasma}. The bottom row shows equilibrium values with only collisional ionization (CIE) while the top two rows show equilibrium 
 values with photo+collisional ionization (PIE) in  Haardt-Madau extragalactic UV background (\citealt{Haardt2012}) at 
 $z =$ 0.2 and 0.9. 
 The strength of the UV background increases with 
 redshift.}}
	\label{fig:ionization-fracs_combined}
\end{figure}

Here we discuss the effects of photo+collisional ionization on different ion fractions, particularly $\rm OVI$ and $\rm MgII$.
We assume the Haardt-Madau extragalactic UV photo-ionizing background (\citealt{Haardt2012}) at a 
redshift of 0.2. 
As illustrated in Fig. \ref{fig:ionization-fracs_combined}, in the absence of any photo-ionizing background radiation, the ionization fractions of both $\rm OVI$ and $\rm MgII$ are independent of gas density. The ionization fraction of $\rm OVI$ and $\rm MgII$ are oppositely affected by the 
photo-ionizing background. $\rm OVI$ is an intermediate 
ion (tracing warm gas $\sim 10^{5.5} \ {\rm K}$) and its ionization fraction is significantly enhanced at temperatures $\lesssim 10^{5.5}\ {\rm K}$ by the photo-ionizing background. On the other hand, since photo-ionizing background increases the overall ionization of the plasma, it depletes the amount of low 
ions like $\rm MgII$ (tracing cold gas $\sim 10^4 \ {\rm K}$), especially at low densities $n_H\lesssim10^{-4}\ {\rm cm^{-3}}$. Further, the strength of the ionizing background radiation increases with redshift, and its effect is also illustrated in Fig. \ref{fig:ionization-fracs_combined} (see \citealt{10.1093/mnras/stac3567} and Fig. 5 from \citealt{Faerman2019ApJ} for more discussion on this).

\begin{figure}
	\includegraphics[width=0.99\columnwidth]{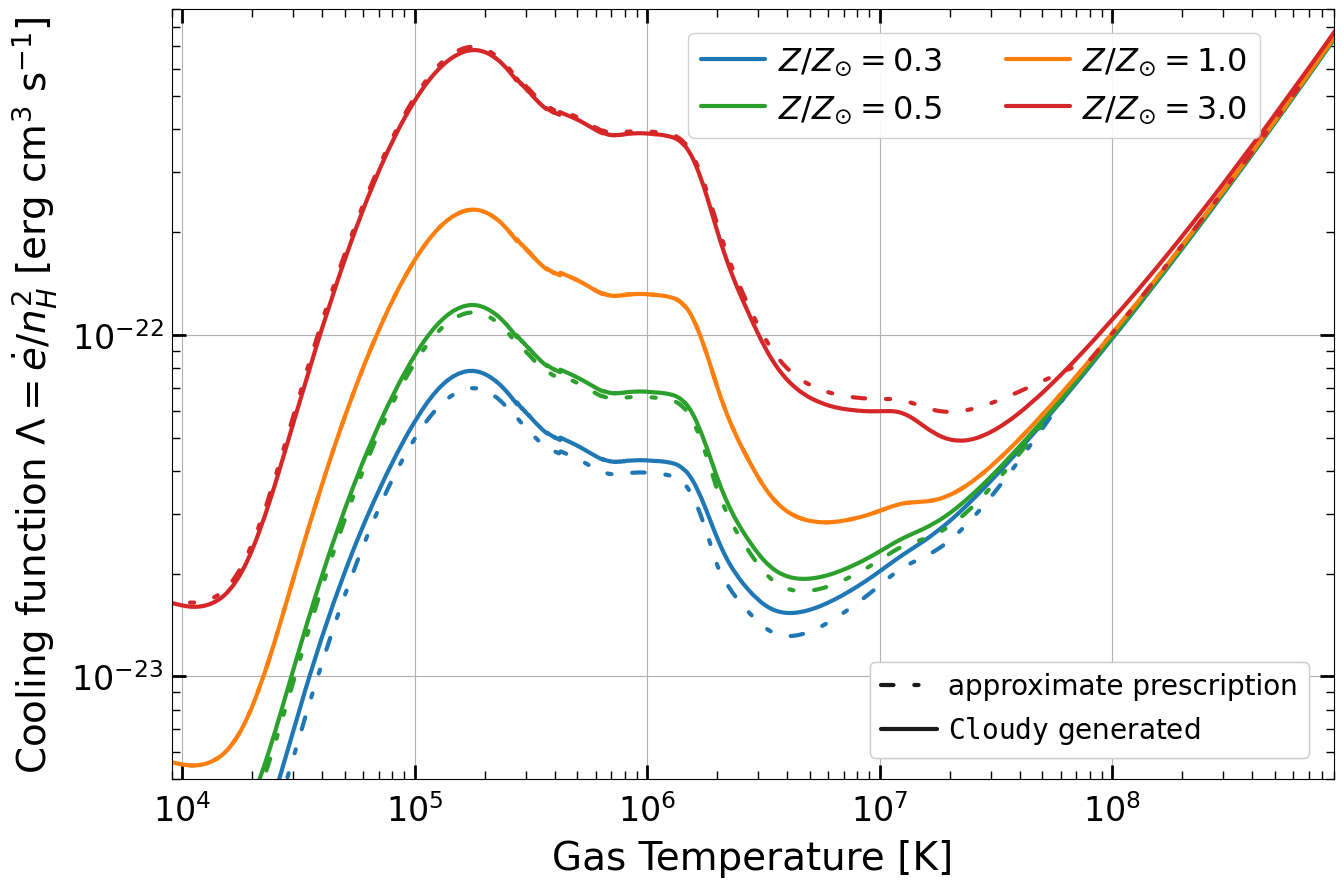}
    \includegraphics[width=0.99\columnwidth]{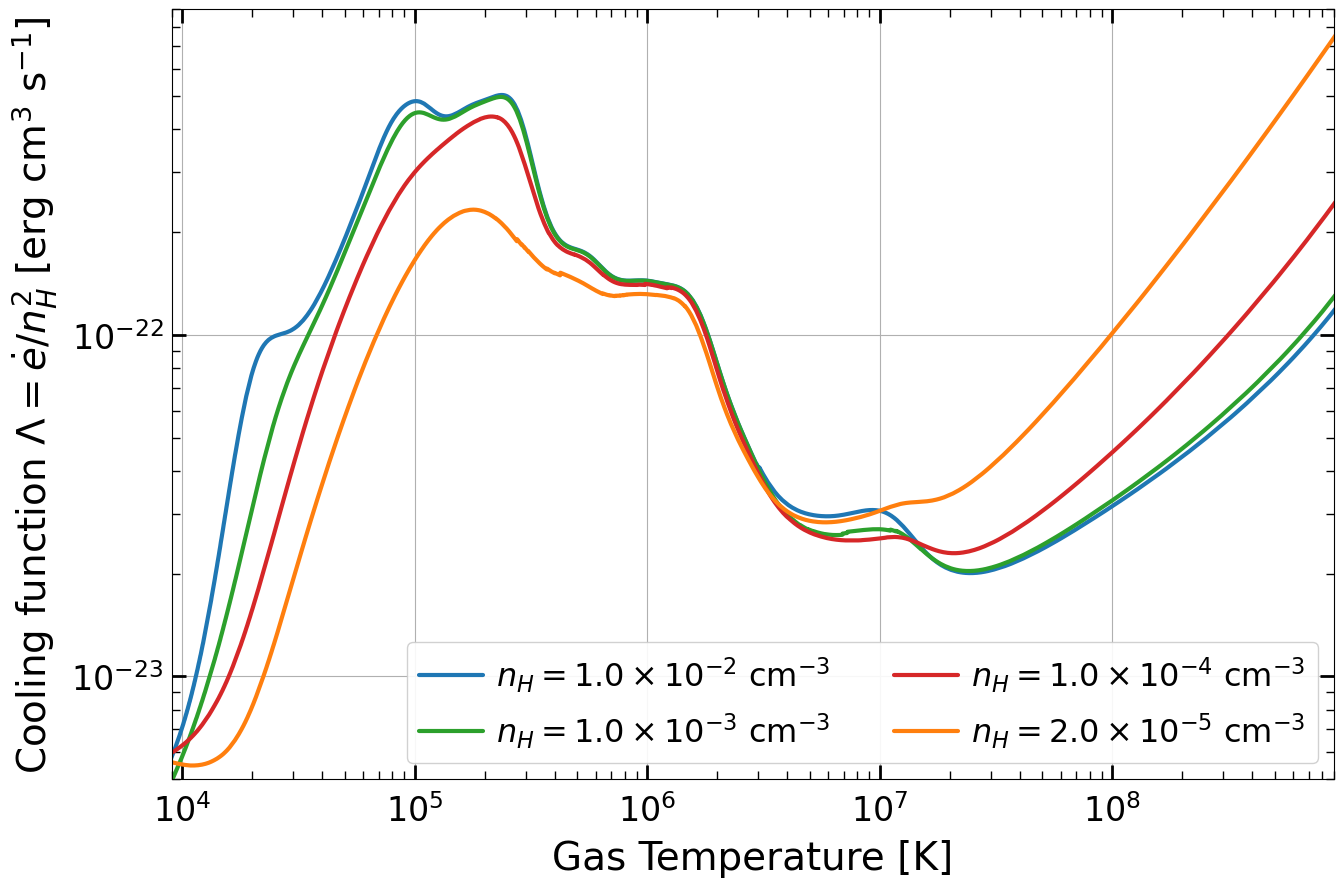}
	\caption{\alankar{
 Equilibrium cooling functions generated using {\tt CLOUDY 2017} spectral synthesis code (\citealt{2017RMxAA..53..385F}). Different cases include collisional+photo-ionization in the presence of Haardt-Madau extragalactic UV 
 background (\citealt{Haardt2012}) at $z = 0.2$. \textit{Top panel}: Demonstration of the robustness of our approximate prescription (Eq. \ref{eq:cool_app}) that scales the cooling function for different metallicities.
 The approximate cooling functions (\textit{dot-dashed lines}) closely follow the actual {\tt CLOUDY} generated cooling curve (\textit{solid lines}). 
 \textit{Bottom panel} shows the equilibrium cooling functions at different hydrogen number densities. Due to the presence of the photo-ionizing background, the cooling functions become weakly dependent on the gas density. 
 For both panels and across this work, we use the cooling function fixed to the average density of our fiducial TNG50 halo, $n_H = 2.0\times 10^{-5}\ {\rm cm^{-3}}$. 
 }}
	\label{fig:cool_curves}
\end{figure}


Now we show the {\tt CLOUDY} (\citealt{2017RMxAA..53..385F}) generated equilibrium cooling function used in this work. 
For 
fast computation, we use the {\tt CLOUDY} generated equilibrium cooling table for plasma at solar metallicity and use an approximate prescription, described in Eq. \ref{eq:cool_app}, to scale the cooling value to different metallicities. The robustness of this approximation (\ref{eq:cool_app}) is illustrated in the top panel of Fig. \ref{fig:cool_curves}.  
The approximate 
cooling 
function at different metallicities (for a fixed density) scaling from 
solar metallicity is given by
\begin{equation}
    \label{eq:cool_app}
    \Lambda \left(T, \frac{Z}{Z_{\odot}}\right) = 
           \left[\mathcal{F}\left(T\right) + 
           \left\{1-\mathcal{F}\left(T\right)\right\}\left(\frac{Z}{Z_{\odot}}\right)\right]
           \; \Lambda\left(T, \frac{Z}{Z_{\odot}}=1\right),
\end{equation}
where
\begin{equation}
    \nonumber
    \mathcal{F}\left(T\right) = 
    \begin{cases}
        0; & T < 2\times 10^{6}\ {\rm K}\\
        \frac{\log_{10}\left(T\right)-\log_{10}\left(2\times 10^6\right)}{\log_{10}\left(8\times 10^7\right)-\log_{10}\left(2\times 10^6\right)}; &   2\times 10^{6}\ {\rm K} \leq T \leq 8\times 10^{7}\ {\rm K} \\
        1; & T > 8\times 10^{7}\ {\rm K}
    \end{cases}.
\end{equation}

Because of 
photo-ionization, the cooling function can be significantly affected by a change in the density of the gas, as illustrated in the bottom panel of Fig. \ref{fig:cool_curves}. For all the cooling curves used in our work, we set 
$n_H = 2.0\times 10^{-5}\ {\rm cm^{-3}}$, roughly the average density of our TNG50 halo (see Fig. \ref{fig:2dPDFs}). 
}

\bsp	
\label{lastpage}
\end{document}